\documentclass[a4paper,12pt]{amsart}
\usepackage{amsthm}
\usepackage{amsmath}
\usepackage[authoryear]{natbib}
\usepackage[colorlinks,citecolor=blue,urlcolor=blue,filecolor=blue,backref=page]{hyperref}
\usepackage{graphicx}

\usepackage{amssymb}
\usepackage{bm}
\usepackage{enumitem}
\usepackage{xcolor}
\usepackage[ruled,vlined]{algorithm2e}
\usepackage[figuresright]{rotating}
\usepackage{pdflscape}

\theoremstyle{plain}
\newtheorem{defn}{Definition}[section]
\newtheorem{thm}{Theorem}[section]
\newtheorem{prop}{Proposition}[section]

\newcommand{\E}{\mbox{\textup{E}}}
\newcommand{\mb}[1]{\mathbf{#1}}
\newcommand{\tr}[1]{#1^{\intercal}}
\newcommand{\norm}[1]{\lVert#1\rVert}
\DeclareMathOperator*{\argmax}{arg\,max}
\DeclareMathOperator*{\argmin}{arg\,min}

\newcommand{\by}{{\mb{y}}}
\newcommand{\bd}{{\mb{d}}}
\newcommand{\bx}{{\mb{x}}}
\newcommand{\bX}{{\mb{X}}}

\newcommand{\bw}{{\mb{w}}}
\newcommand{\bv}{{\mb{v}}}
\newcommand{\balpha}{{\bm{\alpha}}}
\newcommand{\bbeta}{{\bm{\beta}}}
\newcommand{\bgamma}{{\bm{\gamma}}}
\newcommand{\bdelta}{{\bm{\delta}}}
\newcommand{\btheta}{{\bm{\theta}}}

\newcommand{\bthetaeb}{\btheta^{\textsc{eb}}}
\newcommand{\bthetaep}{\btheta^{\textsc{ep}}}

\def\bSig\mathbf{\Sigma}

\usepackage{fancyhdr}
\begin{document}

\title{Inference for multiple treatment effects  using confounder importance learning}

\author{Omiros Papaspiliopoulos $^1$}
\thanks{1: Bocconi University, omiros@unibocconi.it}
\author{David Rossell $^2$}
\thanks{2: Pompeu Fabra University, david.rossell@upf.edu}
\author{Miquel Torrens-Dinar\`es $^3$}
\thanks{3: Centre for Genomic Regulation, miquel.torrens@crg.eu}


\maketitle

\begin{abstract}

We address modelling and computational issues for multiple treatment effect inference under many potential confounders. 
Our main contribution is providing a trade-off between preventing the omission of relevant confounders, while not running into an over-selection of instruments that significantly inflates variance. 
We propose a novel empirical Bayes framework for Bayesian model averaging that learns from data the  prior inclusion probabilities of key covariates.  
Our framework sets a  data-dependent  prior that asymptotically matches the true amount of confounding in the data, as measured by a novel confounding coefficient. 
A key challenge is computational. We develop fast algorithms,  using an exact gradient of the marginal likelihood that has linear cost in the number of covariates, and a variational counterpart. 
Our framework uses widely-used ingredients and largely existing software,  and it is implemented within  the R package mombf. We illustrate our work with two applications. The first  is the association between  salary variation and discriminatory factors.   The second, that has been debated in previous works, is the association between abortion policies and crime. Our approach provides insights that differ from previous analyses especially in situations with weaker treatment effects. 

\end{abstract}


\noindent%
\vfill

\section{Introduction} \label{sec:intro}

We consider a fundamental problem in applied research, that of evaluating the joint  association, if any, of multiple treatments on an outcome when working with observational data and when there are many potential adjustment covariates. 
In such settings, it is common to use generalized linear models (GLMs) or additive models.  
Our discussion also applies to causal inference, whereby relying on the so-called non-interference and no unmeasured confounding assumptions, one may identify causal treatment effects from a regression model, provided one selects the necessary covariates  and models properly their association with the outcome  (see \cite{antonelli_dominici:2021} for a review).
Following standard terminology, we refer by \textit{confounders} to covariates that are truly associated with both treatment(s) and the response (given other covariates), and by \textit{instruments} to covariates that correlate with the treatment(s) but are conditionally independent of the outcome.
 A common approach to  estimate conditional associations  is to learn which covariates are confounders 
using high-dimensional regression, and we pursue this direction in this article.
 As we shall discuss, a key issue that we address is attaining a good trade-off between avoiding omitted variable biases and variance inflation driven by instruments. We achieve this by setting a data-dependent prior via empirical Bayes, and proposing highly efficient algorithms to estimate the required hyper-parameters.  We show that asymptotically the hyper-parameter estimates adapt to the true sparsity in the data, and capture a novel measure of confounding that we introduce in this article. 

We model the dependence of the outcome $y_{i} \sim p(y_{i}; \eta_{i}, \phi)$ on $t=1,\ldots,T$ treatments $d_{i,t}$ and $j=1,\ldots,J$ covariates $x_{i,j}$, via 
\begin{align}
\eta_{i} = \sum_{t=1}^{T} \alpha_{t} d_{i,t} + \sum_{j=1}^{J} \beta_{j} x_{i,j}, i=1,\ldots, n \label{eq:y_eq}
\end{align}
where $p(y_i; \eta_i, \phi)$ defines a GLM with linear predictor $\eta_i$ and dispersion parameter $\phi$ (i.e. the error variance in the Gaussian case, and a known $\phi = 1$ in logistic and Poisson regression). 
Whereas from an interpretational and policy making point of view the distinction between treatments and covariates is clear, statistically the difference is one of priorities: we are primarily interested in inference for treatment effects ($\alpha_t$'s in \eqref{eq:y_eq}), including uncertainty quantification, 
whereas the $\beta_j$'s are considered to avoid omitted variable biases and to allow for flexible regression functions. 
Although our primary interest is in average treatment effects, it is possible to consider heterogeneous effects by incorporating into $d_{i,t}$ interactions between treatments and covariates. In our salary example we illustrate this by considering interactions between the four primary treatments and state. Importantly, such interactions are added with an add-to-zero constraint. This ensures that the $\alpha_t$ associated to a primary treatment quantifies its corresponding average treatment effect, whereas the $\alpha_t$'s associated to interactions quantify deviations from the average treatment effect.

Our main interest is in scenarios where the number of covariates $J$ is large.
This setting spurred significant interest due to the observation that standard shrinkage and selection 
methods for learning \eqref{eq:y_eq}, such as LASSO and Bayesian Model Averaging (BMA), can have an undesirable behavior for treatment effect inference. 
When many confounders are strongly associated with the treatments,  a situation that we refer to as high-confounding,  standard high-dimensional methods may fail to include said confounders (or even the treatments) in \eqref{eq:y_eq}, resulting in significant omitted variable biases. 
Two seminal works are \cite{Belloni14b} and \cite{Wang12}. 
Both set the basis for subsequent literature, and both consider a single treatment setting ($T=1$).
\cite{Belloni14b} proposed a double-LASSO (DL) approach where one regresses separately the outcome and the treatment on the covariates via the LASSO, 
takes the covariates with a non-zero estimated effect either on the treatment or the outcome, and in a second step  fits a model like \eqref{eq:y_eq} by maximum likelihood estimation (MLE) with these selected covariates. Notably, this treatment effect estimator is asymptotically normal and has a variance that can be estimated from data.
In a similar spirit \cite{Wang12} proposed Bayesian adjustment for confounders (BAC), which models jointly the outcome and treatments and uses a prior distribution that encourages covariates to be simultaneously selected in the two regression models.


The main idea in DL, BAC and subsequent literature (reviewed below) is that, by including covariates that are associated to the treatment, one ameliorates omitted variable biases.
 A related notion called regularization-induced confounding (RIC) refers to estimation biases due to not properly accounting for confounders, due to the prior over-shrinking in specific directions \citep{Hahn18,linero:2023}.
This notion is related to the omitted variable bias discussed by \cite{Belloni14b}, i.e. the concern is not properly handling the confounders, and is addressed by linking the outcome and treatment models. 

A key distinction motivating our work is that, by protecting oneself against omitted variables, one may force (or encourage) the inclusion of instruments, i.e. covariates for which truly $\beta_j=0$ in \eqref{eq:y_eq}.
Under such covariate over-selection treatment effects remain identifiable, however there is a problematic \emph{variance inflation}, see \cite{de2011covariate,Lefebvre14,zigler:2014,Talbot15,henckel2022graphical}.
 We argue that one should try to reach a compromise between handling properly the confounders and the instruments. 
Adding instruments can severely inflate the treatment effect mean squared error (MSE), and reduce the power to detect weaker effects. 
To gain intuition, consider a setting with a fixed number of covariates $J$. A classical strategy is to fit one model including all covariates to obtain unbiased treatment effects, potentially at the cost of high variance. Specifically, the variance inflation factor for a least-squares estimator of $\alpha_t$ is given by $(1 - R_t^2)^{-1}$, where $R_t$ is the multiple $R^2$ coefficient for regressing treatment $t$ on the covariates.
 Hence, if one has instruments that accurately predict the treatment, $R_t^2$ is close to 1 and variance inflation is severe. 
\cite{Belloni14b} explain that, when $J$ is fixed, their approach is asymptotically first-order equivalent to fitting a model with all covariates, and hence incurs variance inflation.
A second issue is a more subtle \emph{over-selection bias} that received less attention  (but see \cite{zigler:2014} for a brief mention).  Namely, including covariates in \eqref{eq:y_eq} that are correlated with the treatments and the outcome may lead to biased inference.
In our experience over-selection bias is not a major issue in practice, further the results in \cite{Belloni14b} prove that it vanishes asymptotically, hence we defer further discussion to Section S1.

Related literature includes \cite{Farrell15}, who adapted the DL framework by using a robust estimator to safeguard from mistakes in the double selection step, and \cite{Shortreed17}, who employed a two-step adaptive LASSO approach.
\cite{Victor18} extended DL by introducing a de-biasing step, and cross-fitting to ameliorate false positive inclusion of covariates.
On the Bayesian side,
\cite{Lefebvre14} discussed how to set the BAC hyper-parameter $\omega$ in a data-based manner to improve the treatment effect MSE.
When $\omega= \infty$, the outcome equation includes any covariate associated with the treatment, which akin to DL reduces omitted variable bias at the cost of potential variance inflation.
The authors warn that the results are sensitive to using half of the data in their sample-splitting strategy, and of computational challenges if one wanted to consider $T>1$ treatments, further they use a leaps-and-bounds model search that only accommodates up to 31 covariates.
\cite{Wang15} extended BAC to GLMs and considered pairwise interactions between the treatment and covariates. The authors used the same prior as BAC and focused on the hyper-parameter choice $\omega= \infty$, which as discussed can be problematic.
\cite{Talbot15} propose a 
similar framework to BAC where prior probabilities deter the inclusion of instruments to reduce the inclusion of instruments, however said prior probabilities still require a tuning hyper-parameter playing a role similar to $\omega$ in BAC.
A proposal  that is closest to ours  is the ACPME method of \cite{Wilson18}.
The framework considers $T>1$ treatments and the prior inclusion probability for covariate $j$ depends, via logistic regression, on a measure of dependence between $j$ and the treatments. 
Analogously to BAC, prior inclusion probabilities are controlled by a tuning parameter, which by default sets the same average penalty for covariate inclusion as the Bayesian information criterion.
 A key difference is that \cite{Wilson18} do not use the outcome data to drive prior inclusion probabilities. They assume that any control associated to the treatment(s) is likely to be needed in the outcome equation, to an extent driven by a user-defined hyper-parameter. In low-confounding settings where there are many instruments, this assumption is violated. Instead, we use the outcome to set data-dependent prior inclusion probabilities, by learning whether there truly is high or low confounding (hence the naming {\it Confounding Importance Learning}). See Section \ref{ssec:model} for further comparisons with ACPME. 
\cite{Antonelli19} proposed continuous spike-and-slab Laplace priors on high-dimensional covariates. The framework is designed to reduce the shrinkage to zero for covariates that are associated to the treatment. They  discuss how to elicit hyper-parameters to help shrink the effects of instruments. 
In a different thread, \cite{Hahn18} proposed shrinkage priors based on re-parameterizing a joint outcome and treatment regression. 

\begin{figure}[htbp]
\centering
\begin{tabular}{ccc}
$\alpha = 1$ & $\alpha = 1/3$ & $\alpha = 0$ \\
\includegraphics[width=0.31\textwidth]{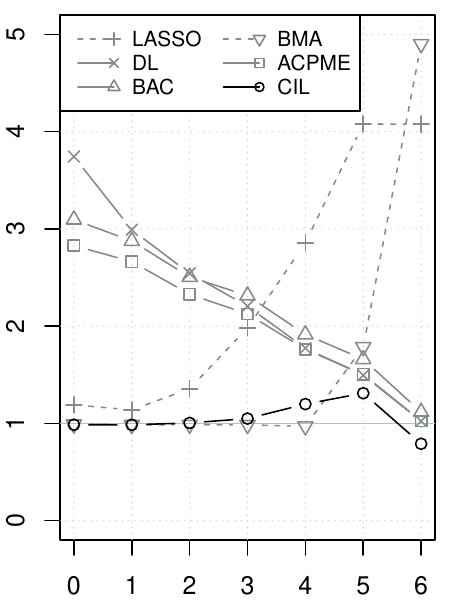} &
\includegraphics[width=0.31\textwidth]{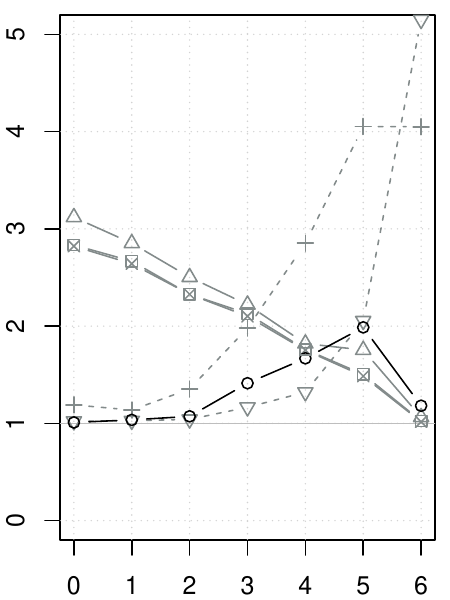} &
\includegraphics[width=0.31\textwidth]{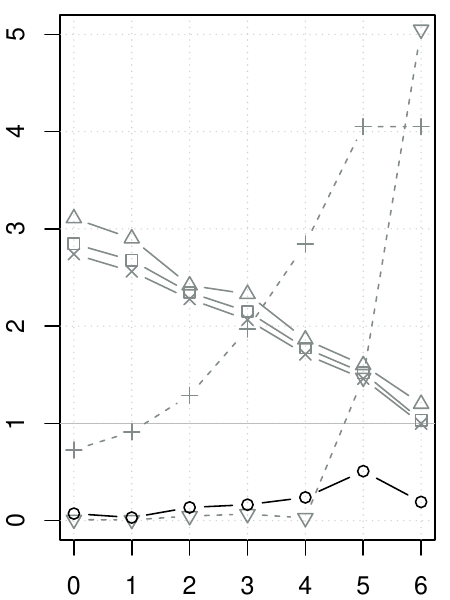} \\
\vspace{0.3cm}
\end{tabular}
\caption{Parameter root MSE relative to an oracle OLS, for a single treatment, considering strong ($\alpha=1$), weak $(\alpha=1/3)$ and  no effect $(\alpha=0)$. 
In all panels, $n=100$, $J=49$ and the outcome and treatment are simulated from a linear regression model based on 6 active covariates each. 
 The $x$-axis is the overlap between the two sets of active covariates varies from 0 (no confounding) to 6 (full confounding). 
DL is double LASSO, BMA is Bayesian model averaging, BAC is Bayesian Adjustment for Confounding and CIL is Confounder Importance Learning}
\label{fig:intro2}
\end{figure}

Overall, a recurrent issue 
is how to set hyper-parameters to avoid omitted-variable biases but also prevent 
variance inflation due to selecting instruments.
Our main contribution is a novel framework that sets
 (data-dependent)  prior inclusion probabilities to balance these two competing goals.
 We prove that our framework sets prior inclusion probabilities which, using empirical Bayes, asymptotically reflect a novel confounding coefficient introduced here. Said coefficient reflects whether one is in a situation with many confounders (high confounding), many instruments (low confounding), or neither (neutral confounding). 
Our framework, which we call Confounder Importance Learning (CIL),
is designed to deal with both over- and under-selection, in both high and low confounding situations. 
Figure \ref{fig:intro2} is a first illustration of its merits (see Section \ref{subsec:singleT} for details).
As discussed, 
due to omitted-variable bias standard LASSO and BMA suffer from high MSE in high-confounding settings, whereas DL and BAC attain much lower MSE. In low-confounding settings however the reverse is true, here DL and BAC have high MSE due to over-selection variance.
CIL attains low MSE across the high-to-low confounding spectrum.
CIL can also consider multiple treatments, a setting that has received less attention in the literature.
Although our model has similarities to ACPME, 
Figure \ref{fig:intro2} shows that the two methods behave quite differently, 
as ACPME closely mimics the behavior of BAC.
Relative to \cite{Lefebvre14}, we learn hyper-parameters using the marginal likelihood associated to a Bayesian model rather than a training-test data split,  which is integral in showing that our prior probabilities asymptotically match the (unknown) true confounding coefficient.  Another important contribution are two scalable computational algorithms, based on MCMC and on a variational approximation.
In principle evaluating the (log) marginal likelihood requires a costly sum over $2^{T+J}$ models. We show that, under our proposed prior model, its gradient only requires a sum over $J$ terms that involves only marginal posterior inclusion probabilities. We further propose an expectation-propagation (EP) approximation that bypasses the need to re-estimate said marginal posterior probabilities, which would typically require MCMC.
For example, BAC and ACPME failed to return a solution in our salary example after 2 days, whereas our CIL could complete the task in 8 hours and 33 minutes on the largest dataset of 2010.
CIL can be easily implemented using existing software, and we provide an implementation within the \texttt{R} package \texttt{mombf} \citep{Rossell20}.


This paper is structured as follows.
Section \ref{sec:method} details our proposed approach, a Bayesian model averaging where prior inclusion probabilities vary across covariates.
 It also introduces a confounding coefficient that plays a pivotal role in interpreting our methodology, and how it differs from current literature. 
Section \ref{sec:comput} describes our computational methods,  and how empirical Bayes seeks to match the prior mean of the confounding coefficient to its posterior mean.
Section \ref{sec:asymptotics} shows that the latter converges to the data-generating confounding coefficient, allowing for model misspecification, in finite-dimensional settings. 
Section \ref{sec:results} shows simulations, and a salary and a crime case study. 
All proofs and additional empirical results are provided as a supplement.

\section{Framework} \label{sec:method}

\subsection{Model}
\label{ssec:model}

We model the dependence of the outcome $y_{i}$ on treatments $\mb{d}_{i} = (d_{i,1}, \ldots, d_{i,T})$ and covariates $\mb{x}_{i} = (x_{i,1}, \dots, x_{i,J})$, according to \eqref{eq:y_eq}.
We are primarily interested in inference for the treatment effects $\balpha = (\alpha_{1}, \dots, \alpha_{T})$. 
We adopt a Bayesian framework where we   introduce variable inclusion indicators   $\gamma_{j} = \text{I}(\beta_{j} \neq 0)$ and $\delta_{t} = \text{I}(\alpha_{t} \neq 0)$, and define a prior 
\begin{align}
p(\balpha, \bbeta, \bdelta, \bgamma, \phi \mid \btheta) = p(\balpha, \bbeta \mid \bdelta, \bgamma, \phi) p(\bgamma \mid \btheta) p(\bdelta) p(\phi), \label{eq:probmodel}
\end{align}
where $\btheta $ are hyper-parameters discussed below, and $p(\phi)$ is dropped for models with known dispersion parameter (e.g. logistic or Poisson regression). For the regression coefficients, we assume prior independence,
\begin{align*}
p(\balpha, \bbeta \mid \bdelta, \bgamma, \phi) = \prod_{t=1}^{T} p(\alpha_{t} \mid \delta_{t}, \phi) \prod_{j=1}^{J} p(\beta_{j} \mid \gamma_{j}, \phi).
\end{align*}

We remark that much of the Bayesian treatment effects literature does not consider treatment inclusion indicators, rather they are forced into the model. While that strategy is easily accommodated in our framework by setting $\delta_t=1$ for all $t$, we consider that one often wishes to assess whether the treatment effects exist in the first place, and otherwise shrink their estimates towards zero.
Accordingly with this goal, we adopt the so-called product moment (pMOM) non-local prior of \cite{Johnson12}.
Briefly, non-local priors improve the rates at which one discards the truly zero parameters, see \cite{Johnson12, wu2016nonlocal, Shin18, Rossell21}.
Under the pMOM prior, one has $\alpha_{t} = 0$ almost surely if $\delta_{t} = 0$, and
\begin{align*}
p(\alpha_{t} \mid \delta_{t} = 1, \phi) = \frac{\alpha_{t}^{2}}{\phi \tau/ v_t} \text{N}(\alpha_{t}; 0, \phi \tau / v_t),
\end{align*}
with the analogous setting for every $\beta_{j}$. 
Figure S3 
illustrates its density.
Above $v_t$ is the sample variance of treatment $t$, to ease notation we assume that treatments and covariates have unit variance and take $v_t=1$.
The pMOM prior involves a prior dispersion parameter $\tau > 0$, that by default we set to $\tau = 1/3$ following \cite{Rossell20a}, which leads to a minimally informative prior akin to the unit information prior underlying the Bayesian information criterion.
As for the dispersion parameter, where unknown,  we also place a minimally informative $\phi \sim \text{IGam}(0.01, 0.01)$ prior.

For the inclusion indicators, we assume prior independence, and set
\begin{align}
p(\bdelta,\bgamma \mid \btheta) &= \prod_{t=1}^{T} \text{Bern}(\delta_{t}; 1/2) \prod_{j=1}^{J}  \text{Bern}(\gamma_{j}; \pi_{j}(\btheta)). \label{eq:incl} 
\end{align}
All treatments get a fixed marginal prior inclusion probability $P(\delta_t=1)=1/2$, as we do not want to favor their exclusion a priori, considering that there is at least some suspicion that any given treatment has an effect. 
This choice is a practical default when the number of treatments $T$ is not too large, else one may set $P(\delta_t=1)<1/2$ to avoid false positive inflation due to multiple hypothesis testing \citep{Scott10, Rossell21}.
Our software allows such possibilities. 

The main modelling novelty in this article is the choice of covariate prior inclusion probabilities $\pi_{j}(\btheta)= P(\beta_j \neq 0 \mid \btheta)$, 
where $\btheta = (\theta_{0}, \theta_{1}, \dots, \theta_{T})$ is a key prior hyper-parameter relating $\pi_j(\btheta)$ to (positive) measures of association ${\bf f}_j=(f_{j,1},\ldots,f_{j,T})$ between covariate $j$ and the $T$ treatments.
Specifically, prior probabilities are given by a logistic regression equation with success probability
\begin{align}
\tilde{\pi}_{j}(\btheta) = \left( 1 + \exp \left\{ - \theta_{0} - \sum_{t=1}^{T} \theta_{t} f_{j,t} \right\} \right)^{-1} \label{eq:cilprior}
\end{align}
truncated to lie in a pre-specified interval $[\underline{\rho},\bar{\rho}]$. That is,
\begin{align}
\pi_j(\btheta)= 
\begin{cases}
 \underline{\rho}  \mbox{, if } \tilde{\pi}_j(\btheta) \leq \bar{\rho}
\\
\tilde{\pi}_j(\btheta) \mbox{, if } \tilde{\pi}_j(\btheta) \in (\underline{\rho}, \bar{\rho})
\\
\bar{\rho} \mbox{, if } \tilde{\pi}_j(\btheta) \geq \bar{\rho}
\end{cases}.
\nonumber
\end{align}
 The truncation to $[\underline{\rho},\bar{\rho}]$ ensures that one does not include/exclude covariates a priori, and is required for the asymptotic properties discussed in Section \ref{sec:asymptotics}. 
We propose default $\underline{\rho} = 1/J$ and $\bar{\rho}=0.95$. The former allows enforcing sparsity, while ensuring that the prior expected model size is non-decreasing in $J$. The latter avoids assigning overly strong evidence a priori that a covariate is needed, which helps prevent covariate over-selection. See Section \ref{sec:asymptotics} for further discussion on $(\underline{\rho},\bar{\rho})$. 

\begin{figure}[h]
\centering
\includegraphics[width=0.6\textwidth,height=0.5\textwidth]{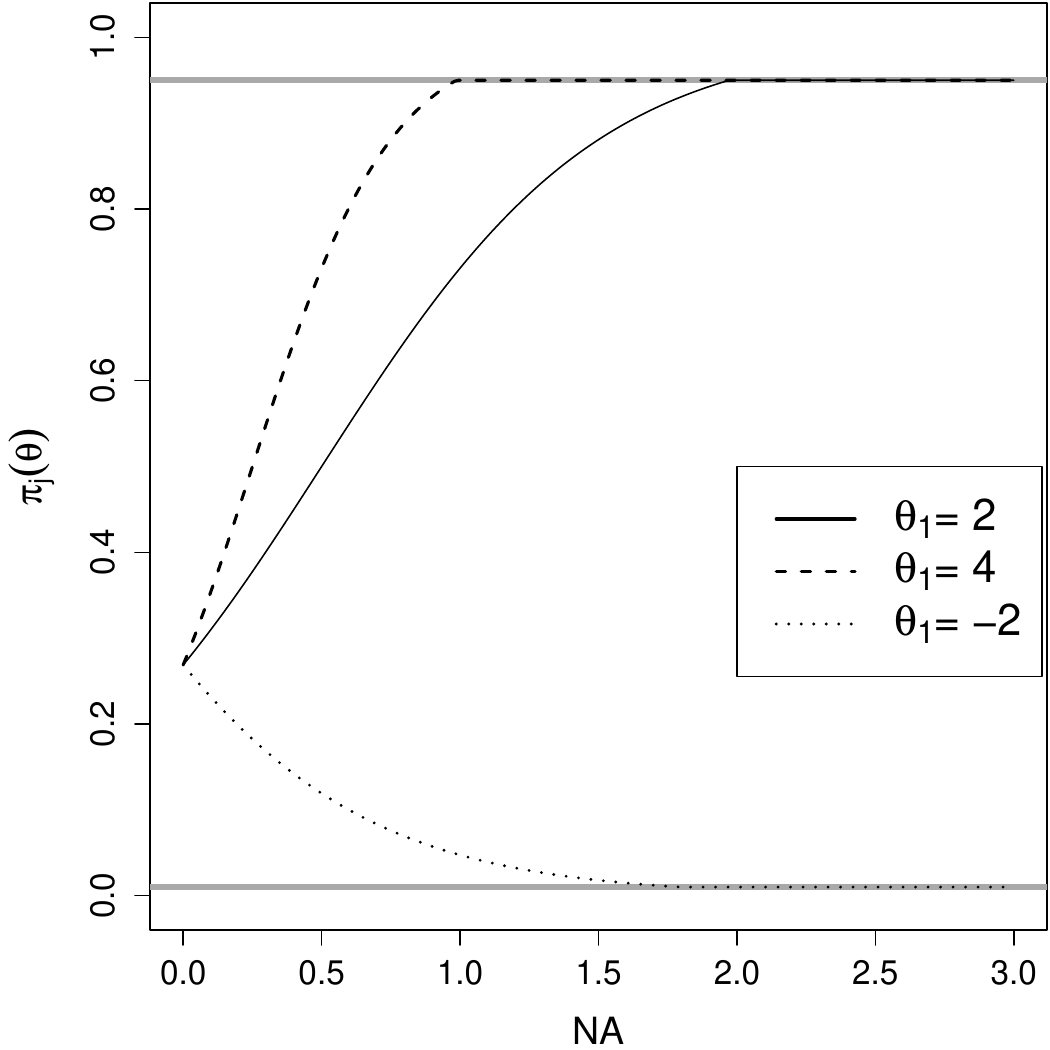} 
\caption{Prior inclusion probability \eqref{eq:cilprior} as a function of $f_{j,1}$, a feature measuring correlation between covariate $j$ and treatment $t=1$, for $\theta_{0}=-1$, and $J=99$ covariates. The grey lines show the lower and upper bounds, $\underline{\rho}=1/J$ and $\bar{\rho}=0.95$, respectively.}
\label{fig:theta1}
\end{figure}

Akin to DL, BAC and related methods, the idea is that if covariate $j$ is highly associated to treatment $t$ then $f_{j,t}$ will be large, and if $\theta_t>0$ then one favors the inclusion of such a covariate. 
In contrast, if $\theta_t=0$ then said inclusion is not encouraged, and if $\theta_t<0$ it is discouraged.
Figure \ref{fig:theta1} illustrates $\pi_{j}(\btheta)$ for three different values of $\theta_{1}$. Setting $\btheta$ is critical for the performance of our inferential paradigm, and in Section \ref{sec:comput} we introduce data-driven criteria and algorithms for its choice. 
Intuitively, in high-confounding scenarios where covariates associated to treatment $t$ are also associated to the outcome, one expects to learn $\theta_t>0$. In contrast, in low-confounding scenarios where most covariates associated to treatment $t$ are instruments, one expects to learn $\theta_t<0$.

Our generic approach is to take $f_{j,t} = |w_{j,t}|$, where $\bw_t = (w_{1,t},\ldots, w_{J,t})$ are regression coefficients obtained via a high-dimensional regression of $\bd_t$ on the covariates. 
The idea is that covariates with large $f_{j,t}$ are likely to be parents (in a generative directed graphical model that describes the whole system) of treatment $t$, and that including parents of the treatment ensures satisfying Pearl's backdoor criterion and hence identifying the treatment effects.
Although our framework allows the user to specify any suitable $f_{j,t}$, here we highlight two possible choices. First, a LASSO regression,
\begin{align}
\bw_{t} := \argmin_{(v_{t,1}, \dots, v_{t,J})} \left\{ \sum_{i=1}^{n} \log p\left( d_{i,t}; \sum_{j=1}^{J} x_{i,j} v_{t,j} \right) + \lambda \sum_{j=1}^{J} |v_{t,j}| \right\}, \label{eq:lasso}
\end{align}
where $\lambda > 0$ is a regularization parameter, which we set by minimizing the BIC (we obtained similar results when using cross-validation). 
The choice in \eqref{eq:lasso} balances speed with reasonable point estimate precision, and is the option that we used in all our examples.
A second option, available when dealing with continuous treatments, is to use the minimum norm ridge regression,
\begin{align}
\mb{w}_{t} =  \left( \tr{\bX} \bX \right)^{+} \bd_{t}, \label{eq:pseudoinv}
\end{align}
where $\left( \tr{\bX} \bX \right)^{+}$ is the Moore-Penrose pseudo-inverse, and $\bX$ the $n \times J$ design matrix. For $J<n$ this is the familiar OLS estimator, but \eqref{eq:pseudoinv} is also well-defined when $J>n$, and it has been recently investigated in terms of the so-called benign overfitting property in \cite{Bartlett20}. \cite{Wang16} showed that when $J>n$, \eqref{eq:pseudoinv} ranks the coefficients consistently under theoretical conditions slightly broader than the LASSO.
Therefore, one expects that all parents of treatment $t$ have larger values of $f_{j,t}$.
Similarly, by the screening property of the LASSO one expects that all parents of treatment $t$ have $f_{j,t}= |w_{j,t}| > 0$.
This is appealing in our context, since $\pi_j(\btheta)$ are mainly driven by the relative magnitudes of $f_{j,t}$, the prior inclusion probabilities are allowed to favor or discourage the parents of treatments (depending on whether $\theta_t>0$ or $\theta_t<0$). 

We remark that the ACPME framework of \cite{Wilson18} also pre-computes features relating treatments to covariates.
The main difference is that we  use the outcome data to  estimate $\btheta$, whereas in ACPME it is a fixed hyper-parameter.
By fixing $\btheta$, ACPME cannot adapt the prior behavior depending on whether there is low or high confounding (or neither), which is critical to prevent variance inflation.
 This occurs by design, if one does not use the outcome data, one cannot assess whether controls are confounders or instruments (i.e. associated or not with the outcome). 
In contrast CIL seeks to estimate $\theta_t>0$ when there is high confounding between treatment $t$ and covariates, as measured by our novel confounding coefficient (Section \ref{ssec:confounding_coef}).
Critically, CIL can also set $\theta_t=0$ or even $\theta_t<0$ when there is no confounding to help exclude instruments, a feature that is not provided by ACPME (nor other competing methods, to our knowledge).
The other main difference is that our features $f_{j,t}$ are obtained by regressing the treatments on the covariates, whereas in ACPME features are obtained by regressing the covariates on the treatments. Large $f_{j,t}$ suggests that covariate $j$ is a parent of treatment $t$, 
which is critical to interpret the CIL solution as setting prior probabilities that reflect the true value of the confounding coefficient (Section \ref{sec:asymptotics}).
Such an interpretation is not possible for ACPME.
In Section \ref{sec:results} we show comparisons with BAC, ACPME, and other methods.

\subsection{Confounding coefficient}
\label{ssec:confounding_coef}

For each treatment $t=1,\ldots,T$, we define a confounding coefficient $\kappa_t^*$ quantifying the extent to which the treatment is truly confounded with the $J$ covariates in \eqref{eq:y_eq}. We also define a sample-based counterpart $\kappa_t$.
Intuitively, large positive $\kappa_t^*$ indicates high confounding, i.e. that covariates truly associated with treatment $t$ are mostly confounders ($\beta_j \neq 0$ in \eqref{eq:y_eq}), whereas large negative $\kappa_t^*$ indicates no confounding, i.e. that said covariates are mostly instruments ($\beta_j=0$).
$\kappa_t^*=0$ indicates neutral confounding, i.e. being associated to the treatment is unrelated to being a confounder or instrument.

To provide the definition, we introduce two elements.
First, let $\beta_j^*$ be the Kullback-Leibler optimal parameter in \eqref{eq:y_eq} quantifying the effect of covariate $j$ on the outcome (if (1.1) is correctly specified, then $\beta_j^*$ is the true covariate effect).
Let $\bgamma^*= (\gamma_1^*, \ldots, \gamma_d^*)$, where $\gamma_j= \mbox{I}(\beta_j^* \neq 0)$, so that $|\bgamma^*|_0/J$ is the proportion of truly active covariates.
Second, let $f_{j,t}^*$ be a measure of the (unknown) true effect of covariate $j$ on treatment $t$.
For example, consider a generalized linear model where the mean of treatment $t$ is truly driven by $\sum_{j=1}^J w_{j,t}^* x_{i,j}$ and let $f_{j,t}^*=|w_{j,t}^*|$ (under model misspecification, $w_{j,t}^*$ can be defined as the Kullback-Leibler optimal parameter).
Another interesting example is to define $f_{j,t}^* = \mbox{I}(w_{j,t}^* \neq 0)$, an indicator for covariate $j$ truly having an effect on treatment $t$.
Without loss of generality, we assume that the vector $(f_{1,t}^*,\ldots,f_{j,t}^*)$ has zero mean and unit variance.

\begin{defn}
The confounding coefficient $\kappa_t^*$ is 
\begin{align}
\kappa_t^*= \frac{1}{J} \sum_{j=1}^J f_{j,t}^* \left( \gamma_j^* - \frac{|\bgamma^*|_0}{J} \right)
\label{eq:confounding_coef}
\end{align}
\end{defn}
Note that $\kappa_t^*/ \sqrt{V(\bgamma^*)}$ is the correlation between covariate effect indicators $\bgamma^*$ and treatment-covariate association features $(f_{1,t}^*, \ldots, f_{j,t}^*)$.
Hence, when $\kappa_t^*>0$, large $f_{j,t}^*$ (covariate $j$ is associated with treatment $t$) indicates that it is likely that $\gamma_j^*=1$ (covariate $j$ is a confounder).
In contrast, when $\kappa_t^*<0$ then it is more likely that $\gamma_j^*=0$ (covariate $j$ is an instrument).
Finally, we let $\kappa_t= \frac{1}{J} \sum_{j=1}^J f_{j,t}( \gamma_j - |\bgamma|_0/J)$ be the sample confounding coefficient, i.e. replacing $f_{j,t}^*$ by their estimates $f_{j,t}$ and acknowledging that $\bgamma^*$ is unknown.


\section{Computational Methodology} \label{sec:comput}

\subsection{Bayesian model averaging (BMA)} \label{ssec:bma}

All expressions in this section are conditional on the observed $(\bx_i,\bd_i)$, we drop them from the notation for simplicity.
Inference for our approach relies on posterior model probabilities
\begin{align}
\nonumber
p(\bgamma, \bdelta \mid \by, \btheta) \propto p(\by \mid \bgamma, \bdelta) p(\bgamma \mid \btheta) p(\bdelta),
\nonumber
\end{align}
where
\begin{align}
p(\by \mid \bgamma, \bdelta) = \int p(\by \mid \balpha, \bbeta, \phi, \bdelta, \bgamma) p(\balpha, \bbeta \mid \bdelta, \bgamma, \phi) p(\phi) \text{d}\balpha \text{d}\bbeta \text{d}\phi \label{eq:marglik2}
\end{align}
is the marginal likelihood of model $(\bgamma,\bdelta)$.
We set the hyper-parameter $\btheta$ to a point estimate $\hat{\btheta}$ described in Sections \ref{sec:ml}-\ref{sec:ep}.
Conditional on $\btheta$, our model prior $p(\bgamma \mid \btheta)$ is a product of independent Bernouilli's with asymmetric success probabilities defined by \eqref{eq:cilprior}. As a simple variation of standard BMA, one can exploit existing computational algorithms, which we outline next.


Outside particular cases such as Gaussian regression under Gaussian priors, \eqref{eq:marglik2} does not have a closed-form expression.
To estimate \eqref{eq:marglik2} under our pMOM prior we adopt the approximate Laplace approximations of \cite{Rossell20a},
see Section S6.1 
for an overview.
We then obtain point estimates using BMA,
\begin{align}
\hat{\balpha} := \sum_{\bgamma, \bdelta}  \text{E}(\balpha \mid \by, \bgamma, \bdelta) p(\bgamma, \bdelta \mid \by, \btheta), \label{eq:teffest}
\end{align}
and similarly we employ the BMA posterior density $p(\balpha \mid \by, \bgamma, \bdelta, \btheta)$ to provide posterior credible intervals.
To this end we use posterior samples from the pMOM posterior density  using a latent truncation representation described by \cite{Rossell17}. 
Expression \eqref{eq:teffest} is a sum across $2^{T+J}$ models, when it is unfeasible we use Markov Chain Monte Carlo to explore the posterior distribution $p(\bgamma,\bdelta \mid \by,\btheta)$,
see e.g. \cite{Clyde12} for a review. 

We used all the algorithms described above as implemented by the \texttt{cil} function in \texttt{R} package \texttt{mombf} \citep{Rossell20}.

\subsection{Confounder importance learning via marginal likelihood} \label{sec:ml}

Our main computational contribution is a strategy to learn the hyper-parameter $\btheta$, which plays a critical role by determining prior inclusion probabilities. Below we devise an empirical Bayes approach  and a variational approximation thereof. 

The starting point is the marginal likelihood, 
\begin{align*}
p(\by \mid \btheta)= 
\sum_{\bdelta, \bgamma} p(\by \mid \bdelta, \bgamma) p(\bdelta, \bgamma \mid \btheta)\,,
\end{align*}
with the first term inside the sum given in \eqref{eq:marglik2}. The empirical Bayes estimator is $\bm{\theta}^{\text{EB}} = \argmax_{\bm{\theta}} p(\by \mid \btheta)$ and its  use for hyper-parameter learning in variable selection has been well-studied, see \cite{George00, Scott10, petrone:2014}.
We remark that it is possible to add a prior $p(\btheta)$ and use the marginal posterior modal estimate $\bm{\theta}^{\text{EB}} = \argmax_{\bm{\theta}} p(\by \mid \btheta) p(\btheta)$, however in our experiments this did not lead to noticeable differences in the results.

The main challenge in obtaining $\btheta^{\text{EB}}$ is that evaluating $p(\by \mid \btheta)$ requires a costly sum over $(\bdelta,\bgamma)$. Fortunately, it is possible to obtain a simpler expression for the gradient of the log-marginal likelihood, given in Proposition \ref{prop:one}. 
The proof (see Section S2) 
leverages the fact that $\by$ is conditionally independent from $\btheta$ given $(\bgamma,\bdelta)$, that the prior $p(\bdelta,\bgamma \mid \btheta)$ factorizes, and the specific form of $\pi_j(\btheta)$ in \eqref{eq:cilprior}.

\begin{prop} \label{prop:one}
  For our model as defined in \eqref{eq:y_eq}, \eqref{eq:probmodel}, \eqref{eq:incl} and  \eqref{eq:cilprior} we obtain that 
\begin{align}
\nabla_{\btheta} \log p(\by \mid \btheta)= 
\sum_{j: \pi_j(\btheta) \in  (\underline{\rho},\bar{\rho})} \mb{f}_{j} \left[ P(\gamma_{j} = 1 \mid \by, \btheta) - \pi_{j}(\btheta) \right].
\label{eq:prop_two}
\end{align}
where $\mb{f}_{j} = \tr{(1, f_{j,1}, \dots, f_{j,T})}$.
\end{prop}

Proposition \ref{prop:one}  allows using gradient-based optimization to approximate $\btheta^{\text{EP}}$. Notice that we only need to sum over at most $J$ terms, as opposed to $2^{J+T}$ for evaluating the marginal likelihood. Also, the gradient only depends on the data via the marginal inclusion probabilities $P(\gamma_j = 1 \mid \by, \btheta)$, which can typically be estimated more accurately than joint model probabilities. 
However, 
one must still compute $P(\gamma_j = 1 \mid \by, \btheta)$ for every considered $\btheta$, which is intensive when the optimization requires more than a few iterations, since typically an MCMC algorithm will be used to estimate these probabilities.
Section \ref{sec:ep} describes an expectation-propagation variational approximation 
that in most of our experiments provided a good approximation to the global mode.

The empirical Bayes solution given by Proposition \ref{prop:one} has a natural interpretation.
For simplicity we discuss the case where $\pi_j(\btheta) \in (\underline{\rho},\bar{\rho})$ for all $j=1,\ldots,J$.
When some $\pi_j(\btheta) \not\in (\underline{\rho},\bar{\rho})$ a similar interpretation holds, basically one excludes covariates with $\pi_j(\btheta)$ equal to $\underline{\rho}$ or $\bar{\rho}$. 
Setting the first entry of the gradient equal to zero gives
\begin{align}
& \sum_{j=1}^J \pi_{j}(\btheta)= \sum_{j=1}^{J} P(\gamma_{j} = 1 \mid \by, \btheta)
\label{eq:cil_zerograd_intercept}
\end{align}
That is, prior inclusion probabilities are set such that the prior expected model size is equal to the posterior expected model size.
This seems appealing, as the latter converges to the number of truly active covariates under suitable conditions (see Section \ref{sec:asymptotics}).

Simple algebra shows that setting the other entries of the gradient to zero gives that
$E(\kappa_t \mid \btheta)= E(\kappa_t \mid y, \btheta)$ for $t=1,\ldots,T$,
where
$
E(\kappa_t \mid \btheta)= J^{-1} \sum_{j=1}^J f_{j,t} [\pi_{j}(\btheta) - \bar{\pi}]
$
is the prior expectation of the sample confounding coefficient (Definition \eqref{eq:confounding_coef}), and
\begin{align}
E(\kappa_t \mid y, \btheta)=
\frac{1}{J}  \sum_{j=1}^{J} f_{j,t} [P(\gamma_{j} = 1 \mid \by, \btheta) - E(|\bgamma|_0 \mid \by, \btheta) ]
\label{eq:confcoef_postmean}
\end{align}
its posterior expectation, where $E(|\bgamma|_0 \mid \by, \btheta)= J^{-1} \sum_{j=1}^J P(\gamma_{j} = 1 \mid \by, \btheta)$.
Setting the prior mean of $\kappa_t$ to $E(\kappa_t \mid y, \btheta)$ is again appealing, since the latter converges to the true confounding coefficient $\kappa_t^*$ (Section \ref{sec:asymptotics}). Hence, CIL seeks to set prior probabilities that reflect the true amount of confounding, in particular learning from data whether one is in a high, neutral, or no confounding scenario.

\subsection{Confounder importance learning by expectation-propagation} \label{sec:ep}

The use of expectation-propagation (EP) \citep{Minka01a, Minka01b} is common in machine learning, including in variable selection  \citep{Seeger07, HdezLobato13}. 
We propose a computationally efficient approximation to the marginal likelihood optimizer that can be used as is, or serve as initialization for the gradient-based optimization in Section \ref{sec:ml}. 
For simplicity, we denote the posterior inclusion probabilities for the specific value $\btheta=\bm{0}$:
$$
p_{0}(\bdelta, \bgamma \mid \by) = p(\bdelta, \bgamma \mid \by,\btheta=\bm{0}) \propto p(\by \mid \bdelta,\bgamma)  \,. 
$$
The right-hand side arises because, when $\btheta=\bm{0}$, the model space prior $p(\bdelta,\bgamma \mid \btheta=0)=1/2^{J+T}$ is uniform.
We consider a mean-field approximation to $p_{0}(\bdelta, \bgamma \mid \by) $,
\begin{align}
q(\bdelta, \bgamma) = \prod_{t=1}^{T} \text{Bern}(\delta_{t}; s_{t}) \prod_{j=1}^{J} \text{Bern}(\gamma_{j}; r_{j}). \label{eq:mfapp}
\end{align}
where $\mb{s}=(s_1,\ldots,s_T)$ and $\mb{r}=(r_1,\ldots,r_J)$ are variational parameters, which EP chooses by minimizing a Kullback-Leibler divergence, as given in Proposition \ref{prop:EP}.

\begin{prop} \label{prop:EP}
The minimizer of the Kullback-Leibler divergence between  $p_{0}$ and $q$,
\begin{align}
  (\mb{r}^{\textsc{ep}},\mb{s}^{\textsc{ep}}) = \argmax_{\mb{r},\mb{s}} \sum_{\bgamma, \bdelta} p_{0}(\bdelta, \bgamma \mid \by) \log \left({p_{0}(\bdelta, \bgamma \mid \by) \over q(\bdelta, \bgamma)} \right)\,.
\nonumber
\end{align}
is given by
\begin{align}
  r_j^{\textsc{ep}} = P_{0}(\gamma_j =1 \mid \by)\,,\quad s_t^{\textsc{ep}} = P_{0}(\delta_t=1 \mid \by)\,.
\label{eq:EP-par}
\end{align}
\end{prop}

This approximation provides a computationally inexpensive estimator, denoted $\bthetaep$. 
First, note that
\begin{equation}
p(\by \mid \btheta) = 
\sum_{\bdelta,\bgamma} p(\by \mid \bdelta,\bgamma) p(\bdelta,\bgamma \mid \btheta) \propto
\sum_{\bdelta,\bgamma} p_{0}(\bdelta,\bgamma \mid \by) p(\bdelta,\bgamma \mid \btheta). \label{eq:marglik1}
\end{equation}
Our strategy is to replace $p_0$ by $q$, using the variational parameters in Proposition \ref{prop:EP}.
Section Section S4.1 
shows that the above sum over $2^{J+T}$ terms is reduced to one over $J$ terms, specifically
\begin{equation}
  \bthetaep = \argmax_{\btheta} \sum_{j=1}^{J} \log \left( r_{j}^{\textsc{ep}} \pi_j(\btheta) + (1-r_{j}^{\textsc{ep}}) (1-\pi_j(\btheta)) \right) \,.
  \label{eq:theta-ep}
\end{equation}

The gradient of the objective function in \eqref{eq:theta-ep} is
\begin{align}
\sum_{j: \pi_j(\btheta) \in (\underline{\rho},\bar{\rho})} \mb{f}_{j} \left[ P^{EP}(\gamma_j=1 \mid \by, \btheta) - \pi_j(\btheta) \right].
\nonumber
\end{align}
where
$P^{EP}(\gamma_j=1 \mid \by, \btheta)= \pi_{j}(\btheta) r_j^{EP} / [\pi_j(\btheta) r_j^{EP} + (1-\pi_j(\btheta))(1-r_j^{EP})]$.
This expression is analogous to \eqref{eq:prop_two}, and allows to similarly interpret $\bthetaep$ in terms of confounding coefficients (see discussion after \eqref{eq:prop_two}).

Our strategy is to pre-compute  $r_{j}^{\textsc{ep}}= P(\gamma_{j} = 1 \mid \by, \btheta = \mb{0})$, prior to conducting the optimization in \eqref{eq:theta-ep}. This leads to an optimization over $\btheta \in \mathbb{R}^{T+1}$ where, in contrast to the marginal likelihood estimate $\btheta^{\text{EB}}$, the objective function can be cheaply evaluated.
Since the function in \eqref{eq:theta-ep} is not concave, we conduct an initial grid search and subsequently use a quasi-Newton algorithm, see Algorithm 1 
in the supplement.
Although this was not an issue in our examples, when the number of treatments $T$ is large the mentioned grid search becomes too costly. Possible alternatives are either using Bayesian optimization methods or simply initializing at $\btheta=0$ (uniform model prior) and using the gradient 
of the objective function in \eqref{eq:theta-ep}.

In most of our examples $\bthetaeb$ and $\bthetaep$ provided virtually indistinguishable inference, the latter incurring a significantly lower computational cost. Figure S4 
shows a comparison for one of our simulated datasets.  On the other hand, $\bthetaeb$ does provide slight advantages in some settings where the number of parameters $J + T$ was particularly large (see Section \ref{subsec:singleT}).

\section{Asymptotic analysis} \label{sec:asymptotics}

Empirical Bayes seeks $\bthetaeb$ such that the prior mean of the confounding coefficient $E(\kappa_t \mid \bthetaeb)$ equals its posterior mean $E(\kappa_t \mid \by, \bthetaeb)$ (see Section \ref{sec:ml}), and a similar interpretation applies to the expectation-propagation estimator $\bthetaep$.
 Theorem \ref{thm:conv_confounding_coef} shows that 
$E(\kappa_t \mid \by, \btheta)$ converges in probability to the true $\kappa_t^*$ uniformly across $\btheta$, as $n \rightarrow \infty$, which implies that $E(\kappa_t \mid \btheta^{EB}) \stackrel{P}{\longrightarrow} \kappa_t^*$. That is, that empirical Bayes sets the CIL prior to match the true confounding coefficient, asymptotically.
We allow for model \eqref{eq:y_eq} to be misspecified, and focus on the case where $p$ is fixed. High-dimensional settings where $p \gg n$ are also interesting, but they require a delicate treatment beyond our scope.

From \eqref{eq:confcoef_postmean},
to prove that $E(\kappa_t \mid \by, \btheta) \stackrel{P}{\longrightarrow} \kappa_t^*$
it suffices that ${\bf f}_t \stackrel{P}{\longrightarrow} {\bf f}_t^*$ and that $p(\bgamma \mid \by, \btheta)$ converges to a point mass distribution at $\bgamma^*$.
The convergence of ${\bf f}_t$ follows from standard theory. For example, if ${\bf f}_t$ is obtained from the MLE separately regressing each treatment on the covariates, then the result holds under mild conditions, even if the model for the treatments is misspecified (\cite{vandervaart:1998}, Theorem 5.7).
If ${\bf f}_t$ is obtained from LASSO regressions, then consistency also holds under mild conditions, see \cite{buhlman:2011} (Chapters 6-7).
Regarding $p(\bgamma \mid \by, \btheta)$, its convergence to $\bgamma^*$ can be established by showing that the posterior odds between any model $\bgamma$ and $\bgamma^*$ converges to 0.
This requires mild conditions D0-D3, see Section S7.1. These conditions require log-likelihood concavity at the MLE (which holds for full-rank generalized GLMs under the canonical link), that the MLE is consistent as $n \rightarrow \infty$, that the asymptotic hessian is strictly positive definite, and a betamin condition that can be relaxed but simplifies our exposition.

\begin{thm}
 Suppose that Conditions D0-D3 hold and that ${\bf f}_t \stackrel{P}{\longrightarrow} {\bf f}_t^*$ as $n \rightarrow \infty$.
Then, $\sup_{\btheta} |E(\kappa_t \mid \by, \btheta) - \kappa_t^*| \stackrel{P}{\longrightarrow} 0$, as $n \rightarrow \infty$. 
\label{thm:conv_confounding_coef}
\end{thm}

 We next sketch the proof, see Section S7 for further details. 
We consider separately overfitted and non-overfitted models. The former refer to models $(\bdelta,\bgamma)$ that include all parameters in $(\bdelta^*,\bgamma^*)$ plus some truly zero parameters. In contrast, non-overfitted models fail to include some truly non-zero parameters.

We denote by $d= |(\bdelta,\bgamma)|_0 - |(\bdelta^*,\bgamma^*)|_0$ the difference between model dimensions,
by $d_1= \sum_{j=1}^J \gamma_j (1-\gamma_j^*)$ the number of covariates included in $\bgamma$ but not in $\bgamma^*$,
and by $d_2= \sum_{j=1}^J (1-\gamma_j) \gamma_j^*$ that of covariates included in $\bgamma^*$ but not in $\bgamma$.
Section S7 
shows that, for overfitted models, the posterior odds satisfy
\begin{align}
\frac{p(\bdelta, \bgamma \mid \by, \btheta)}{p(\bdelta^*,\bgamma^* \mid \by, \btheta)} \leq
\left(\frac{\bar{\rho}}{(n\tau)^{3/2} (1 - \bar{\rho})}\right)^d \times O_p(1)
,
\label{eq:postodds_overfitted}
\end{align}
as $n \rightarrow \infty$, uniformly across $\btheta$. This polynomial rate in $n$ reflects the effect of using a pMOM prior, for any other prior with a density that does not vanish at zero the rate is slower, specifically $(n\tau)^{3/2}$ is replaced by $(n\tau)^{1/2}$.
Note that for our default upper-bound on the prior inclusion probability $\bar{\rho}=0.95$, \eqref{eq:postodds_overfitted} converges to 0, as desired.

In contrast, for non-overfitted models,
\begin{align}
&\log \left( \frac{p(\bdelta, \bgamma \mid \by, \btheta)}{p(\bdelta^*,\bgamma^* \mid \by, \btheta)} \right)
\leq -\frac{3d}{2} \log(n\tau) - n c 
+ d_1\log \left( \frac{\bar{\rho}}{1 - \bar{\rho}}  \right)
+ d_2\log \left( \frac{1 - \underline{\rho}}{\underline{\rho}}  \right)
+ O_p(1)
\nonumber
\end{align}
again uniformly in $\btheta$.
That is, for overfitted models one obtains a faster rate that is almost exponential in $n$.
For these posterior odds to vanish it suffices that
$$
\lim_{n \rightarrow \infty} -\frac{3(|\bdelta^*|_0 + |\bgamma^*|_0)}{2} \log(n\tau) + n c - d_1\log\left( \frac{\bar{\rho}}{1 - \bar{\rho}} \right) + |\bgamma^*|_0 \log( \underline{\rho})= \infty,
$$
where we used that $d \geq -(|\bdelta^*|_0 + |\bgamma^*|_0)$ and $d_2 \leq |\bgamma^*|_0$, and recall that our default is $\underline{\rho}=1/J$.
Although we do not consider high-dimensional theory here, note that there one often sets sparse prior probabilities $\bar{\rho}<1/2$, then
$- d_1\log( \bar{\rho} / [1 - \bar{\rho}] ) > 0$, which helps ensure that the sufficient condition above is met.

\section{Results} \label{sec:results}

In Section \ref{subsec:singleT} we present a series of simulation studies, where we aim to illustrate the over-selection and under-selection issues discussed earlier across a range of settings. These range from no confounding settings, where all covariates are instruments, to full confounding scenarios where all covariates are confounders. We also consider single and multiple treatments, as well as varying sample sizes and problem dimensions.
We next present two separate case studies.
Section \ref{sec:cps_results} studies the association between certain demographics and the hourly salary, and its evolution between 2010 and 2019 (prior to \textsc{covid}-19, to avoid potential pandemic-related distortions), to assess wage discrimination. 
In Section \ref{subsec:abortion} we analyse a putative association between less favorable environmental conditions at birth and subsequent crime levels some years later, following a study carried out by \cite{donohue:2001}, retaken by \cite{Belloni14b}. 


In Section \ref{subsec:singleT} we compare our CIL (under the EP approximation) to DL based on the LASSO \citep{Belloni14b}, BAC \citep{Wang12}, 
ACPME \citep{Wilson18}, a standard LASSO regression on the outcome equation \eqref{eq:y_eq} (setting the penalization parameter via cross-validation), and standard BMA with a $\text{Beta-Binomial}(1,1)$ model prior and the pMOM prior on the coefficients (Section \ref{sec:method}).
We compare these methods to the oracle OLS, i.e. based on the subset of covariates truly featuring in \eqref{eq:y_eq}. 
In Section \ref{sec:cps_results} we focus on DL, standard BMA and, since $n$ is large relative to the number of parameters, we also consider ordinary least-squares (OLS) under the full model. We did not include BAC and ACPME here, as they failed to return results after 2 days (ACPME also exhausted 96Gb of RAM memory).
Finally, in Section \ref{subsec:abortion} we consider DL, BAC, ACPME and standard BMA.  
These methods are implemented in \texttt{R} packages \texttt{glmnet} \citep{glmnet} for the LASSO, \texttt{mombf} for BMA, \texttt{hdm} \citep{hdm} for DL, \texttt{bacr} \citep{bacr} for BAC and \texttt{regimes} \citep{regimes} for ACPME.
Throughout we set the BAC hyper-parameter to $\omega = +\infty$, which is the default in R package \texttt{bac} and encourages the inclusion of confounders relative to standard BMA.
R code to reproduce all our analyses is at \url{https://github.com/mtorrens/cil_article}.

\subsection{Simulation Studies} \label{subsec:singleT}

\subsubsection{Single treatment}
\label{sssec:singletreatment}

We consider an outcome generated according to \eqref{eq:y_eq} under a Gaussian likelihood, a single treatment $(T=1)$, and an error variance $\phi = 1$. 
The covariates are obtained as independent Gaussian draws $\bx_j \sim \text{N}(\mb{0}, \mb{I})$, with any active covariate affecting $y_{i}$ having an associated coefficient $\beta_{j} = 1$.
The treatment $d_{i}$ is generated to be a linear combination of the covariates, plus a zero-mean Gaussian random error with unit variance. Similarly to $y_{i}$, covariates having an effect on $d_{i}$ have a unit regression coefficient.
In all simulations, we set the total number of covariates that truly have an effect on $d_{i}$ to be equal to $|\bgamma|_0$, the number of covariates that have an effect on the outcome $y_{i}$.
To illustrate issues associated to under- and over-selection of covariates, a key factor we focus on is the \textit{level of confounding}.
Our scenarios range from no confounding (none of the $|\bgamma|_0$ covariates affecting $y_{i}$ have an effect on $d_{i}$) to complete confounding (all $|\bgamma|_0$ covariates affecting $y_{i}$ also affect $d_{i}$).
We measure the square-root MSE (RMSE) of the estimated $\hat{\alpha}$. 

Figure \ref{fig:intro2} summarizes the results when the number of active covariates is $|\bgamma|_0=6$ out of a total of $J=49$, $n=100$, and the treatment effect is either strong ($\alpha=1$), weak ($\alpha=1/3$), or non-existent ($\alpha=0$).
The two main features are as follows.
First, standard high-dimensional methods such as LASSO and BMA incur a high RMSE in high-confounding settings, 
 incurring both high bias and variance (Figure S6)
  whereas methods such as DL and BAC that are designed to prevent omitted variable biases perform much better in this regime. 
Second, in low confounding settings DL, BAC and ACPME incur a high RMSE,  due to high variance (Figure S6). 
 Figure S10 
shows that this is due to selecting all instruments, resulting in a larger model size (Figure S11). 
Figure S7 shows that for BAC this behavior is highly sensitive to the choice of hyper-parameter $\omega$, driving the prior dependence between inclusion in the outcome and treatment equations ($\omega=\infty$ for complete dependence, $\omega=1$ for independence).  
In contrast, our CIL framework performs well at all levels of confounding,  by avoiding the inclusion of instruments. Accordingly, in low-confounding scenarios we obtain hyper-parameter estimates $\hat{\theta}_1<0$, and  $\hat{\theta}_1>0$  under high-confounding (Figure S8). 
 It is also informative to assess how the prior inclusion odds assigned by CIL behave relative to those of ACPME. Figure S9 shows that in high confounding scenarios CIL assigns higher prior inclusion odds to controls that are associated to the treatment (which are mostly confounders) than it does in low confounding (when they are mostly instruments). In contrast, ACPME sets the same prior odds in either high or low confounding, i.e. it does not adapt to the true amount of confounding in the data. Also, the CIL prior inclusion odds are overall smaller, this is because $\theta_0$ in \eqref{eq:cilprior} learns the true amount of sparsity, as shown by our asymptotic study in Section \ref{sec:asymptotics}. For further discussion see Section S8.6.

We remark that, when there truly is no treatment effect, CIL (and BMA, in some instances) attains a much lower RMSE than the oracle OLS. This occurs because CIL effectively shrinks the treatment estimate to zero. Of course, it is possible to modify methods such as BAC or ACPME to also run selection on the treatment and one would then expect a comparable shrinkage, our results simply point out the potential benefits in conducting selection on the treatment effects.
The Empirical Bayes and the expectation-propagation versions of CIL provide nearly indistinguishable results (not shown).
Figure S11 
complements these results by showing the probability of selecting the treatment (bottom panels). Overall, the RMSE inflation incurred by LASSO and BMA in high-confounding settings is due to omitted variable biases,  and that of double LASSO, BAC AND ACPME under low confounding  is due to selecting instruments. 

We next consider two extensions of our simulation study.
First,  Figure S12  
considers a growing number of covariates, 
specifically $J+T=\{25, 100, 200\}$ with corresponding sample sizes $n=\{50,100,100\}$, in all cases under a strong treatment effect ($\alpha=1$). 
As dimensionality grows, the standard LASSO and BMA incur a significantly higher RMSE under strong confounding. Our CIL generally provides a significantly lower RMSE  over BMA in high-confounding scenarios, and a similar RMSE under mild and no confounding. An exception is the larger $J+T=200$, where under mild and no confounding the RMSE for BMA was roughly half that for CIL, although the latter was still significantly better than DL and BAC.
At $J+T=100$, ACPME departs from the behavior pattern of BAC and sensibly improves its relative performance for low levels of confounding,  although  it cannot attain the results of CIL. At $J+T=200$, where $n < J + T$, ACPME cannot be computed.
It is in this latter setting where we observe the only perceptible differences between the EB and EP approximations, with the former attaining better results, pointing to advantages of the EB approach in higher dimensions. 

As a second extension,  Figure S13  
shows the results when considering less sparse settings, specifically with $|\bgamma|_0 = 6$, 12 and 18 active parameters. Overall, the results are similar to Figures \ref{fig:intro2} and  S12. 
Our CIL continues to provide a competitive and more robust behavior across levels of confounding, relative to the other considered methods.
It is worth noting that again ACPME is able to improve the results of BAC, although it still cannot match the performance of CIL, particularly in low confounding scenarios.

\subsubsection{Multiple treatments}
\label{sssec:multitreatment}

To help understand under- and over-selection issues in multiple treatment inference,
we consider an increasing number of treatments, with a maximum of $T=5$. There, every present treatment is active, setting $\alpha_t = 1$ on all treatments. For all levels of $T$, we set $\beta_j = 1$ for $j = 1, \dots, 20$, denoting the set of active covariates by $\bx_{1:20}$, and $\beta_j = 0$ for the rest of covariates $\bx_{21:J}$. Regarding the association between treatments and covariates, $\bx_{1:20}$ are divided into five disjoint subsets with four variables each, and each of these subsets is associated to a different treatment. The treatments depend linearly on the covariates of its associated subset. Additionally, each treatment also depends on a further subset of inactive covariates $\bx_{21:J}$, i.e. instruments. In this case, the size of such subset is increasing by four with each added treatment: treatment 1 is associated to $\bx_{21:24}$, treatment 2 is associated to $\bx_{21:28}$, etc., up to treatment 5, which is associated to $\bx_{21:40}$. 
All covariates that affect a treatment have a regression coefficient equal to 1. 
The idea is that, as one considers a larger number of treatments $T$, one expects that potential ill-effects of over-selecting instruments and under-selecting confounders may become more problematic. Accordingly, as described our simulation considers $4T$ confounders and a growing number of instruments (4, 8, 12, 16, 20) for $T=(1,2,3,4,5)$. 
The rest of the design is as  in Figure \ref{fig:intro2}.

\begin{figure}[h]
\centering
\includegraphics[scale=0.7]{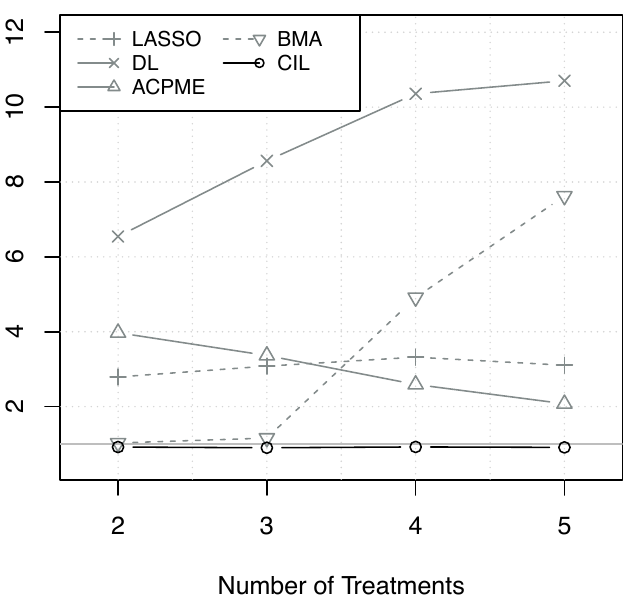} 
\caption{Treatment parameter RMSE (relative to oracle OLS) based on $R=250$ simulated datasets at every value of $T$, for $n=100$, $J=95$, and $T \in \{2, 3, 4, 5 \}$. For every $T$ ($x$-axis), we show the average RMSE across Treatments $1,\ldots,T$.
}
\label{fig:multitreat_uncounfounded}
\end{figure}

Figure \ref{fig:multitreat_uncounfounded} shows the RMSE associated to $\hat{\balpha}$ for the different values of $T$, i.e.
\begin{align}
\nonumber
\mbox{RMSE}_T= \frac{1}{T} \sum_{t=1}^T \mbox{RMSE}(\hat{\alpha}_t, \alpha_{t}^{*}) \nonumber
\end{align}
where $\alpha_t^*$ is the data-generating truth and $\hat{\alpha}_t$ the estimate provided by each method.
We observe similar trends as before. Methods prone to over-selection recover more instruments as $T$ increases. Some of these covariates are increasingly influential with $T$ as they are associated to more treatments, and so they become harder to discard. 
Interestingly, ACPME was designed to ameliorate the over-selection of instruments in multiple treatment settings, and indeed we observe an improved behavior as $T$ grows. Still, its RMSE  ranged from 2 to 4 times larger than that of CIL.

It is also interesting to remark that under-selection issues (here suffered by BMA) are also problematic. As $T$ grows the model becomes highly confounded, as a subset of the covariates account for a larger proportion of the variance in the outcome, as well as for that of the treatments. This leads to BMA discarding with high probability confounders that are truly active but are highly correlated to the treatments. 
Our CIL proposal is able to achieve oracle-type performance for every considered $T$.

\subsection{Salary variation and discriminating factors} \label{sec:cps_results}

We analyze the USA Current Population Survey (CPS) microdata \citep{ipums}, which records many social, economic and job-related factors.
We download data 
from 2010 and 2019 and analyze each year separately (see Section S8.2 
for details on data acquisition and pre-processing). We select individuals aged over 18, with a yearly income over \$1,000 and working 20-60 hours per week, giving $n=64,380$ and $n=58,885$ in 2010 and 2019, respectively. 
The covariates include characteristics of the place of residence, education, labor force status, migration status, household composition, housing type, health status, financial records, reception of subsidies, and sources of income (beyond wage). 
Overall, there are $J=278$ covariates, 228 given in the raw data plus 50 indicators for state.
The outcome is the individual log-hourly wage, rescaled by the consumer price index of 1999, and we consider $T=4$ main treatments of interest:  sex, black race, Hispanic ethnicity and Latin America as place of birth. 
 Specifically, we introduce a female indicator for individuals who declared female as their single sex, and a black indicator for those who declared black as their single race. 
These treatments are highly correlated to sociodemographic and job characteristics that can impact salary, i.e. there are many potential confounders.
Since every state has its own regulatory, sociodemographic and political framework, we capture heterogeneous state effects by adding interactions for each pair of treatment and state. On these interactions, we apply a sum to zero constraint, so that the coefficients associated to the four treatments remain interpretable as average treatment effects across the USA, and the interactions as deviation from the average. Hence, overall we have $4+4 \times 50 = 204$ parameters quantifying treatment effects, and our main interest is in the first four. 
To simplify computation in our CIL prior we assume a common $\theta_t$
shared between each main treatment and all its interactions with state, so that $\mbox{dim}(\btheta)=5$.
 Since there are 4 treatments and their $4 \times 50=200$ interactions with states, in principle $\mbox{dim}(\btheta)=205$ (including $\theta_0$). We view this as undesirable because optimizing over a 205-dimensional state would create a significant bottleneck, and because one does not expect to estimate precisely so many parameters.
Further, one expects some structure in $\btheta$. If there is high confounding for a treatment in one state (positive entry in $\btheta$ for that state), then the same may hold in many other states. Setting $\mbox{dim}(\btheta)=5$ assumes that such measure of confounding is the same across all states.

\begin{figure}[htbp]
\centering
\begin{tabular}{cc}
\includegraphics[width=0.48\textwidth]{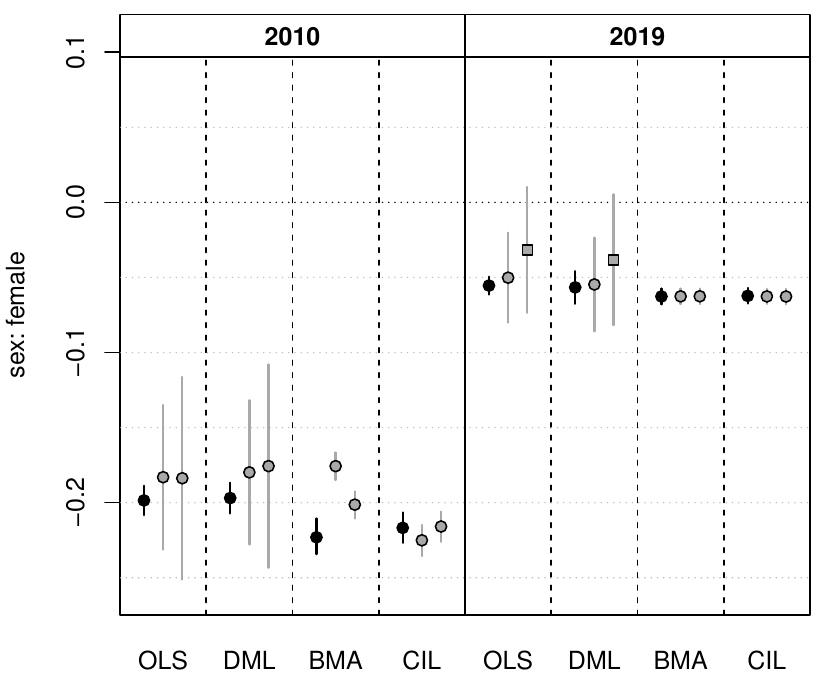} &
\includegraphics[width=0.48\textwidth]{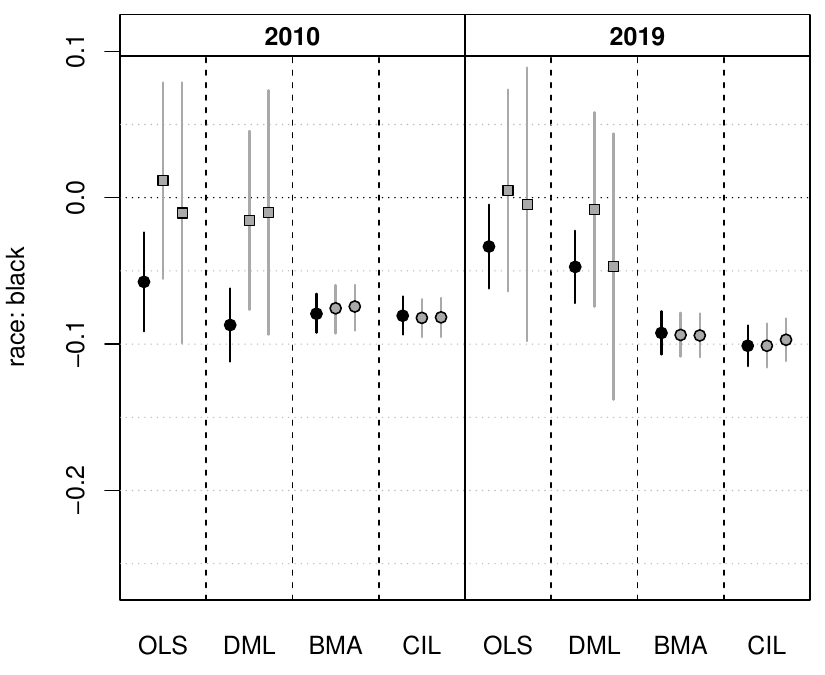} \\
\vspace{0.3cm}
\end{tabular}
\caption{Inference for treatments ``female'' (left) and ``black'' (right) in 2010 and 2019. We analyze Current Population Survey data with $J=278$ covariates (left black point and bar in each panel) but also adding 100 (middle) and 200 (right) artificial instruments. Names of methods as in the caption of Figure \ref{fig:intro2}.}
\label{fig:intro1}
\end{figure}

Figure \ref{fig:intro1} reports the results for  sex  
and race. More detailed results in Figure S5 
show that none  of the methods finds  an association between salary variation and ethnicity or place of birth. The treatment effect for sex is picked up by all methods in both years with similar point estimates. All methods suggest a slight decrease of this effect in 2019.
For race  the methods vary in their findings.   All methods find a negative association between black race and salary, but in 2019 OLS and DL estimate a fairly lower effect.

In order to understand this difference better, and explore whether it is due to over-selection, we analyze two additional augmented datasets where we add artificial instruments. 
Specifically, we incorporate 100 instruments in the first scenario, and 200 in the second one. The instruments are split into four equal subsets, each of which is designed to correlate to one of the four main treatments, see Section S8.3 
for full details.
The resulting average correlation between sex and its associated artificial instruments is $0.83$, and analogously $0.69$, $0.76$ and $0.67$ for black race, ethnicity and place of birth.
Upon adding said instruments, the confidence intervals for OLS and DL become notably wider, whereas CIL and BMA results remain particularly robust. This variance inflation is particularly pronounced for the effect of black race, which in 2019 lead to a loss of statistical significance according to OLS and DL. These findings suggest that the smaller racial gap estimated in 2019 in the original data may be due to variance inflation rather than an actual improvement in the racial gap.

The full scope of a Bayesian inferential framework is materialized when, additionally to quantifying treatment effects, one also considers more complex functions of the parameters. As an illustration, we study a measure of overall treatment contribution to deviations from the average salary. The idea is that the conditional associations between salary and the four treatments ( sex , race, Hispanic ethnicity and birth in Latin America) may reflect salary discrimination associated to these demographics, and it is hence interesting to quantify the overall effect of all four treatments.
 For a new observation $n+1$, with observed treatments $\mb{d}_{n+1}$ and covariates $\mb{x}_{n+1}$, let
\begin{align}
  &h_{n+1}(\mb{d}_{n+1}, \balpha, \mb{x}_{n+1}) 
  = | \E(y_{n+1} \mid \mb{d}_{n+1}, \mb{x}_{n+1}, \balpha, \bbeta) - \E(y_{n+1} \mid \mb{x}_{n+1}, \balpha, \bbeta) | \nonumber \\
&= | \tr{[\mb{d}_{n+1} - \E(\mb{d}_{n+1} \mid \mb{x}_{n+1})]} \balpha | \label{eq:app2}
\end{align}
be its expected salary minus the expected salary averaged over possible $\mb{d}_{n+1}$, given equal covariate values $\mb{x}_{n+1}$.
Since $y_{n+1}$ is a log-salary, we examine the posterior predictive distribution of $\exp \left\{ h_{n+1}(\mb{d}_{n+1}, \balpha, \mb{x}_{n+1}) \right\}$ as a measure of salary variation associated to the treatments. A value of 1 indicates no deviation from the average salary, relative to another individual with the same covariates $\mb{x}_{n+1}$.

\begin{figure}[h]
\centering
\includegraphics[width=0.48\textwidth]{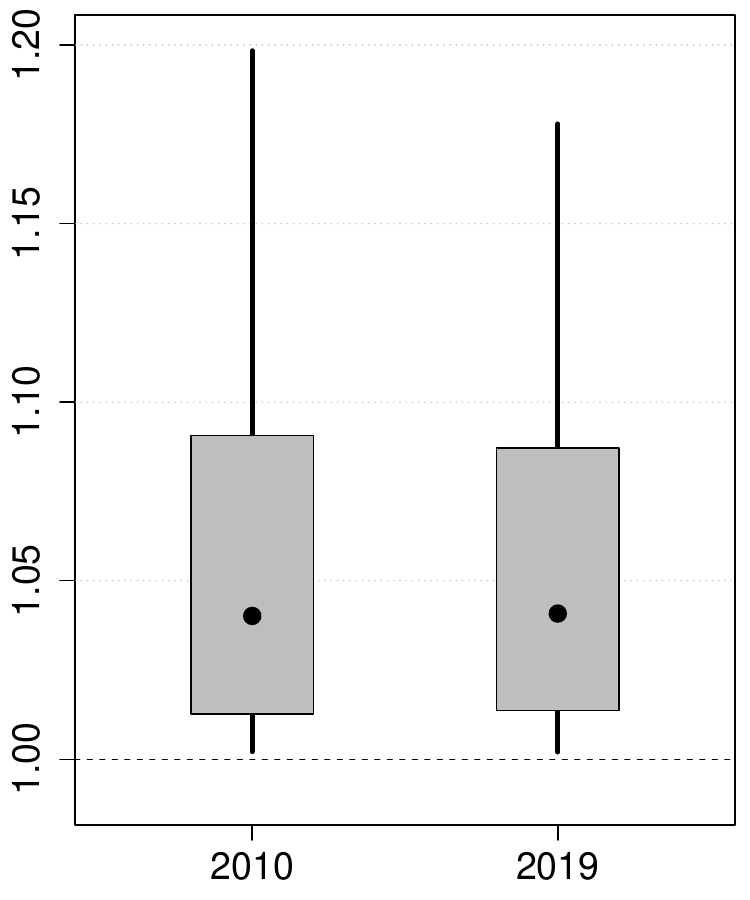} 
\includegraphics[width=0.48\textwidth]{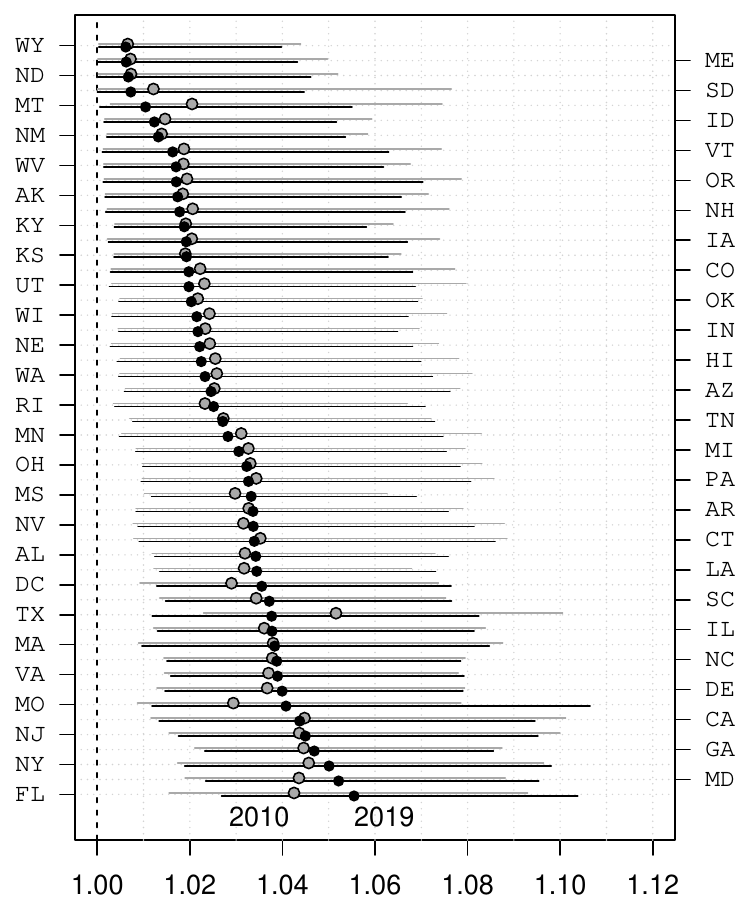}
\caption{The left panel shows the posterior predictive distribution of deviations from average salary as given by $\exp \left\{ h_{n+1}(\mb{d}_{n+1}, \balpha, \mb{x}_{n+1}) \right\}$ in \eqref{eq:app2}, for 2010 and 2019. The gray boxes represent 50\% posterior intervals and the black lines are 90\% intervals. The black dot is the posterior median. The right panel shows the posterior median of these deviations for every U.S. state in 2010 and 2019 on the horizontal axis, ordered by their value in 2019, with the corresponding 50\% posterior intervals for both years.}
\label{fig:salary_variation}
\end{figure}

To evaluate the posterior predictive distribution of \eqref{eq:app2} given $\by$, the observed $\bd$ and the set of covariates, we obtain posterior samples from the model averaged posterior $p(\balpha \mid \by)$ associated to CIL (Section \ref{ssec:bma}). Given that we do not have an explicit model for $(\bd_{n+1}, \mb{x}_{n+1})$, we sample pairs $(\mb{d}_{n+1}, \mb{x}_{n+1})$ from their empirical distribution, and estimate $\E(\mb{d}_{n+1} \mid \mb{x}_{n+1})$ from a logistic regression of $\bd$ on the set of covariates.
Figure \ref{fig:salary_variation} shows the results. 
There is fairly little progress in reducing the joint association between the outcome and treatments, i.e. the potentially discriminatory factors, both at nation- and state-wide level (upper and lower panels in Figure \ref{fig:salary_variation}, respectively).
In 2010, joint variation in the treatments is associated to an average 6.2\% salary variation (90\% predictive interval [0.2\%, 19.8\%]). The posterior mean in 2019 drops slightly to 5.9\% and the 90\% predictive interval is [0.2\%, 17.8\%]. That is, the treatments play a similar role in the 2019 both in average and in the whole predictive distribution.

It is also of interest to study differences between states. This is possible in our model, which features 200 interaction terms for the 4 treatments and 50 states. Figure \ref{fig:salary_variation} (right) shows the results. The most salient feature is a slightly lower heterogeneity across states in 2019 relative to 2010.
The three states whose median improves the most are Texas (reducing it by 1.4\%), Montana (1.0\%) and South Dakota (0.5\%), while the three in which it worsens the most are Florida (increasing it by 1.3\%), Missouri (1.1\%) and Maryland (0.9\%), which are already among the bottom-ranking states in 2010. 

\subsection{Abortion and Crime Study} \label{subsec:abortion}

\begin{table*}
 \centering
\caption{Estimated association between three crime types and abortion}
\label{tab:abortion}
\begin{tabular}{|l|cccc|} \hline
 & \multicolumn{4}{|c|}{{\bf Violent crime}} \\
 & Estimate & 95\% interval & $p$-value & $P(\alpha \neq 0 \mid \by)$ \\
Donohue III and Levitt & $-0.13$ & $(-0.18,-0.08)$ & $<0.001$ & $-$ \\
OLS (all covariates)   & $-0.04$ & $(-1.37, 1.30)$  & $0.958$ & $-$ \\
  DL                   & $-0.21$ & $(-0.46, 0.04)$ & $0.105$ & $-$ \\
BAC                    & $0.37$  & $(-0.67, 1.28)$ & $-$     & $-$ \\
ACPME                  & $-0.42$ & $(-0.54, -0.31)$ & $-$     & $-$ \\
BMA (Normal)           & $-0.07$ & $(-0.29, 0)$    & $-$     & $0.216$ \\
BMA (MOM)              & $-0.01$ & $(-0.16, 0)$    & $-$     & $0.032$ \\
CIL (Normal)           & $-0.17$ & $(-0.31, 0)$    & $-$     & $0.977$ \\
CIL (MOM)              & $-0.11$ & $(-0.24, 0)$    & $-$     & $0.657$ \\
\hline
 & \multicolumn{4}{|c|}{{\bf Property crime}} \\
Donohue III and Levitt & $-0.90$ & $(-0.94,-0.87)$ & $<0.001$ & $-$ \\
OLS (all covariates)   & $-0.19$ & $(-0.56, 0.19)$ & $0.327$ & $-$ \\
DL                    &$-0.04$  & $(-0.12,0.04)$ & $0.369$ & $-$ \\
BAC                   &$-0.16$  & $(-0.42,0.11)$ & $-$     & $-$ \\
ACPME                 & $-0.14$ & $(-0.22, -0.07)$ & $-$     & $-$ \\
BMA (Normal)          & $0$     & $(0,0)$        & $-$     & $0.063$  \\
BMA (MOM)             & $0.00$  & $(0,0)$        & $-$     & $0.001$ \\
CIL (Normal)          & $-0.05$ & $(-0.15,0)$    & $-$     & $0.593$ \\
CIL (MOM)             & $-0.02$ & $(-0.12,0)$    & $-$     & $0.122$ \\
\hline
 & \multicolumn{4}{|c|}{{\bf Murder}} \\
Donohue III and Levitt  & $-0.12$ & $(-0.21,-0.03)$ & $0.010$ & $-$ \\
OLS (all covariates)    & $1.73$ & $(-3.70,7.15)$  & $0.531$ & $-$ \\ 
DL                    & $-0.12$ & $(-0.95,0.716)$ & $0.785$  & $-$ \\
BAC                   & $-0.25$ & $(-3.47, 3.01)$ & $-$     & $-$ \\
ACPME                 & $-0.51$ & $(-0.91, -0.11)$ & $-$     & $-$ \\
BMA (Normal)          & $0$     & $(0,0)$         & $-$     & $0.009$   \\
BMA (MOM)             & $0$     & $(0,0)$         & $-$     & $<0.001$\\
CIL (Normal)          & $-0.03$ & $(-0.61,0.10)$  & $-$     & $0.136$   \\
CIL (MOM)             & $0$     & $(0,0)$         & $-$     & $0.003$   \\
\hline
\end{tabular}
\end{table*}

\cite{Belloni14b} revisit a study by \cite{donohue:2001} that assesses the association between yearly state-level abortion rates and crime rates 15--25 years later. A hypothesis is that, if parents choose a moment of birth to raise children in a favorable environment, the latter might be less likely to commit crimes when they reach ages 15--25. 
The authors consider three crime types (violent, property, murder) and a measure of abortion associated to each type (a weighted average of abortion rates across age groups, where the weights are the fraction of that crime type committed by each age group).

To avoid confounding when estimating the association between abortion and crime, it is important to account for state and time effects for the period 1985 to 1997, and various other covariates.
\cite{donohue:2001} consider the log of prisoners per capita and of police per capita, unemployment and poverty rates, income and beer consumption per capita, the aid to families with dependent children (AFDC) program generosity, the existence of a concealed guns law, and the one-year lagged versions of these variables.
\cite{Belloni14b} extend the analysis by adding quadratic covariate effects, interactions and linear and quadratic interactions with time. 
To account for state effects, they define the outcome as the increase in crime rates between two consecutive years, the treatment as the increase in abortion rates, and they include as covariates the within-state crime averages and the initial crime rates at 1985.
They also force the inclusion of the year indicators in the model to avoid their estimates being driven by time dynamics in crime and abortion.
The additions of \cite{Belloni14b} are done to reduce the misspecification of the outcome model, which could hamper causal interpretations. Our analysis keeps these covariates and overall we have  $n=576$ observations, 1 treatment and $J=295$ covariates.
See Section S8.5 for further details. 

Table \ref{tab:abortion} summarizes the results obtained with different approaches, the previous ones and the one we develop in this article.
The least-squares analysis of \cite{donohue:2001} based on a pre-defined set of covariates find a statistically significant association between abortion and the three crime types.
For comparison we also run a least-squares regression using all covariates considered by \cite{Belloni14b}, which returns no statistically significant results and very wide confidence intervals. This is as expected in situations where covariates are strongly correlated with the treatment.
The DL analysis of \cite{Belloni14b} also returns no statistically significant associations, using the latest version (0.3.1) of their \texttt{R} package \texttt{hdm}.
The main difference between DL and \cite{donohue:2001} is that the former selected numerous covariates that are associated to the treatment (abortion) but not to the outcome, i.e. likely instruments.
For violent crime, a LASSO analysis of the outcome equation selected no covariates, whereas 9 covariates are selected in the abortion equation. DL then proceeds by regressing violent crime on those 9 covariates.
As shown in Table \ref{tab:abortion}, adding said covariates that are highly correlated  with abortion causes a variance inflation in the estimated effect for the latter.
Also note that there is little evidence that the covariates are needed in the outcome equation, e.g. only one of their naive $p$-values (not accounting for post-selection inference) are below 0.05 (Table S1). 
A similar situation occurs for property crime and murder.
For property crime 13 covariates are selected (1 $p$-value below 0.05), and for murder it is 8 covariates (no $p$-value below $0.05$).
Applying BAC to these data gives qualitatively similar results to DL, in that no significant treatment effect is detected. 
Again a potential issue is that many covariates have a non-negligible contribution, which can cause variance inflation, e.g. 98 covariates have marginal posterior inclusion probability above 0.5 for violent crime, 97 for property crime, and 80 for murder.
ACPME did find a significant association for all three crime types. This is interesting because, as discussed, ACPME presents similarities to BAC but attempts to ameliorate variance inflation due to selecting instruments.

We re-analyze the data of \cite{Belloni14b} with standard BMA (Beta-Binomial model prior) and our CIL methodology, also forcing the inclusion of the year indicators in the model, following \cite{Belloni14b}.
To explore sensitivity of the results to the prior, we obtain results under a default normal prior on the parameters with diagonal covariance and a MOM prior with default dispersion.
As shown in Table \ref{tab:abortion}, the results of our CIL approach lie somewhere in between the significant results found by \cite{donohue:2001} and by ACPME, and the non-significant results found by DL, BAC and BMA.
In the violent crime analysis, CIL finds moderate evidence for a negative association between abortion and crime. 
Under CIL all covariates have a negligible posterior inclusion probability. This is contrast to BMA where several covariates (6 for BMA normal, 2 for BMA MOM) have posterior inclusion probabilities above 0.1, and to DL which selects 8 covariates (which, as discussed, are likely to be instruments).
We observe a similar situation for property crime, where CIL produces higher posterior probabilities for the existence of an association than its BMA counterpart. However, these are only moderate and the estimated effect is small.
For murder CIL finds no evidence for an association with abortion.
See Section S8.5 
for a discussion on the covariates selected by BMA and CIL in each analysis.

Overall the CIL results provide moderate, but not overwhelming, evidence for the existence of an association between abortion and crime (violent crime in particular). On the basis of CIL's analysis the applied researcher might try and obtain further evidence to evaluate the assumed association, whereas the results with DL and BMA could be construed as fairly strong evidence against said association.

\section{Discussion} \label{sec:discuss}

The two main ingredients in our proposal are learning from data whether and to what extent covariate inclusion/exclusion should be encouraged to improve multiple treatment inference, and a convenient computational strategy to render the approach practical.
 Our framework learns from data whether the data-generating truth is of a high, neutral or low confounding nature, as measured by our novel confounding coefficient. 
One then hopes to obtain a better balance between over-selection of instruments and under-selection of confounders.

These issues are practically relevant, e.g. in the salary data we show that one may underestimate the association black race and salary. Further, the proposed Bayesian framework naturally allows for posterior predictive inference on functions that depend on multiple parameters, such as the variation in salary jointly associated with multiple treatments. Interestingly, our analyses reveal little progress in the association between salary and potentially discriminatory factors such as  sex  or race in 2019 relative to 2010, nation- and state-wise.
These results are conditional on covariates that include education, employment and other characteristics that affect salary. That is, our results reveal lower salary discrepancies in 2019 between races/sex, provided that two individuals have the same characteristics (and that they were hired in the first place).
This analysis offers a complementary view to analyses that are unadjusted by confounders, and which may reveal equally interesting information.
For example, if females migrated towards lower-paying occupational sections in 2019 and received a lower salary as a consequence, this would not be detected by our analysis, but would be revealed by an unadjusted analysis.

We remark that our methodology can be extended to other settings where one wishes to treat the inclusion of covariates non-exchangeably a priori. For example, an interesting avenue for future research are settings where one has meta-covariates distinguishing covariate subsets (e.g. clinical variables, genomic markers, diagnostic tests), where it is natural to consider that different subsets may warrant different inclusion probabilities.

\section{Acknowledgments}

DR gratefully acknowledges support from grant \textit{Consolidaci\'on investigadora} CNS2022-135963 by the AEI (Government of Spain), \textit{Ayudas Fundaci\'on BBVA Proyectos de Investigación Científica en Matemáticas 2021}, Europa Excelencia EUR2020-112096 from the AEI/10.13039/501100011033 and European Union NextGenerationEU/PRT, and grant PID2022-138268NB-I00 financed by MCIN/AEI/10.13039/501100011033 and the FSE+.

\newpage

\section*{Supplementary material}

\section{Over-selection bias}
\label{ssec:oversel_bias}

\subsection{Discussion}

As explained in the main text, selecting variables that are truly not associated to the outcome can introduce a bias in the estimated treatment effect.
We remark that the issue does not occur when the selected variables are pre-specified, for example it's immediate to show that the least-squares estimator is unbiased for any model containing the truly active covariates plus a pre-defined set of extra covariates.
The over-selection bias issue arises when variables are selected in a data-based fashion.

\begin{figure}[htbp]
\centering
\begin{tabular}{cc}
\includegraphics[width=0.48\textwidth]{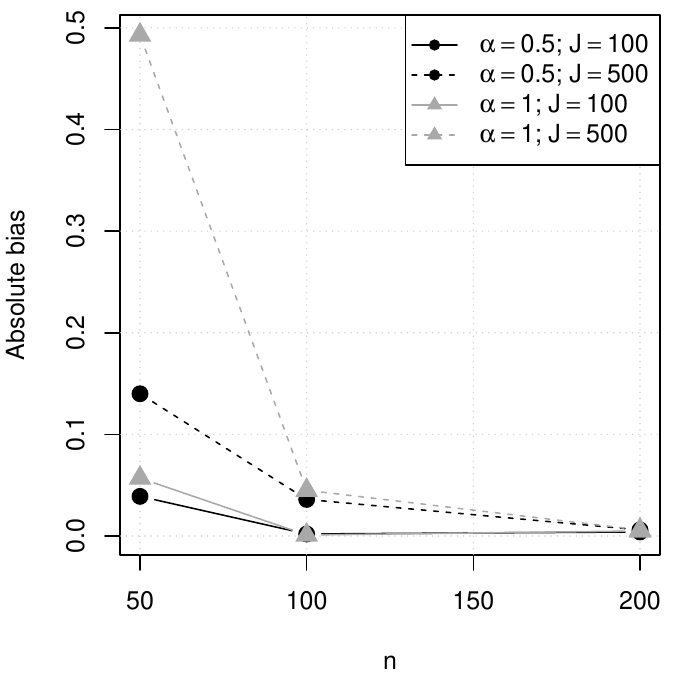}  &
\includegraphics[width=0.48\textwidth]{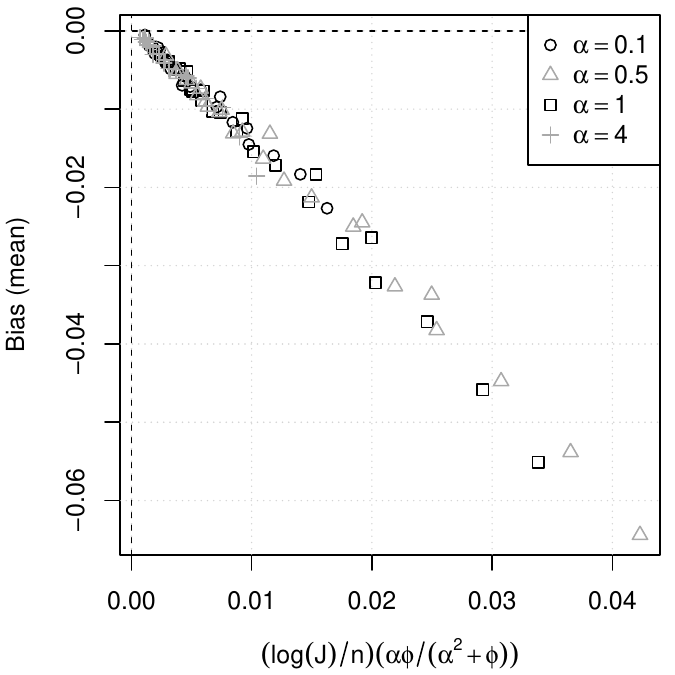} 
\end{tabular}
\caption{Over-selection bias simulations. The generative model for $y_i$ is as in  \eqref{eq:y_eq} with a Gausssian density, $\mb{x}_i \sim \mathrm{N}(\mb{0},\mb{I})$ and $d_i \mid \mb{x}_i \sim \text{N}(\tr{\mb{x}}_i \bv, 1)$. 
Left: only three elements of $\bbeta$ and three different ones of $\bv$ are non-zero, and all equal to  2, which makes the $y_i$'s indirectly correlated to three covariates whose corresponding elements in $\bbeta$ are equal to zero.
Said correlation becomes stronger as $|\alpha|$ grows. At the same time, screening out these covariates becomes harder as $J/n$ increases, difficulting post-selection inference.
Here $\hat{\alpha}$ is estimated with double-lasso, and we report absolute bias over 200 indepdendent simulations for every condition. 
Right: $\bbeta$ and $\bv$ are zero-vectors, the Gaussian observation variance is $0.5^2$,  $\hat{\alpha}$ is estimated from OLS of $y_{i}$'s on $d_{i}$'s and the covariate selected among $J$ available ones to have the highest correlation in absolute value with the $y_{i}$'s; the estimation is carried out for different $J \in \{10,20,40,80,160\}$ and $n \in \{30,50,100,150,200\}$ and the figure plots the estimated bias (over $15 \times 10^{3}$ independent experiments) versus $(\alpha \phi \log J)/(n(\alpha^2+\phi))$, for $\phi = 0.5^2$,  which is the approximation of the size of the bias suggested by the argument we develop in the Supporting material. Different colours correspond to different values of $\alpha \in \{1,0.1,0.5,4\}$. } 
\label{fig:sel-bias}
\end{figure}

To illustrate this point we use two examples where there truly is no confounding. 
Consider a data-generating truth as in \eqref{eq:y_eq} where there is one treatment, $T=1$, and the generative model for the covariates and the treatment is as follows: $\mb{x}_i \sim \text{N}(\mb{0}, \mb{I})$,  $d_i \mid \mb{x}_i \sim \text{N}(\tr{\mb{x}}_i \bv, 1)$,  where unknown to the data analyst $\bbeta$ has 3 non-zero entries and $\bv$ has 3 different non-zero entries, all equal to 2.
Figure \ref{fig:sel-bias} (left) shows that the DL-based $\hat{\alpha}$ has a bias that grows with $\alpha$ and $J$ and decreases with $n$ (\cite{Belloni14b} showed that the bias vanishes as $n \rightarrow \infty$, under suitable conditions).
The issue arises because the selection of covariates depends on the observed outcome.
To obtain further insight, the right panel considers a setting where covariate selection is also outcome-dependent, in a simpler fashion.
All entries in $\bbeta$ and $\bv$ are truly 0, the analyst selects the covariate with the highest absolute correlation with the $y_{i}$'s and estimates $\alpha$ by OLS on the $d_{i}$'s and the selected covariate. The resultant estimate of $\alpha$ has a negative bias, which an analysis we carry out in the Supporting material approximates it to be 
$$
- c {\alpha \phi \over \alpha^2+\phi} {\log J \over n},
$$
for some constant $c>0$. The simulation experiment in Figure \ref{fig:sel-bias} provides strong numerical evidence towards this approximation. 
This over-selection bias is fairly subtle, notice that both  small and large signal-to-noise ratios $\alpha/\sqrt{\phi}$ lead to small bias but intermediate ones to large.
In our experience said bias has little impact in most examples, unless $J$ is really large relative to $n$.
The take-home message is that, whereas for $J<n$ one may add all covariates to the model to obtain a (high-variance) unbiased estimator, when $J > n$ and one applies some shrinkage or selection,  inference can be subject to bias. 

The resultant over-selection mean squared error  worsens as the number of treatments increases and as the the proportion of covariates that are relevant both for the response and the treatments decreases.

\subsection{Derivations}

We sketch an argument that is based on some explicit mathematical derivations, some careful numerics and educated guesses (based on intuitions from properties of maxima of Gaussians) and quantifies the over-selection bias in a simple yet instructive example. As one can see from the following analysis, even in this simplified case it is not straightforward to obtain clean results, therefore we see this example as one that strikes a good balance between making an interesting point and being sufficiently tractable.

The setting is as follows. The data generating process is  $\mb{y} \mid \mb{d} \sim \text{N}( \alpha \mb{d}, \phi)$ where $\mb{d}$ is random with mean 0 and variance 1. 
The analyst has further available covariates $\mb{x}_j$ that have been centered and scaled and unknown to them are independent of each other and $\mb{d}$ and $\mb{y}$. Let
$$
S = \arg \max_{1\leq j \leq J} \tr{\mb{x}}_j \mb{y}
$$
where, obviously, due to our setting $S$ is marginally a uniformly distributed integer from 1 to $J$. (In the numerical experiment reported in Figure \ref{fig:sel-bias} we screen the predictor using $|\tr{\mb{x}}_j \mb{y}|$ but for the analysis here we omit the absolute value to simplify the problem. There are good reasons why this does not change the obtained result materially, which is why Figure \ref{fig:sel-bias} is in agreement with the result obtained using the analysis below). 
Let $\hat{\alpha}_S$ be the OLS estimate of $\alpha$ by regressing $\mb{y}$ on $\mb{d}$ and $\mb{x}_S$ (without intercept). Application of standard OLS results show that
$\hat{\alpha}_S = \alpha + \xi_S$ where
$$
\xi_S = {\tr{\mb{d}}(\mb{I} - \mb{x}_S \tr{\mb{x}}_S/n)(\mb{y} - \alpha \mb{d}) \over \tr{\mb{d}}(\mb{I} - \mb{x}_S \tr{\mb{x}}_S/n) \mb{d}}.
$$
We take $\mb{e} = (\mb{y} -\alpha \mb{d})/\sqrt{\phi}$, which by construction has a standard Gaussian distribution. Direct calculation gives further that
$$
\xi_S =  {\sqrt{\phi}{\tr{\mb{d}}\mb{e}\over n}  \over 1 -  (\tr{\mb{d}}\mb{x}_S/n)^2}  -  {\sqrt{\phi} \over n} { {\tr{\mb{d}} \mb{x}_S \over \sqrt{n}} {\tr{\mb{e}}\mb{x}_S \over \sqrt{n}} \over  1 - {1\over n} (\tr{\mb{d}}\mb{x}_S/\sqrt{n})^2}.
$$
The first term  is fairly symmetric around 0. We concentrate on the second term, and in particular the expectation of its numerator which will determine the bias, if any, to the highest order.

\begin{figure}[h]
\centering
\includegraphics[width=0.48\textwidth]{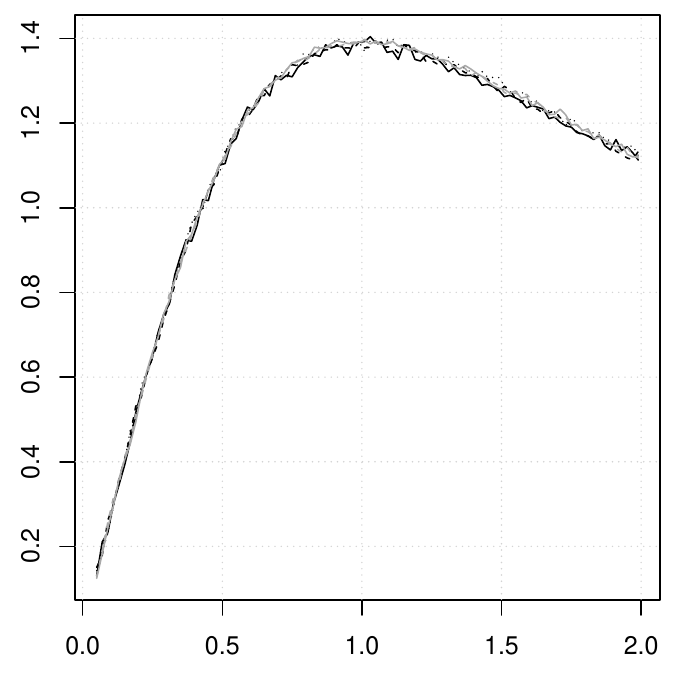} 
\includegraphics[width=0.48\textwidth]{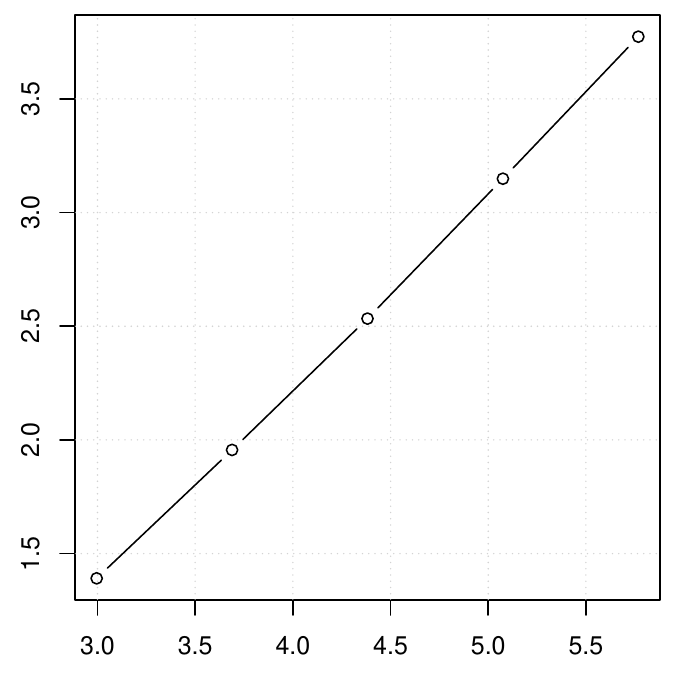} 
\caption{We show results of a large simulation study where  we estimate $\text{E}[\gamma_S \delta_S] $, for $S = \arg \max_j (r \gamma_j + \delta_j)$ and $\gamma_j,\delta_j$ are i.i.d. standard Gaussian,  for various values of $J$ and $r$. On the left plot the estimates against the values of $r$, for the different values of $J$ shown with different combinations of line type and color. We have scaled the curves to have the same value at $r=1$ and the figure confirms the conjecture that the curves are the same up to multiplication of a function of $J$ alone. On the right we plot the estimates that correspond to $r=1$ against $\log J$ for the different values of $J$.}
\label{fig:sel-bias-partial}
\end{figure}

Notice that $\tr{\mb{x}}_j \mb{y} = \alpha \tr{\mb{x}}_j \mb{d} + \sqrt{\phi}\tr{\mb{x}}_j \mb{e}$, where $ \tr{\mb{x}}_j \mb{d}$ and $\tr{\mb{x}}_j \mb{e}$ are uncorrelated, zero mean and have variance $n$. Hence, for obtaining   an estimate of the bias we concentrate now on the simplified problem of approximating $\text{E}[\gamma_S \delta_S]$ for 
$$
S = \arg \max_{1\leq j \leq J} \left ({\alpha \over \sqrt{\phi}} \gamma_j +  \delta_j \right)
$$
for  $\gamma_j,\delta_j$ are i.i.d standard Gaussian. A basic exchangeability argument shows that for given $J$, as a function of $r$, when $S = \arg \max_j (r \gamma_j + \delta_j)$,  $\text{E}[\gamma_S \delta_S] = h(r)$ where $h(r)=h(1/r)$ and it is maximized at $r=1$; such a function is $h(r) = 1/(r+1/r)$.  The intuition behind this result  is that for very large or very small values of the ratio one of the two terms dominates the choice of $S$ and the other is independent of that choice. An educated guess which builds upon results for maxima of Gaussian sequences is that to the highest order
$$
\text{E}[\gamma_S \delta_S]  \approx c {1 \over r + 1/r} \log J
$$
for some constant $c$ (that does not depend on $p$ or $J$).  The results in  Figure \ref{fig:sel-bias-partial} provide strong numerical support for this conjencture. 

Putting all the steps together, we obtain an approximation of the bias to be 
$$
- c {\alpha \phi \over \alpha^2+\phi} {\log J \over n}\,.
$$

\section{Proof of Proposition 3.1} \label{sec:proof_prop1}

We first state and prove an auxiliary result regarding the gradient of $p(\gamma_j \mid \btheta)$, and subsequently prove the proposition.
To ease notation, let $\mb{f}_{j} = \tr{(1, f_{j,1}, \dots, f_{j,T})} $ be the vector of features for covariate $j$, including the intercept.

\subsection{Auxiliary result}

Recall that the prior inclusion probability for covariate $j$ is 
\begin{align}
\pi_j(\btheta)= 
\begin{cases}
\bar{\rho} \mbox{, if } \tilde{\pi}_j(\btheta) \leq \bar{\rho}
\\
\tilde{\pi}_j(\btheta) \mbox{, if } \tilde{\pi}_j(\btheta) \in (\underline{\rho}, \bar{\rho})
\\
\bar{\rho} \mbox{, if } \tilde{\pi}_j(\btheta) \geq \bar{\rho}
\end{cases}
\nonumber
\end{align}
where $\tilde{\pi}_j(\btheta)= (1 + e^{-{\bf f}_j^\top \btheta})^{-1}$.

Let $p(\gamma_j \mid \btheta)= \pi_j(\btheta)^{\gamma_j} [1 - \pi_j(\btheta)]^{1 - \gamma_j}$ be the corresponding prior probability mass function.
We prove that 
\begin{align}
\nabla_\btheta p(\gamma_j \mid \btheta)= \begin{cases}
0 \mbox{, if } \tilde{\pi}_j(\btheta) < \underline{\rho} \mbox{ or } \tilde{\pi}_j(\btheta) > \bar{\rho}
\\
\mbox{undefined, if } \tilde{\pi}_j(\btheta) = \underline{\rho} \mbox{ or } \tilde{\pi}_j(\btheta) = \bar{\rho}
\\
{\bf f}_j p(\gamma_j \mid \btheta) [\gamma_j - \pi_j(\theta)] \mbox{, if } \tilde{\pi}_j(\btheta) \in (\underline{\rho}, \bar{\rho})
\end{cases}
\label{eq:deriv_priorinclprob}
\end{align}

The first line in \eqref{eq:deriv_priorinclprob} holds trivially, since in that case $p(\gamma_j \mid \btheta)$ does not depend on $\btheta$.
The second line in \eqref{eq:deriv_priorinclprob} follows immediately by proving the third line, since then the directional derivatives at $\underline{\rho}$ and $\bar{\rho}$ do not match.
To prove the third line in \eqref{eq:deriv_priorinclprob}, note that when $\tilde{\pi}_j(\btheta) \in (\underline{\rho}, \bar{\rho})$ we have
\begin{align}
\nabla_\btheta \pi_j(\btheta)= \nabla_\btheta \tilde{\pi}_j(\btheta)=
\frac{{\bf f}_j e^{-{\bf f}_j^\top \btheta}}{(1 + e^{-{\bf f}_j^\top \btheta})^2}=
{\bf f}_j \pi_j(\btheta) [1 - \pi_j(\btheta)].
\nonumber
\end{align}

Finally, we obtain $\nabla_\btheta p(\gamma_j \mid \btheta)$ separately for the $\gamma_j=1$ and $\gamma_j=0$ cases. For the case $\gamma_j=1$,
\begin{align}
\nabla_\btheta p(\gamma_j=1 \mid \btheta)= \nabla_\btheta \pi_j(\btheta)= {\bf f}_j \pi_j(\btheta) [\gamma_j - \pi_j(\btheta)]=
{\bf f}_j p(\gamma_j \mid \btheta) [\gamma_j - \pi_j(\btheta)],
\nonumber
\end{align}
proving the result. For the case $\gamma_j=0$,
\begin{align}
\nabla_\btheta p(\gamma_j=0 \mid \btheta)= \nabla_\btheta [1 - \pi_j(\btheta)]= -{\bf f}_j \pi_j(\btheta) [1 - \pi_j(\btheta)]=
{\bf f}_j p(\gamma_j \mid \btheta) [\gamma_j - \pi_j(\btheta)],
\nonumber
\end{align}
since $p(\gamma_j=0 \mid \btheta)= 1 - \pi_j(\btheta)$, again proving the result.

\subsection{Proof of Proposition 3.1}

The empirical Bayes estimate writes
\begin{align*}
\bthetaeb &= \argmax_{\btheta \in \mathbb{R}^{T+1}} \log p(\by \mid \btheta)
= \argmax_{\btheta \in \mathbb{R}^{T+1}} \log \sum_{(\bgamma, \bdelta)} p(\by \mid \bgamma, \bdelta) p(\bgamma, \bdelta \mid \btheta).
\end{align*}
For short, denote $H(\btheta) = p(\by \mid \btheta)$ 
where generically $\nabla_{\btheta} \log H(\btheta) = \nabla_{\btheta} H(\btheta) / H(\btheta)$. Under the assumptions of Proposition \ref{prop:one}
\begin{eqnarray}
&&\nabla_{\btheta} H(\btheta) = \sum_{(\bgamma, \bdelta)} p(\by \mid \bgamma, \bdelta) p(\bdelta) \nabla_{\btheta} \prod_{j=1}^{J} p(\gamma_j \mid \btheta) \nonumber \\
&=& \sum_{(\bgamma, \bdelta)} p(\by \mid \bgamma, \bdelta) p(\bdelta) \sum_{j=1}^{J} \left( \nabla_{\btheta} p(\gamma_j \mid \btheta) \prod_{j \neq l} p(\gamma_l \mid \btheta) \right).  \label{eq:prop2_eq1}
\end{eqnarray}

Replacing \eqref{eq:deriv_priorinclprob} into \eqref{eq:prop2_eq1}
\begin{multline*}
\nabla_{\btheta} H(\btheta) = \sum_{(\bgamma, \bdelta)} p(\by \mid \bgamma, \bdelta) p(\bdelta) \sum_{j: \pi_j(\btheta) \in  (\underline{\rho},\bar{\rho})} \mb{f}_{j} (\gamma_j - \pi_{j}(\btheta)) \prod_{l=1}^{J} h_{l}(\btheta) \nonumber \\
= \sum_{j: \pi_j(\btheta) \in  (\underline{\rho},\bar{\rho})} \mb{f}_{j} \sum_{(\bgamma, \bdelta)} (\gamma_j - \pi_{j}(\btheta)) p(\by \mid \bgamma, \bdelta) p(\bdelta, \bgamma \mid \btheta) \nonumber \\
= \sum_{j: \pi_j(\btheta) \in  (\underline{\rho},\bar{\rho})} \mb{f}_{j} \left[ (1-\pi_{j}(\btheta)) \sum_{(\bgamma, \bdelta): \gamma_{j}=1} p(\by, \bdelta, \bgamma \mid \btheta) \right.  \left. - \pi_{j}(\btheta) \sum_{(\bgamma, \bdelta): \gamma_{j}=0} p(\by, \bdelta, \bgamma \mid \btheta) \right]. \nonumber
\end{multline*}
Finally
\begin{multline*}
\nabla_{\btheta} \log H(\btheta) = \frac{\nabla_{\btheta} H(\btheta)}{H(\btheta)}= \nonumber \\
\sum_{j: \pi_j(\btheta) \in  (\underline{\rho},\bar{\rho})} \mb{f}_{j} \left[ (1-\pi_{j}(\btheta)) \frac{\sum_{(\bgamma, \bdelta): \gamma_{j}=1} p(\by, \bdelta, \bgamma \mid \btheta)}{\sum_{(\bgamma, \bdelta)} p(\by, \bdelta, \bgamma \mid \btheta)} - \right. \left. \pi_{j}(\btheta) \frac{\sum_{(\bgamma, \bdelta): \gamma_{j}=0} p(\by, \bdelta, \bgamma \mid \btheta)}{\sum_{(\bgamma, \bdelta)} p(\by, \bdelta, \bgamma \mid \btheta)} \right] \nonumber \\
= \sum_{j: \pi_j(\btheta) \in  (\underline{\rho},\bar{\rho})} \mb{f}_{j} [ (1-\pi_{j}(\btheta)) P(\gamma_{j} = 1 \mid \by, \btheta) - \pi_{j}(\btheta) (1 - P(\gamma_{j} = 1 \mid \by, \btheta)) ] \nonumber \\
= \sum_{j: \pi_j(\btheta) \in  (\underline{\rho},\bar{\rho})} \mb{f}_{j} \left[ P(\gamma_{j} = 1 \mid \by, \btheta) - \pi_{j}(\btheta) \right]. \nonumber
\end{multline*}
\hfill $\blacksquare$

\section{Proof of Proposition 3.2}  \label{sec:proof_prop3}

Consider the optima of the marginal likelihood,
\begin{align}
\argmax_{\btheta \in \mathbb{R}^{T+1}} \sum_{(\bdelta, \bgamma)} p_{0}(\bdelta, \bgamma \mid \by) p(\bdelta, \bgamma \mid \btheta)
\label{eq:l1eq1}
\end{align}
where $p_{0}(\bdelta, \bgamma \mid \by)$ are the posterior probabilities under a uniform prior $p_{0}(\bdelta, \bgamma) \propto 1$. We seek to set the parameters $s_t$ and $r_j$ in the approximation
\begin{align}
q(\bdelta, \bgamma \mid \by) = \prod_{t=1}^{T} \text{Bern}(\delta_{t}; s_{t}) \prod_{j=1}^{J} \text{Bern}(\gamma_{j}; r_{j})
\nonumber
\end{align}
using Expectation Propagation. That is, setting and $\mb{r} = (r_1, \dots, r_J)$ such that
\begin{multline*}
\mb{r}^{\textsc{ep}} = \argmax_{\mb{r} \in [0,1]^{J}} \sum_{(\bgamma, \bdelta)} p_{0}(\bdelta, \bgamma \mid \by)  \log \left(\prod_{t=1}^{T} s_{t}^{\delta_{t}} (1-s_{t})^{1-\delta_{t}} \prod_{j=1}^{J} r_{j}^{\gamma_{j}} (1-r_{j})^{1-\gamma_{j}}  \right).
\end{multline*}
and analogously for $\mb{s} = (s_1, \dots, s_T)$. Proceeding elementwise, we derive
\begin{multline*}
r_{j}^{\textsc{ep}} := \argmax_{r_j \in [0,1]} \sum_{(\bgamma, \bdelta)} p_{0}(\bdelta, \bgamma \mid \by) \times \nonumber \\
( \sum_{j=1}^{J} [\gamma_j \log r_{j} + (1 - \gamma_{j}) \log(1-r_{j})] +  \sum_{t=1}^{T} [\delta_t \log s_{j} + (1 - \delta_{t}) \log(1-s_{t})] ) \nonumber \\
= \argmax_{r_j \in [0,1]} \sum_{(\bgamma, \bdelta)} p_{0}(\bdelta, \bgamma \mid \by) \left[ \sum_{j=1}^{J} [\gamma_j \log r_{j} + (1 - \gamma_{j}) \log(1-r_{j})] \right] \nonumber \\
= \argmax_{r_j \in [0,1]} \sum_{j=1}^{J} \sum_{(\bgamma, \bdelta)} p_{0}(\bdelta, \bgamma \mid \by) \left[ \gamma_j \log r_j + (1 - \gamma_j) \log (1 - r_j) \right]. \nonumber
\end{multline*}
Optimizing this expression yields
\begin{eqnarray}
&&\frac{\partial}{\partial r_j} = 0 \Leftrightarrow \sum_{(\bgamma, \bdelta)} p_{0}(\bdelta, \bgamma \mid \by) \left( \frac{\gamma_j}{r_{j}^{\textsc{ep}}} - \frac{1 - \gamma_j}{1 - r_{j}^{\textsc{ep}}} \right) = 0 \nonumber \\
&\Leftrightarrow& \frac{1}{r_{j}^{\textsc{ep}}} \sum_{(\bgamma, \bdelta): \gamma_j = 1} p_{0}(\bdelta, \bgamma \mid \by)  
 - \frac{1}{1 - r_{j}^{\textsc{ep}}} \sum_{(\bgamma, \bdelta): \gamma_j = 0} p_{0}(\bdelta, \bgamma \mid \by) = 0 \nonumber \\
&\Leftrightarrow& \frac{P_{0}(\gamma_j = 1 \mid \by)}{r_{j}^{\textsc{ep}}} - \frac{P_{0}(\gamma_j = 0 \mid \by)}{1 - r_{j}^{\textsc{ep}}} = 0 \nonumber \\
&\Leftrightarrow& r_{j}^{\textsc{ep}} = P_{0}(\gamma_j = 1 \mid \by) = P(\gamma_j = 1 \mid \by, \btheta = \mb{0}). \label{eq:prop3_eq2}
\end{eqnarray}
With the same exact procedure one analogously obtains $s_{t}^{\textsc{ep}} = P_{0}(\delta_t = 1 \mid \by)$.
\hfill $\blacksquare$

\section{Derivations}
\label{ssec:proofs}

\subsection{Derivation of Equation 3.9}
\label{ssec:proof_eq_thetaep}

Let
\begin{align*}
h(\bdelta) := \prod_{t=1}^{T} \left[ s_{t}^{\textsc{ep}} \pi_t \right]^{\delta_t} \left[ (1-s_{t}^{\textsc{ep}}) (1-\pi_t) \right]^{1-\delta_t},
\end{align*}
which is independent of $\btheta$, and where $\pi_{t}$ is the marginal prior inclusion probability for treatment $t$ (by default, $\pi_t=1/2$). 
Then, taking \eqref{eq:l1eq1} and replacing $p_0(\bdelta,\bgamma \mid \by)$ by the approximation given by \eqref{eq:prop3_eq2} gives
\begin{multline}
\bthetaep := \argmax_{\btheta \in \mathbb{R}^{T+1}} \sum_{(\bgamma, \bdelta)} h(\bdelta) \prod_{j=1}^{J} \left[ r_{j}^{\textsc{ep}} \pi_j(\btheta) \right]^{\gamma_j} \left[ (1-r_{j}^{\textsc{ep}}) (1-\pi_j(\btheta)) \right]^{1-\gamma_j}. 
\nonumber
\end{multline}
The terms inside the sum in the right-hand side defines a probability distribution on $(\delta_1, \dots, \delta_{T}, \gamma_1, \dots, \gamma_{J})$ with independent Bernoulli components, hence their sum is the normalizing constant of said distribution. Since the distribution has independent components, the normalizing constant is just the product of the univariate normalizing constants. The univariate normalizing constant of each Bernouilli is then
\begin{align*}
r_{j}^{\textsc{ep}} \pi_j(\btheta) + (1-r_{j}^{\textsc{ep}}) (1-\pi_j(\btheta))
\end{align*}
for every $r_{j}$, and similarly $s_{t}^{\textsc{ep}} \pi_t + (1-s_{t}^{\textsc{ep}}) (1-\pi_t)$ for every $s_{t}$.
Hence, we directly obtain
\begin{equation}
\bthetaep := \argmax_{\btheta \in \mathbb{R}^{T+1}} \sum_{j=1}^{J} \log \left( r_{j}^{\textsc{ep}} \pi_j(\btheta) + (1-r_{j}^{\textsc{ep}}) (1-\pi_j(\btheta)) \right). \label{eq:eb_res}
\end{equation}
\hfill $\blacksquare$

\subsection{Gradient of Equation 3.9} \label{sec:gradEP}

Note first that, since $\pi_j(\btheta)$ is constant when $\tilde{\pi}_j(\btheta) \not\in (\underline{\rho},\bar{\rho})$, the gradient for such terms is zero. We hence focus on $j: \pi_j(\btheta) \in (\underline{\rho},\bar{\rho})$, since in that case $\pi_j(\btheta)= \tilde{\pi}_j(\btheta)$.

Denote $h_{j}(\btheta) := r_{j} \pi_{j}(\btheta) + (1-r_{j}) (1-\pi_{j}(\btheta))$ for short. Simple algebra provides
\begin{align*}
\nabla_{\btheta} h_{j}(\btheta) = (2 r_j - 1) \nabla_{\btheta} \pi_{j}(\btheta).
\end{align*}
From \eqref{eq:deriv_priorinclprob} we recover the remaining gradient in the last expression and derive
\begin{align*}
\nabla_{\btheta} \log h_{j}(\btheta) &= \frac{\nabla_{\btheta} h_{j}(\btheta)}{h_{j}(\btheta)} \\ &= \frac{2 r_{j} - 1}{h_{j}(\btheta)} \left[ (1-2\rho) \mb{f}_{j}  \pi_{j}(\btheta) (1-\pi_{j}(\btheta)) \right],
\end{align*}
where $\mb{f}_{j} = \tr{(1, f_{j,1}, \dots, f_{j,T})}$, and so the gradient for the expression in \eqref{eq:eb_res} is then
\begin{align}
\nabla_{\btheta} \sum_{j: \pi_j(\btheta) \in (\underline{\rho},\bar{\rho})} \log h_{j}(\btheta) = \sum_{j: \pi_j(\btheta) \in (\underline{\rho},\bar{\rho})} \mb{f}_{j} \frac{\pi_{j}(\btheta) (1-\pi_{j}(\btheta))}{h_{j}(\btheta)}.
\label{eq:grad_ep_alternative}
\end{align}

Finally, note that
\begin{align}
\pi_j(\btheta) (1-\pi_j(\btheta))=
\pi_j(\btheta) (r_j - r_j \pi_j(\btheta) + (1-r_j) - \pi_j(\btheta) + r_j \pi_j(\btheta)  )=
\pi_j(\btheta) r_j - \pi_j(\btheta) h_j(\btheta),
\nonumber
\end{align}
giving that \eqref{eq:grad_ep_alternative} is
\begin{align}
\sum_{j: \pi_j(\btheta) \in (\underline{\rho},\bar{\rho})} \mb{f}_{j} 
\left[ \frac{\pi_{j}(\btheta) r_j}{h_{j}(\btheta)} - \pi_j(\btheta) \right]
= \sum_{j: \pi_j(\btheta) \in (\underline{\rho},\bar{\rho})} \mb{f}_{j} 
\left[P^{EP}(\gamma_j=1 \mid \by, \btheta) - \pi_j(\btheta) \right].
\nonumber
\end{align}
where
\begin{align}
P^{EP}(\gamma_j=1 \mid \by, \btheta)= \frac{\pi_{j}(\btheta) r_j}{h_{j}(\btheta)},
\nonumber
\end{align}

\hfill $\blacksquare$

\section{Product MOM non-local prior}
\label{ssec:pmom}
 
\begin{figure}[h]
\centering
\includegraphics[scale=0.6]{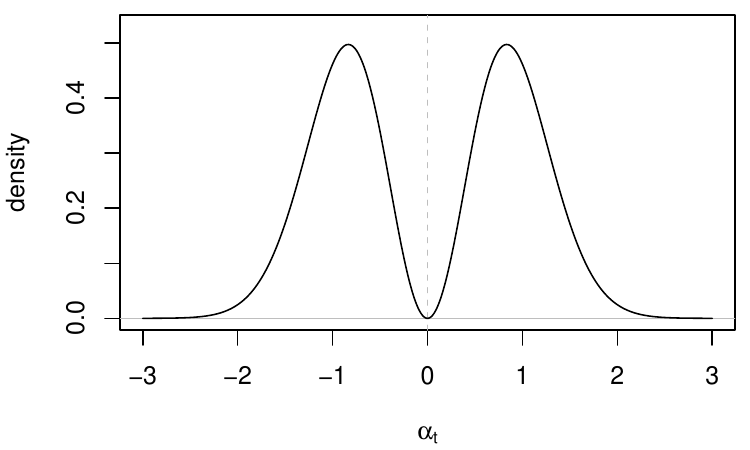} 
\caption{Prior density $p(\alpha_{t} \mid \delta_{t} = 1, \phi = 1)$ of the MOM non-local prior, with $\tau = 0.348$.}
\label{fig:nlpmom}
\end{figure}

Figure \ref{fig:nlpmom} illustrates the density of the product MOM non-local prior of \cite{Johnson12}.
 
\section{Computational Methods}
\label{ssec:computatinal_methods}

\subsection{Numerical computation of the marginal likelihood for non-local priors} \label{subsec:nlp_approx}

Briefly, denote by $p^\textsc{n}(\alpha_t \mid \delta_t=1, \phi) = \text{N}(\alpha_t; 0, \tau \phi)$ independent Gaussian priors for $t=1,..,T$, and similarly $p^\textsc{n}(\beta_j \mid \gamma_j=1, \phi) = \text{N}(\beta_j; 0, \tau \phi)$ for $j=1,\ldots, J$. Proposition 1 in \cite{Rossell17} shows that the following identity holds exactly
\begin{align*}
p(\by \mid \bgamma, \bdelta)= p^\textsc{n}(\by \mid \bgamma, \bdelta) \E^\textsc{n} \left[ \prod_{t=1}^{T} \frac{\alpha_t^2}{\tau\phi} \prod_{j=1}^{J} \frac{\beta_j^2}{\tau\phi} \mid \by, \bgamma, \bdelta \right]
\end{align*}
where $p^\textsc{n}(\by \mid \bgamma, \bdelta)$ is the integrated likelihood under $p^\textsc{n}(\balpha, \bbeta)$, and $\E^\textsc{n}[\cdot]$ denotes the posterior expectation under $p^\textsc{n}(\balpha, \bbeta \mid \by, \bgamma, \bdelta)$.
To estimate $p^\textsc{n}(\by \mid \bgamma, \bdelta)$ for non-Gaussian outcomes we use a Laplace approximation. Regarding the second term, we approximate it by a product of expectations, which \cite{Rossell20a} showed leads to the same asymptotic properties and typically enjoys better finite-$n$ properties than a Laplace approximation.

\subsection{Numerical optimization for empirical Bayes hyper-parameters} \label{sec:algorithm}

Algorithm \ref{alg:one} describes our method to estimate $\bthetaep$ and $\bthetaeb$. We employ the quasi-Newton BFGS algorithm to optimize the objective function. For $\bthetaeb$, we use the gradients from Proposition \ref{prop:one}, while the Hessian is evaluated numerically using line search, with the \texttt{R} function \texttt{nlminb}. Note, however, that obtaining $\bthetaeb$ requires sampling models from their posterior distribution for each $\btheta$, which is impractical, to then obtain posterior inclusion probabilities required by \eqref{eq:prop_two}. Instead, we restrict attention to the models $M$ sampled for either $\btheta = \mb{0}$ or $\btheta = \bthetaep$ in order to avoid successive MCMC runs at every step, relying on the relative regional proximity between the starting point $\bthetaep$ and $\bthetaeb$. This proximity would ensure that $M$ contains the large majority of models with non-negligible posterior probability under $\bthetaeb$. For $\bthetaep$, we use employ the same BFGS strategy using gradient computed in \ref{sec:gradEP}, with numerical evaluation of the Hessian. This computation requires only one MCMC run at $\btheta = \mb{0}$, which allows us to use grid search to avoid local optima. As for the size of the grid, we let the user specify what points are evaluated. For $K$ points in the grid one must evaluate the log objective function $K^{T+1}$ times, so we recommend to reduce the grid density as $T$ grows. By default, we evaluate every integer in the grid assuming $T$ is not large, but preferably we avoid coordinates greater than 10 in absolute value, as in our experiments it is very unlikely that any global posterior mode far from zero is isolated, i.e. not reachable by BFGS by starting to its closest point in the grid. Additionally, even if that were the case, numerically it makes no practical difference, considering that marginal inclusion probabilities are bounded away from zero and one regardless.

\begin{algorithm} \label{alg:one}
\textbf{Output:} Estimates for $\bthetaep$ and $\bthetaeb$

\textbf{1:} Obtain $B$ posterior samples $(\bgamma, \bdelta)^{(b)} \sim p(\bgamma, \bdelta \mid \by, \btheta = \mb{0})$ for $b = 1,\ldots,B$. Denote by $M^{(0)}$ the corresponding set of unique models.\\
\textbf{2:} Compute $s_{t} = P(\delta_{t} = 1 \mid \by, \btheta = \mb{0})$ and $r_{j} = P(\gamma_{j} = 1 \mid \by, \btheta = \mb{0})$.\\
\textbf{3:} Conduct a grid search for $\bthetaep$ around $\btheta={\bf 0}$. Optimize \eqref{eq:theta-ep} with the BFGS algorithm initialized at the grid's optimum.\\
\textbf{4:} Obtain $B$ posterior samples $(\bgamma, \bdelta)^{(b)} \sim p(\bgamma, \bdelta \mid \by, \btheta = \bthetaep)$. Denote by $M^{(1)}$ the corresponding set of unique models. Set $M = M^{(0)} \cup M^{(1)}$.\\
\textbf{5:} Initialize search for $\bthetaeb$ at $\bthetaep$. Use the BFGS algorithm to optimize \eqref{eq:marglik1}, restricting the sum to $(\bdelta,\bgamma) \in M$.
\caption{Obtaining estimates for $\bthetaep$ and $\bthetaeb$}
\end{algorithm}

\section{Derivation of Bayes factor asymptotics}
\label{ssec:bf_asymptotics}

Proposition 1(i) in \cite{Rossell17} gives that the Bayes factor under the pMOM prior
$p(\alpha_{t} \mid \delta_{t} = 1, \phi) = \frac{\alpha_{t}^{2}}{\phi \tau/ v_t} \text{N}(\alpha_{t}; 0, \phi \tau / v_t)$
and
$p(\beta_j \mid \gamma_j = 1, \phi) = \frac{\beta_j^{2}}{\phi \tau/ v_j} \text{N}(\beta_j; 0, \phi \tau / v_j)$ is equal to
\begin{align}
\frac{p(\by \mid \bdelta, \bgamma)}{p(\by \mid \bdelta^*,\bgamma^*)}=
\frac{E^N\left(\prod_{\delta_t=1} \alpha_t^{2} v_t/[\phi \tau] \prod_{\gamma_j=1} \beta_j^{2} v_j/[\phi \tau] \mid \by, \bdelta,\bgamma \right)}{E^N\left(\prod_{\delta_t^*=1} \alpha_t^{2} v_t/[\phi \tau] \prod_{\gamma_j^*=1} \beta_j^{2} v_j/[\phi \tau] \mid \by, \bdelta^*,\bgamma^* \right)}
\frac{p^N(\by \mid \bdelta, \bgamma)}{p^N(\by \mid \bdelta^*,\bgamma^* \mid \by)}
\label{eq:bf_pmom}
\end{align}
where
\begin{align}
p^N(\by \mid \bdelta, \bgamma)=
\int p(\by \mid \balpha_\delta, \bbeta_\gamma) N(\balpha_\delta; 0, \phi \tau V_\delta) N(\bbeta_\gamma; 0, \phi \tau V_\gamma)
d\balpha_\delta d\bbeta_\gamma
\nonumber
\end{align}
is the marginal likelihood under the Normal prior featuring in the pMOM density, 
$E^N()$ denotes a posterior expectation under said Normal prior,
$V_\delta$ is diagonal with entries given by the $v_t$'s and $V_\gamma$ is diagonal with entries given by the $v_j$'s.

Asymptotics for \eqref{eq:bf_pmom} are obtained by studying separately the first term involving the ratio of posterior means, and the second term featuring the Bayes factor obtained under the Normal prior.
Before presenting the arguments, we outline further notation and assumptions needed for the results to hold.

In this section, for simplicity we denote by $\bm{\zeta}=(\balpha, \bbeta)$ the whole set of parameters and by $\bm{\zeta}_{\delta,\gamma}$ the subset of non-zero values under model $(\bdelta,\bgamma)$.

Assume that $y_1,\ldots,y_n$ arise independently from an unknown data-generating distribution $F^*$, which may depend on treatments and covariates.
For any model $(\bdelta,\bgamma)$, let the Kullback-Leibler optimal parameter value be
$$
\bm{\zeta}_{\delta,\gamma}^*= \arg\max_{\bm{\zeta}_{\delta,\gamma}} {\mathbb E}_{F^*} \left[ \log p(y_1 \mid \bm{\zeta}_{\delta,\gamma}, \delta,\gamma) \right],
$$
and $H^*_{\delta,\gamma}$ its hessian evaluated at $\bm{\zeta}_{\delta,\gamma}=\bm{\zeta}_{\delta,\gamma}^*$.

\subsection{Technical conditions}
\label{ssec:tech_conditions}

Let $\widehat{\bm{\zeta}}_{\delta,\gamma}$ be the MLE under model $(\bdelta,\bgamma)$ and
$$
A_n({\bf s})= \log p(\by \mid \widehat{\bm{\zeta}}_{\delta,\gamma}, \bdelta,\bgamma) - \log p(\by \mid \widehat{\bm{\zeta}}_{\delta,\gamma} + {\bf s}/\sqrt{n}, \bdelta,\bgamma)= - \frac{{\bf s}^{\top} H(\bm{\zeta}^*_{\delta,\gamma}) {\bf s}}{2 n} + r_n({\bf s}/\sqrt{n}),
$$
where $r_n$ is the error in the second-order Taylor log-likelihood expansion at $\bm{\zeta}_{\delta,\gamma}^*$.
Let $H(\bm{\zeta}_{\delta,\gamma})$ be the log-likelihood hessian evaluated at $\bm{\zeta}_{\delta,\gamma}$.
We assume that
\begin{enumerate}[leftmargin=*, label={\bf D\arabic*.}]
\setcounter{enumi}{-1}
\item $P\left( A_n({\bf s}) \text{ is convex in } {\bf s}\right ) \to 1$, as $n\to \infty$

\item $\widehat{\bm{\zeta}}_{\delta,\gamma} \stackrel{P}{\longrightarrow} \bm{\zeta}_{\delta,\gamma}^*$ under $F^*$, as $n \rightarrow \infty$.
 
\item $H(\bm{\zeta}_{\delta,\gamma}^*)/n \stackrel{P}{\longrightarrow} H_{\delta,\gamma}^*$ under $F^*$, as $n \rightarrow \infty$, for a strictly positive-definite $H^*_{\delta,\gamma}$.
 

\item $\min_{\alpha_t^* \neq 0} |\alpha_t^*| \geq a$ and $\min_{\gamma_j^* \neq 0} |\beta_j^*| \geq b$ for some constants $a,b>0$.
\end{enumerate}

Condition D0 requires that the log-likelihood is convex around the MLE, which holds with probability 1 at any ${\bm \zeta}_{\delta,\gamma}$ for generalized linear models with the canonical link.
Conditions D1-D2 are minimal. If one assumes that $\bm{\zeta}_{\delta,\gamma}$ has bounded support, then D1 holds \citep{hjort:2011} and D2 also holds provided $|H(\bm{\zeta}_{\delta,\gamma}^*)|$ has finite mean, by the continuous mapping theorem and strong law of large numbers.
More generally one may show the asymptotic validity of a Taylor log-likelihood expansion around $\bm{\zeta}_{\delta,\gamma}^*$
and establish asymptotic normality of $\widehat{\bm{\zeta}}_{\delta,\gamma}$, see Theorem 4.1 in \cite{hjort:2011}.
Condition D3 is a beta-min condition that can be relaxed to allow for vanishing $(a,b)$, as long as $(a,b)$ are larger than $\sqrt{n}^{-1/2}$ times logarithmic terms, but here we assume fixed $(a,b)$ to obtain simpler expressions for the asymptotic Bayes factor rates.

\subsection{Laplace approximation to Bayes factors}

Assuming Conditions D0-D2, plus a prior boundedness condition that is satisfied in our setting, Proposition S6 in \cite{rossell:2021b} gives that
\begin{align}
\frac{p^N(\bdelta, \bgamma \mid \by, \btheta)/p^N(\bdelta^*,\bgamma^* \mid \by, \btheta)}
{[\hat{p}^N(\by \mid \bdelta, \bgamma, \btheta)/\hat{p}^N(\by \mid \bdelta^*,\bgamma^*, \btheta)] \frac{p(\bdelta) p(\bgamma \mid \btheta)}{p(\bdelta^*) p(\bgamma^* \mid \btheta)}} \stackrel{P}{\longrightarrow} 1
\label{eq:asymp_posteriorodds}
\end{align}
as $n \rightarrow \infty$, where 
\begin{align}
\frac{\hat{p}^N(\by \mid \bdelta, \bgamma, \btheta)}{\hat{p}^N(\by \mid \bdelta^*,\bgamma^*, \btheta)}=
e^{L(\bdelta, \bgamma)/2} \frac{p(\bm{\zeta}_\gamma^* \mid \bgamma,\bdelta)}{p(\bm{\zeta}_{\gamma^*}^* \mid \bgamma^*,\bdelta^*)}
\left(\frac{2 \pi}{n}\right)^{d/2}
\frac{|H^*_{\delta^*,\gamma^*}|^{1/2}}{|H^*_{\delta,\gamma}|^{1/2}},
\label{eq:asymp_bf}
\end{align}
is a Laplace approximation to the Bayes factor between $(\bdelta,\bgamma)$ and $(\bdelta^*,\bgamma^*)$ under the Normal prior,
$L(\bdelta, \bgamma)$ is the corresponding likelihood-ratio test statistic,
and $d$ is the difference between the number of non-zero parameters in $(\bdelta,\bgamma)$ and $(\bdelta^*,\bgamma^*)$.

Therefore,
in our asymptotic study we may replace the Bayes factor under Normal priors on the right-hand side of \eqref{eq:bf_pmom} by its Laplace approximation in \ref{eq:asymp_bf}.

\subsection{Bayes factor rates}

The frequentist properties of the ratio of posterior expectations in \eqref{eq:bf_pmom} and the Laplace approximation to the Bayes factor in \eqref{eq:asymp_bf} have been well-studied, e.g. see \cite{rossell:2018b} (Proposition 5) for Gaussian outcomes and \cite{rossell:2021b} (Propositions 3-4) for certain survival and generalized linear models.
We now summarize the results.
When $(\bdelta,\bgamma)$ is an overfitted model,
combining Expressions \eqref{eq:bf_pmom} and \eqref{eq:asymp_posteriorodds}, one may show that
$p(\by \mid \bdelta, \bgamma) / p(\by \mid \bdelta^*,\bgamma^*)= (n\tau)^{-3d/2} \times O_p(1)$,
where $d= |\bdelta|_0 + |\bgamma|_0 - |\bdelta^*|_0 - |\bgamma^*|_0$ is the difference between model dimensions.
In contrast, when $(\bdelta,\bgamma)$ is a non-overfitted model, then
\begin{align}
\log \frac{p(\by \mid \bdelta, \bgamma)}{p(\by \mid \bdelta^*, \bgamma^*)}= - \frac{3d}{2} \log(n\tau) - n c + O_p(1),
\label{bf:non_overfitted}
\end{align}
where $c>0$ is a constant that depends on $(\balpha^*, \bbeta^*)$. Under Condition (D3), $c$ can be taken to be a fixed constant (i.e. not depending on $n$).

These Bayes factor rates, combined with the bounds on $\pi_j(\btheta) \in [\underline{\rho},\bar{\rho}]$, give Bayes factor rates for CIL under any given hyper-parameter $\btheta$.
Below, we denote by $d_1= \sum_{j=1}^J \gamma_j (1-\gamma_j^*)$ the number of covariates included in $\bgamma$ but not in $\bgamma^*$,
and by $d_2= \sum_{j=1}^J (1-\gamma_j) \gamma_j^*$ that of covariates included in $\bgamma^*$ but not in $\bgamma$.

Consider first overfitted models. Using that
\begin{align}
\frac{p(\by \mid \bdelta, \bgamma)}{p(\by \mid \bdelta^*,\bgamma^*)}=
(n\tau)^{-3d/2} \times O_p(1),
\nonumber
\end{align}
and that $p(\bdelta)=p(\bdelta^*)$ under our prior, we obtain
\begin{align}
\frac{p(\bdelta, \bgamma \mid \by, \btheta)}{p(\bdelta^*,\bgamma^* \mid \by, \btheta)} &=
(n\tau)^{-3d/2}
\prod_{\gamma_j=1, \gamma_j^*=0} \frac{\pi_j(\btheta)}{1 - \pi_j(\btheta)}
\prod_{\gamma_j=0, \gamma_j^*=1} \frac{1-\pi_j(\btheta)}{\pi_j(\btheta)}
 \times O_p(1)
\nonumber \\
&\leq 
(n\tau)^{-3d/2}
\left( \frac{\bar{\rho}}{1 - \bar{\rho}} \right)^{d_1}
\left( \frac{1-\underline{\rho}}{\underline{\rho}} \right)^{d_2}
 \times O_p(1).
\nonumber
\end{align}
Note that for over-fitted models $d_1=d$ and $d_2=0$, giving
\begin{align}
\frac{p(\bdelta, \bgamma \mid \by, \btheta)}{p(\bdelta^*,\bgamma^* \mid \by, \btheta)} \leq
(n\tau)^{-3d/2} \left( \frac{\bar{\rho}}{1 - \bar\rho} \right)^d \times O_p(1)
,
\nonumber
\end{align}
as we wished to show.

Consider now non-overfitted models. Using \eqref{bf:non_overfitted} and that $p(\bdelta)=p(\bdelta^*)$ gives
\begin{align}
&\log \left( \frac{p(\bdelta, \bgamma \mid \by, \btheta)}{p(\bdelta^*,\bgamma^* \mid \by, \btheta)} \right)=
-\frac{3d}{2} \log(n\tau) - n c 
\nonumber \\
&+ \log \left( \prod_{\gamma_j=1, \gamma_j^*=0} \frac{\pi_j(\btheta)}{1 - \pi_j(\btheta)} \right)
+ \log \left( \prod_{\gamma_j=0, \gamma_j^*=1} \frac{1-\pi_j(\btheta)}{\pi_j(\btheta)} \right)
+ O_p(1).
\nonumber
\end{align}
Noting that $\pi_j(\btheta) \in [\underline{\rho}, \bar{\rho}]$,
that there are $d_1$ terms such that $(\gamma_j=1, \gamma_j^*=0)$,
and that there are $d_2$ terms such that $(\gamma_j=0, \gamma_j^*=1)$, gives
\begin{align}
&\log \left( \frac{p(\bdelta, \bgamma \mid \by, \btheta)}{p(\bdelta^*,\bgamma^* \mid \by, \btheta)} \right) \leq
-\frac{3d}{2} \log(n\tau) - n c 
+ d_1\log \left( \frac{\bar{\rho}}{1 - \bar{\rho}}  \right)
+ d_2\log \left( \frac{1 - \underline{\rho}}{\underline{\rho}}  \right)
+ O_p(1)
\nonumber
\end{align}
as we wished to prove.

\section{Supplementary Results}
\label{ssec:suppl_results}

\subsection{Illustration of the EB and EP objective functions}

\begin{figure}[h]
\centering
\begin{tabular}{cc}
\includegraphics[width=0.48\textwidth]{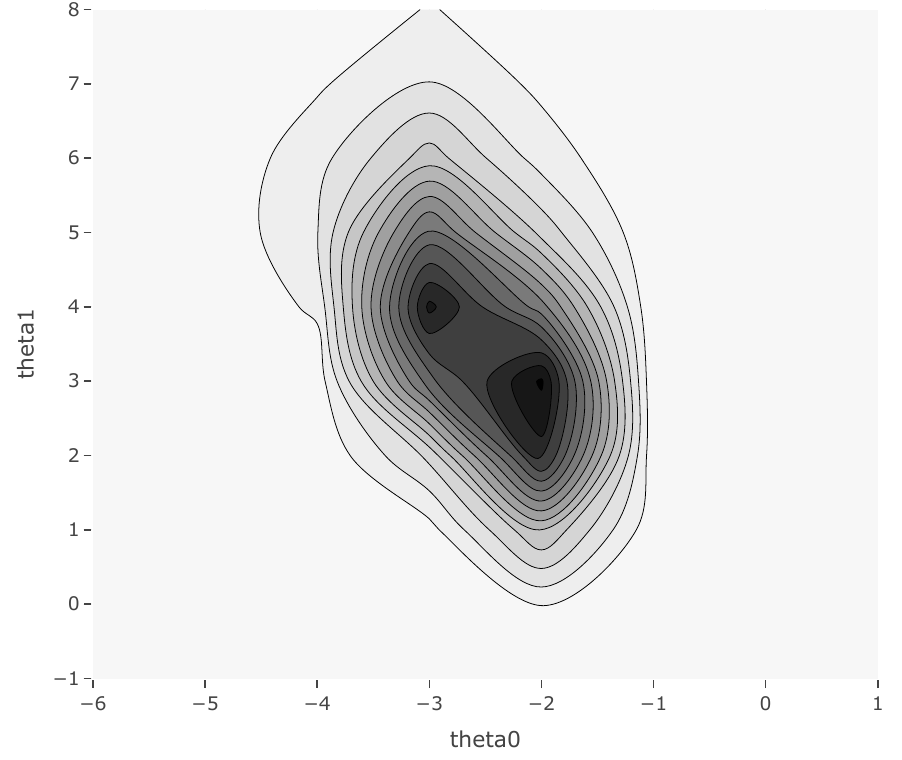}  &
\includegraphics[width=0.48\textwidth]{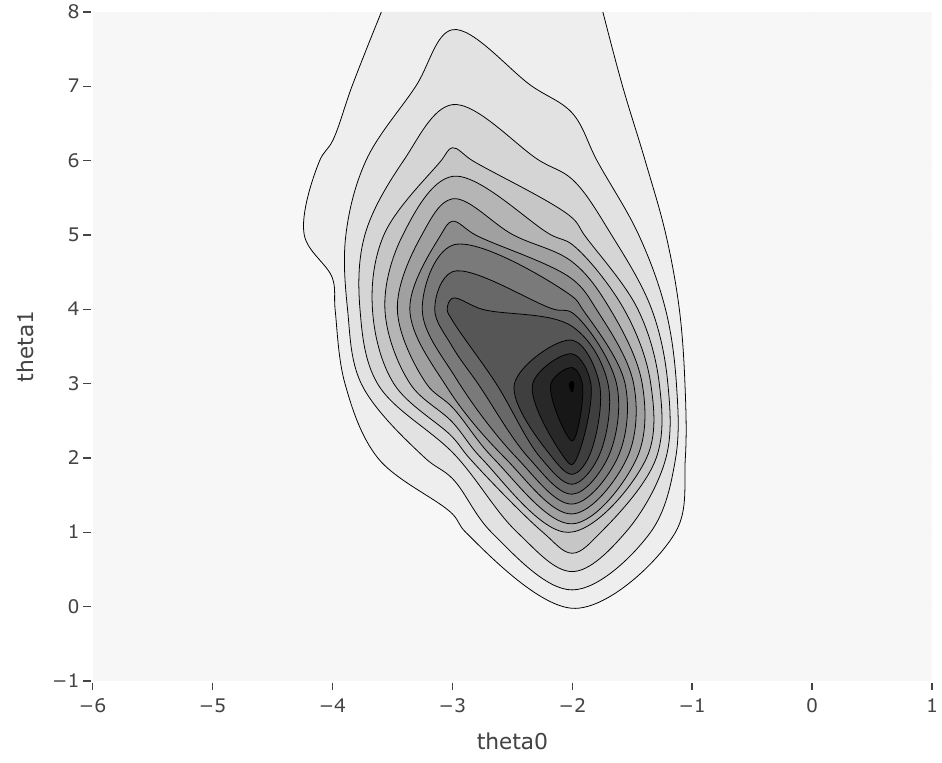} 
\end{tabular}
\caption{Empirical Bayes (left) and Expectation-Propagation (right) objective functions (3.8) 
and (3.9) 
in the single treatment case ($T=1$). Here, $\bthetaeb = (-2.43, 3.19)$ and $\bthetaep = (-2.34, 3.09)$, for $n=100$ and $J=49$, for the first data realization for the simulation design displayed in the center-left panel of Figure 1 
with three confounders. See Section 5.1 
for further details.}
\label{fig:thEPEB}
\end{figure}

Figure \ref{fig:thEPEB} shows the Empirical Bayes objective function in (3.8) 
and (3.9) 
in a simulated dataset with a single treatment. A bimodality is appreciated in the left panel.

\subsection{Salary survey: obtention and pre-processing of CPS microdata} \label{sec:cps_preproc}

Current Population Surveys are administered monthly by the U.S. Bureau of the Census to over 65,000 households. The resulting microdata is made freely available to the public by the Integrated Public Use Microdata Series (IPUMS) website upon registration at:
\begin{itemize}
\item \texttt{https://cps.ipums.org/cps/}
\end{itemize}
We manually download the data including all indicators available for \texttt{03-2010} and \texttt{03-2019}, which include data from the Annual Social and Economic Supplement. All transformations necessary to undertake the different analyses presented in this article are openly accessible at:
\begin{itemize}
\item \texttt{https://github.com/mtorrens/cil\_article}
\end{itemize}
The user is advised to carefully read the \texttt{README.md} file before replicating the analyses. For the CPS raw data pre-processing, we refer to the two scripts created to perform said tasks (in the appropriate order):
\begin{itemize}
\item \texttt{source/04a\_cps\_format.R}
\item \texttt{source/04b\_cps\_transform.R}.
\end{itemize}

\subsection{Salary survey: generation of augmented datasets} \label{sec:fakepreds_supp}

For both amounts $K_1=100$ and $K_2=200$ of artificial predictors, the simulation protocol is the same. Every artificial covariate $\mb{z}_{k} \in \mathbb{R}^{n}$, for $k=1,\dots,100$ or $k=1,\dots,200$ respectively, is simulated to correlate to one individual treatment, according to which subset said covariate is assigned to, correlating only indirectly to the rest of treatments. In particular, we drew elements of $\mb{z}_{k}$ from $z_{i,k} \mid d_{i,t} = 1 \sim \text{N}(1.5, 1)$, and $z_{i,k} \mid d_{i,t} = 0 \sim \text{N}(-1.5, 1)$, where $\bd_{t}$ denotes the corresponding column in the treatment matrix associated to the given $\mb{z}_{k}$. 

\subsection{Further results on salary survey} \label{sec:salary_supp}

\begin{figure}[h]
\centering
\begin{tabular}{cc}
\includegraphics[width=0.48\textwidth]{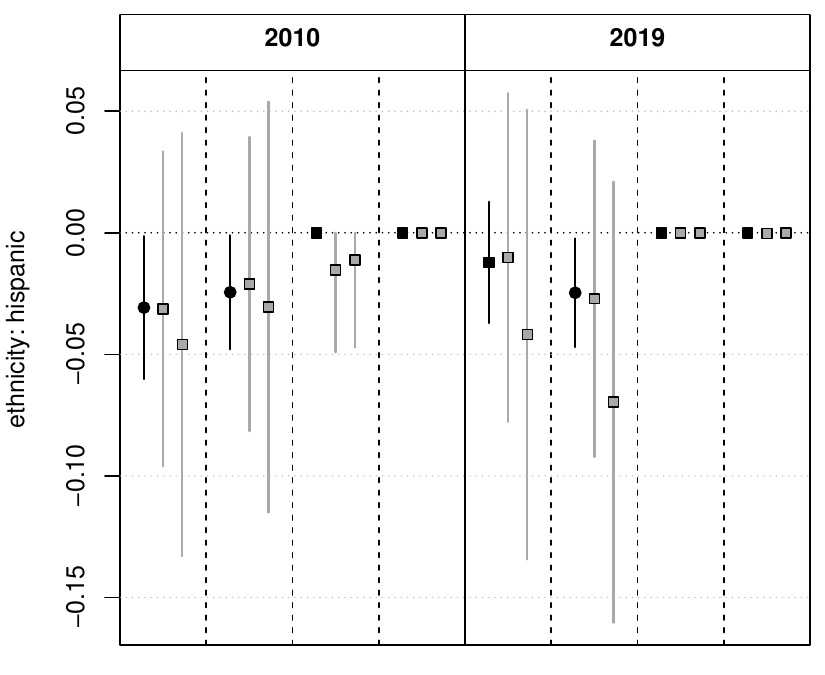} &
\includegraphics[width=0.48\textwidth]{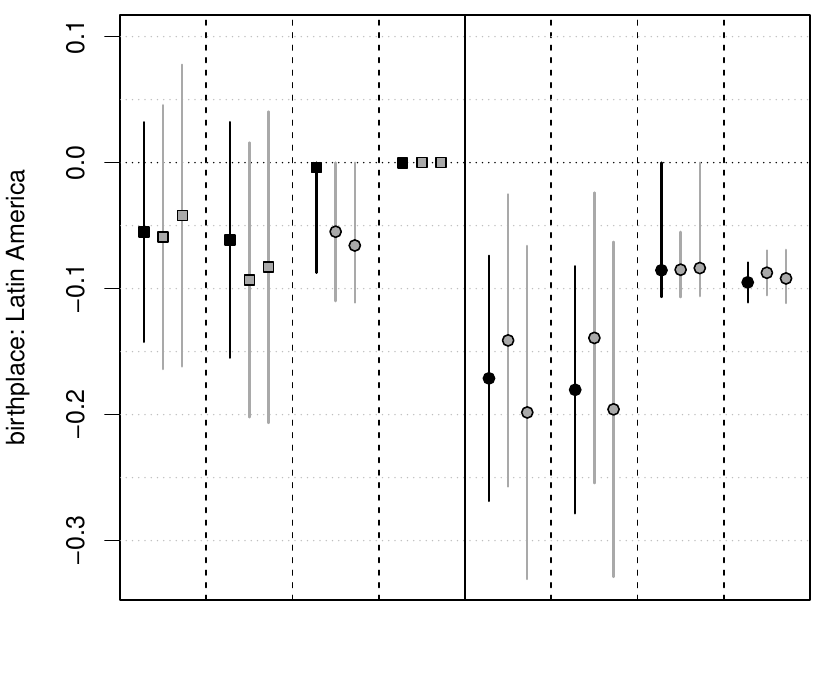}
\end{tabular}
\caption{Inference for treatment variables ``hispanic'' (top) and ``born in Latin America'' (bottom) in 2010 and 2019. Read caption to Figure 4 
to read this figure, including method labels (from left to right: OLS, DL, BMA, CIL).}
\label{fig:figApp1B}
\end{figure}

Figure \ref{fig:figApp1B} follows Figure 1 
by showing the results for the other two treatments: Hispanic ethnicity, and birthplace in Latin America.

Table \ref{tab:cps_y2y_diffs} provides a descriptive analysis of the covariates in the salary data that changed the most between 2010 and 2019. These are covariates with a $p$-value$<0.05$ when assessing their marginal correlation with year (based on a linear for non-binary outcomes, and a chi-squared test for binary outcomes). Further, we only report covariates whose average changed by at least 5\% (in absolute value) between 2010 and 2019. For binary covariates, we also required that their average in 2010 was $>0.05$, to discard covariates that were very rare (or not collected) in 2010. The variable codes used match those supplied by the CPS database.

\newpage

\begin{landscape}

\begin{table}[htbp]
\centering
\begin{tabular}{llccc}
  \hline
Covariate                     & Label      & Type       & Prop. change           & Mean (2010) \\ 
                              &            &            & (2010-19)              &             \\
  \hline
Change of industry from last year                   & \texttt{chindly}              & binary     & $0.093$ & $0.105$ \\    
Change of occupation from last year                 & \texttt{choccly}              & binary     &  $0.051$ & $0.066$ \\   
Has pension plan at work                            & \texttt{pension}              & binary     & $-0.130$ & $0.507$ \\  
Employment coverage (family plan)                   & \texttt{grptyply\_family}     & binary     & $-0.177$ & $0.362$ \\  
Employment coverage (self plan)                     & \texttt{grptyply\_self}       & binary     & $-0.141$ & $0.247$ \\   
Medical out-of-pocket \& Medicare B subsidy (log)   & \texttt{logspmmedxpns}        & non-binary & $0.105$ & $6.645$ \\  
Mortage                                             & \texttt{mortgage}             & binary  &  $-0.101$ & $0.588 $ \\ 
Size of firm where employed (25-99 people)          & \texttt{firmsize\_2(25-99)}   & binary     &  $-0.055$ & $0.131$ \\  
Lunch subsidy                                       & \texttt{lunchsub}             & binary     &  $0.052$ & $0.078$ \\ 
Metropolitan area size $>5$                         & \texttt{areasize\_6(>5)}      & binary     &  $0.063$ & $0.159$ \\ 
Proportion of income from interest                  & \texttt{propoi\_incint}       & non-binary & $0.529$ & $0.295$ \\
Proportion of income from dividends                 & \texttt{propoi\_incdivid}     & non-binary &  $-0.423$ & $0.075$ \\
Weeks unemployed last year (log)                    & \texttt{wksunemly}            & non-binary &  $-0.510$ & $1.296$ \\  
Amount of child tax credit (log \$)                 & \texttt{logctccrd}            & non-binary &  $0.265$ & $1.566$ \\ 
Number of children with  school lunch subsidy       & \texttt{frelunch}             & non-binary &  $0.570$ & $0.142$ \\ 
Log of other person's income                        & \texttt{logotherpersincome}   & non-binary &  $0.117$ & $3.326$ \\  
Number of children who ate school lunch             & \texttt{atelunch}             & non-binary &  $-0.125$ & $0.523$ \\
Proportion of tax income over wages (log)           & \texttt{logproptaxincwageinc} & non-binary &  $-0.062$ & $1.124$ \\
Number of children (log)                            & \texttt{logspmnchild}         & non-binary &  $-0.073$ & $0.972$ \\ 
Number of own children                              & \texttt{nchild}               & non-binary &  $-0.062$ & $1.083$ \\ 
Number of own children under age 5                  & \texttt{nchlt5}               & non-binary &  $-0.107$ & $0.215$ \\
Family market value of school lunch (log)           & \texttt{logschllunch}         & non-binary &  $-0.068$ & $1.426$ \\ 
Number of siblings                                  & \texttt{nsibs}                & non-binary &  $0.193$ & $0.080$ \\  
Number of months receiving food stamps              & \texttt{stampmo}              & non-binary &  $0.147$ & $0.463$ \\
   \hline
\end{tabular}
\caption{Descritive analysis for covariates that changed the most in the salary data between 2010 and 2019. These all displayed a $p$-value$<0.05$ for their marginal association with year, and changed at least by 5\% in absolute value between 2010 and 2019. Label indicates the name of the variable in our processed dataset} 
\label{tab:cps_y2y_diffs}
\end{table}

\end{landscape}

\newpage

\subsection{Further results for abortion data} \label{sec:abortion_supp}

This section contains supplementary results for the abortion data analysis.
The codes for the variable names follow those of \cite{Belloni14b}.
Recall that the main covariates are:
\begin{itemize}
\item \texttt{prison}: log prisoners per capita 
\item \texttt{police}: log police per capita 
\item \texttt{ur}: unemployment rate 
\item \texttt{inc}: income per capita 
\item \texttt{pov}: poverty rate 
\item \texttt{afdc}: Aid to Families with Dependent Children generosity 
\item \texttt{beer}: beer consumption per capita 
\item \texttt{gun}: presence of concealed weapons law (binary)
\end{itemize}

The full list of variable names is given in supplementary file vnames.csv. The nomenclature can be interpreted as follows:

\begin{itemize}
\item Prefix \texttt{D} indicates taking the difference between two consecutive years, e.g. \texttt{Dprison} is the difference of log prisoners per capita in the current vs. previous year
\item Prefix \texttt{L} indicates taking a 1-year lagged value, e.g. \texttt{Lprison} is last year's log prisoners per capita
\item Suffix \texttt{0} indicates the initial value of the variable, e.g. \texttt{prison0} is the initial log prisoners per capita
\item Concatenating variable names with a \texttt{.} indicates interactions, e.g. \texttt{Dprison.Dur} is the interaction (product) between \texttt{Dprison} and \texttt{Dur}
\item Linear interactions are indicated by \texttt{*}, e.g. Dprison*t is the interaction (product) between \texttt{Dprison} and \texttt{time}
\item Suffix \texttt{Bar} indicates the state-level average, e.g. \texttt{prisonBar} is the state's average log prisoners per capita
\item \texttt{xV0}, \texttt{xP0}, \texttt{xM0}: initial violent crime, property crime and murder (respectively)
\item \texttt{DxV0}, \texttt{DxP0}, \texttt{DxM0}: initial difference in violent crime, property crime and murder (respectively)
\end{itemize}

We first discuss violent crime. Tables \ref{tab:abortion_bma_normal} and \ref{tab:abortion_bma_mom} show BMA inference for covariates with marginal posterior inclusion probability above $>0.1$ in the standard BMA analysis with normal and MOM priors (respectively).
In the CIL analyses with normal and MOM priors there are no such covariates.
The top model in the BMA-normal, BMA-MOM, CIL-normal and CIL-MOM analysis contained no confounding covariates, and has posterior probabilities of 0.095, 0.305, 0.923 and 0.670 respectively.

Regarding property crime, the middle panels in Tables \ref{tab:abortion_bma_normal} and \ref{tab:abortion_bma_mom} show results for BMA-Normal and BMA-MOM respectively.
Tables \ref{tab:abortion_cil_normal} and \ref{tab:abortion_cil_mom} show analogous results for CIL-Normal and CIL-MOM.

Finally, for the murder outcome the results from our CIL methodology are very similar to those from standard BMA.
The bottom panels in Tables \ref{tab:abortion_bma_normal} and \ref{tab:abortion_bma_mom} show results for murder under the BMA-Normal and BMA-MOM analyses. The covariate with highest posterior inclusion probability is a term related to the quadratic effect of income.
Said covariate is also the only one receiving non-negligible posterior inclusion probability under the CIL analyses (Tables \ref{tab:abortion_cil_normal}-\ref{tab:abortion_cil_mom}).
In fact, the top model under all analyses contained only this covariate and has a posterior probability of 0.465 for BMA-normal, 0.612 for BMA-MOM, 0.437 for CIL-normal and 0.802 for CIL-MOM.

\begin{table}
\caption{Final DL outcome model for violent crime including covariates found to be related to either the treatment or/outcome. For brevity the intercept and dummy year indicators are omitted}
\begin{center}
\begin{tabular}{|l|ccc|} \hline
 & Estimate & Std. Error & $p$-value \\
\texttt{abortion} & $-0.21$ & $0.13$ & $0.099$ \\ 
\texttt{Lpolice} & $-0.03$ & $0.02$ & $0.184$ \\ 
\texttt{Dinc0*t} & $-26.25$ & $38.51$ & $0.496$ \\ 
\texttt{Dbeer0*t} & $1.24$ & $0.92$ & $0.181$ \\ 
\texttt{Linc0*t} & $-20.31$ & $23.42$ & $0.386$ \\ 
\texttt{Lprison0}$^2$\texttt{*t}$^2$ & $0.03$ & $0.02$ & $0.234$ \\ 
\texttt{prisonBar*t} & $-0.01$ & $0.02$ & $0.453$ \\ 
\texttt{incBar*t} & $23.15$ & $24.13$ & $0.338$ \\ 
\texttt{DxV0*t}$^2$ & $-0.81$ & $0.38$ & $0.035$ \\ 
\texttt{xV0} & $0.36$ & $0.20$ & $0.065$ \\ 
\hline
\end{tabular}
\end{center}
\label{tab:violent_dml}
\end{table}

\begin{table}
\caption{Final DL outcome model for property crime including covariates found to be related to either the treatment or/outcome. For brevity the intercept and dummy year indicators are omitted}
\begin{center}
\begin{tabular}{|l|ccc|} \hline
  & Estimate & Std. Error & $p$-value \\
  \texttt{abortion} & $-0.04$ & $0.04$ & $0.404$ \\ 
  \texttt{Lpolice} & $-0.02$ & $0.01$ & $0.123$ \\ 
  \texttt{Linc} & $41.62$ & $9.36$ & $<0.001$ \\ 
  \texttt{Linc0} & $-18.25$ & $9.12$ & $0.046$ \\ 
  \texttt{Dinc0*t} & $-19.77$ & $21.13$ & $0.350$ \\ 
  \texttt{Dbeer0*t} & $-0.73$ & $0.59$ & $0.214$ \\ 
  \texttt{Linc0*t} & $222.64$ & $256.72$ & $0.386$ \\ 
  \texttt{Lprison0}$^2$\texttt{*t} & $0.04$ & $0.04$ & $0.314$ \\ 
  \texttt{Linc0}$^2$\texttt{*t} & $-1144.36$ & $1295.88$ & $0.378$ \\ 
  \texttt{Lprison0}$^2$\texttt{*t}$^2$ & $-0.02$ & $0.04$ & $0.567$ \\ 
  \texttt{Lbeer0}$^2$\texttt{*t}$^2$ & $-0.03$ & $0.21$ & $0.878$ \\ 
  \texttt{incBar} & $-22.22$ & $11.98$ & $0.064$ \\ 
  \texttt{afdcBar} & $-0.02$ & $0.02$ & $0.125$ \\ 
  \texttt{xP0} & $0.00$ & $0.03$ & $0.967$ \\ 
\hline
\end{tabular}
\end{center}
\label{tab:property_dml}
\end{table}

\begin{table}
\caption{Final DL outcome model for murder including covariates found to be related to either the treatment or/outcome. For brevity the intercept and dummy year indicators are omitted}
\begin{center}
\begin{tabular}{|l|ccc|} \hline
  & Estimate & Std. Error & $p$-value \\
\texttt{abortion} & $-0.12$ & 0.46 & 0.800 \\ 
\texttt{Lur} & $-0.35$ & 0.78 & 0.648 \\ 
\texttt{Dur0}$^2$ & 1.11 & 131.86 & 0.993 \\ 
\texttt{Lprison0*t} & 0.02 & 0.04 & 0.696 \\ 
\texttt{Linc0*t} & 0.52 & 62.67 & 0.993 \\ 
\texttt{Dbeer0*t}$^2$ & $-0.32$ & 4.07 & 0.938 \\ 
\texttt{incBar*t} & $-7.48$ & 62.34 & 0.905 \\ 
\texttt{xM0} & 2.76 & 3.76 & 0.464 \\ 
\texttt{xM0*t} & $-4.74$ & 5.85 & 0.418 \\ 
  \hline
\end{tabular}
\end{center}
\label{tab:murder_dml}
\end{table}

\begin{table*}
 \centering
 \def\~{\hphantom{0}}
 \begin{minipage}{175mm}
\begin{tabular}{rrrrc}  \hline
\multicolumn{5}{c}{Violent crime} \\
 & $\E(\beta_j \mid \by)$ & 2.5\% & 97.5\% & $P(\beta_j \neq 0 \mid \by)$ \\ \hline
  \texttt{Lpolice} & $0.00$ & 0.00 & 0.00 & 0.12 \\ 
  \texttt{Dprison*Dur} & $-0.99$ & $-14.96$ & 0.00 & 0.12 \\ 
  \texttt{Dprison*Dur*t} & $-3.67$ & $-34.15$ & 0.00 & 0.19 \\ 
  \texttt{Dprison*Dpov*t} & 1.05 & 0.00 & 10.56 & 0.19 \\ 
  \texttt{Dprison*Dpov*t}$^2$ & 2.30 & 0.00 & 13.96 & 0.25 \\ 
  \texttt{Dprison0} & $-0.03$ & $-0.19$ & 0.00 & 0.22 \\ 
\hline
\multicolumn{5}{c}{Property crime} \\
  \texttt{Lur} & $-0.06$ & $-0.88$ & 0.00 & 0.18 \\ 
  \texttt{Linc} & 10.61 & 0.00 & 48.54 & 0.33 \\ 
  \texttt{Linc}$^2$ & 50.43 & 0.00 & 241.44 & 0.34 \\ 
  \texttt{Linc0} & $-2.15$ & $-24.57$ & 0.00 & 0.18 \\ 
  \texttt{Linc0}$^2$ & $-9.79$ & $-121.24$ & 0.00 & 0.17 \\ 
  \texttt{incBar} & $-11.27$ & $-42.48$ & 0.00 & 0.26 \\ 
  \texttt{afdcBar} & $0.00$ & $-0.05$ & 0.00 & 0.16 \\ 
  \texttt{incBar}$^2$ & $-42.53$ & $-208.84$ & 0.00 & 0.22 \\ 
  \texttt{afdcBar}$^2$ & $0.00$ & $-0.03$ & 0.00 & 0.12 \\ 
   \hline
\multicolumn{5}{c}{Murder} \\
  \texttt{Dinc}$^2$ & $-385429.57$ & $-645172.36$ & 0.00 & 0.80 \\ 
\hline
\label{tab:abortion_bma_normal}
\end{tabular}
\end{minipage}
\vspace*{-6pt}
\end{table*}

\begin{table*}
 \centering
 \def\~{\hphantom{0}}
\caption{BMA inference (posterior mean, 0.95 interval and inclusion probability) under MOM prior for abortion data. Covariates with posterior marginal inclusion probability $>0.1$}
\begin{tabular}{rrrrc}  \hline
\multicolumn{5}{c}{Violent crime} \\
 & $E(\beta_j \mid \by)$ & 2.5\% & 97.5\% & $P(\beta_j \neq 0 \mid \by)$ \\ \hline
  \texttt{Lprison0*t} & 0.02 & 0.00 & 0.15 & 0.19 \\ 
  \texttt{prisonBar*t}$^2$ & $-0.03$ & $-0.21$ & 0.00 & 0.19 \\ 
\hline
\multicolumn{5}{c}{Property crime} \\
  \texttt{Dinc}$^2$ & $-305131.21$ & $-674405.53$ & 0.00 & 0.62 \\ 
  \texttt{Dinc}$^2$\texttt{*t} & $-409899.71$ & $-3635820.19$ & 0.00 & 0.15 \\ 
  \texttt{Dinc}$^2$\texttt{*t}$^2$ & 396962.96 & 0.00 & 3846788.32 & 0.12 \\ 
\hline
\multicolumn{5}{c}{Murder} \\
  \texttt{Dinc}$^2$ & -305131.21 & $-674405.53$ & 0.00 & 0.62 \\ 
  \texttt{Dinc}$^2$\texttt{*t} & $-409899.71$ & $-3635820.19$ & 0.00 & 0.15 \\ 
  \texttt{Dinc}$^2$\texttt{*t}$^2$ & 396962.96 & 0.00 & 3846788.32 & 0.12 \\ 
   \hline
\label{tab:abortion_bma_mom}
\vspace*{-6pt}
\end{tabular}
\end{table*}

\begin{table*}
 \centering
 \def\~{\hphantom{0}}
\caption{CIL inference (posterior mean, 0.95 interval and inclusion probability) under normal prior for abortion data. Covariates with posterior marginal inclusion probability $>0.1$ for property crime and murder (there are none for violent crime)}
\label{tab:abortion_cil_normal}
\begin{tabular}{rrrrc}  \hline
\multicolumn{5}{c}{Property crime} \\
 & $E(\beta_j \mid \by)$ & 2.5\% & 97.5\% & $P(\beta_j \neq 0 \mid \by)$ \\ \hline
  \texttt{Linc0*t} & $-2.03$ & $-14.13$ & 0.00 & 0.22 \\  \hline
\multicolumn{5}{c}{Murder} \\
  \texttt{Dinc}$^2$ & $-227669.02$ & $-614607.65$ & 0.00 & 0.48 \\ 
   \hline
\end{tabular}
\vspace*{-6pt}
\end{table*}

\begin{table*}
 \centering
 \def\~{\hphantom{0}}
\caption{CIL inference (posterior mean, 0.95 interval and inclusion probability) under MOM prior for abortion data. Covariates with posterior marginal inclusion probability $>0.1$ for property crime and murder (there are none for violent crime)}
\begin{tabular}{rrrrr}  \hline
\multicolumn{5}{c}{Property crime} \\
 & $E(\beta_j \mid \by)$ & 2.5\% & 97.5\% & $P(\beta_j \neq 0 \mid \by)$ \\ \hline
  \texttt{afdcBar}$^2$\texttt{*t} & $-0.01$ & $-0.07$ & 0.00 & 0.38 \\ 
  \texttt{incBar}$^2$\texttt{*t}$^2$ & $-40.58$ & $-97.31$ & 0.00 & 0.40 \\ 
\hline
\multicolumn{5}{c}{Murder} \\
  \texttt{Dinc}$^2$ & $-91217.96$ & $-604869.53$ & 0.00 & 0.18 \\ 
   \hline
\end{tabular}
\label{tab:abortion_cil_mom}
\vspace*{-6pt}
\end{table*}

\subsection{Further results for the simulation study}

\begin{figure}[htbp]
\centering
\begin{tabular}{ccc}
\multicolumn{3}{c}{Bias$^2$} \\
$\alpha = 1$ & $\alpha = 1/3$ & $\alpha = 0$ \\
\includegraphics[scale=0.54]{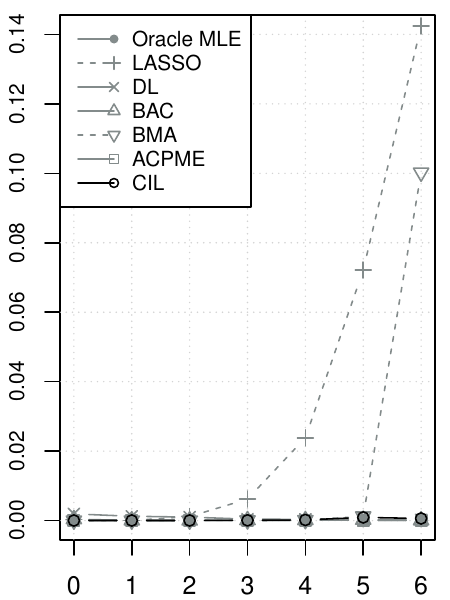} &
\includegraphics[scale=0.54]{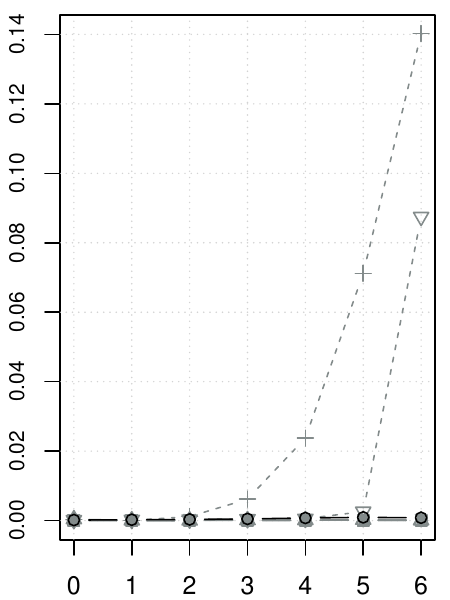} & 
\includegraphics[scale=0.54]{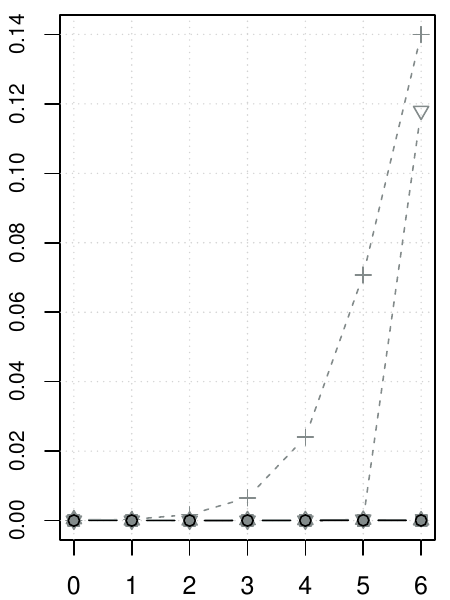}\\
\multicolumn{3}{c}{Variance} \\
$\alpha = 1$ & $\alpha = 1/3$ & $\alpha = 0$ \\
\includegraphics[scale=0.54]{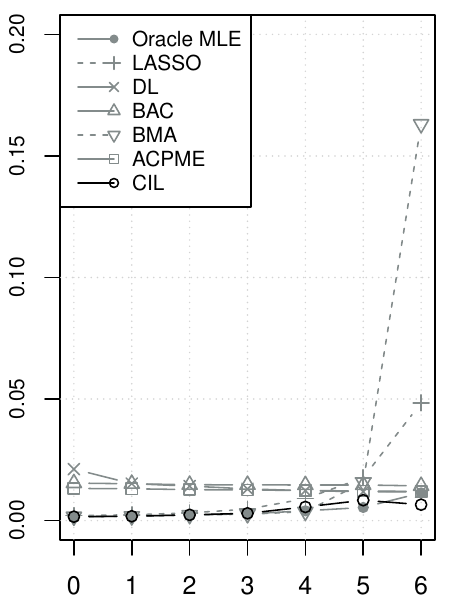}  & 
\includegraphics[scale=0.54]{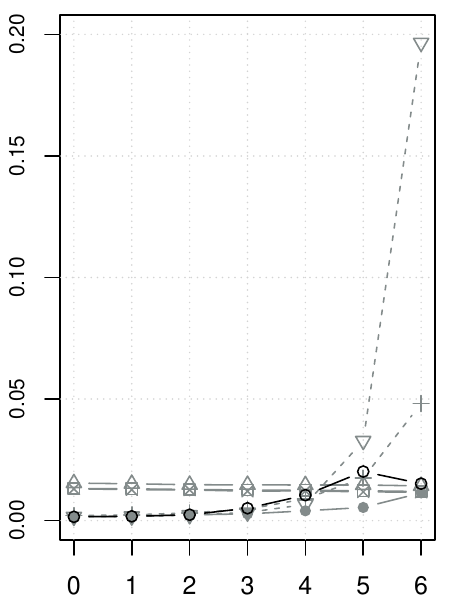}   & 
\includegraphics[scale=0.54]{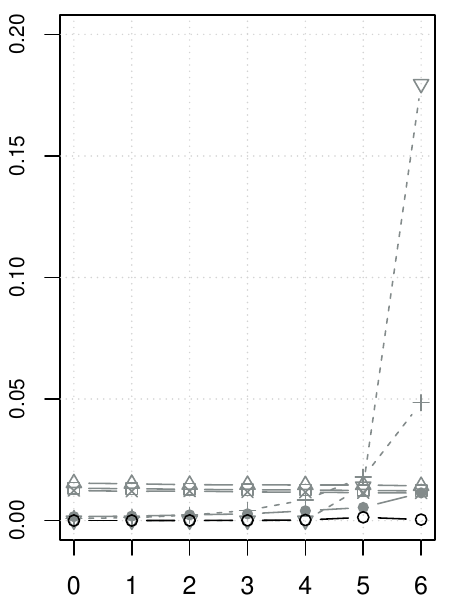} \\
\end{tabular}
\caption{ Squared bias and variance for the simulation scenarios described in Figure 1 
considering strong ($\alpha=1$), weak $(\alpha=1/3)$ and  no effect $(\alpha=0)$.
 The x-axis quantifies the amount of confounding, measured by the number of covariates that are truly related to both the outcome and the treatment} 
\label{fig:sim_bias_variance}
\end{figure}

\begin{figure}[htbp]
\centering
\begin{tabular}{ccc}
$\alpha = 1$ & $\alpha = 1/3$ & $\alpha = 0$ \\
\includegraphics[scale=0.54]{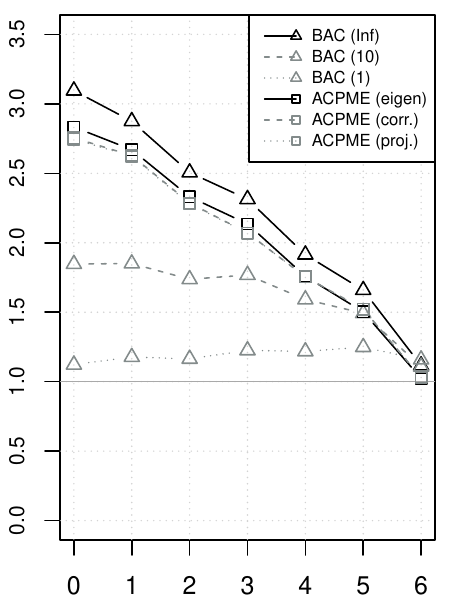}  & 
\includegraphics[scale=0.54]{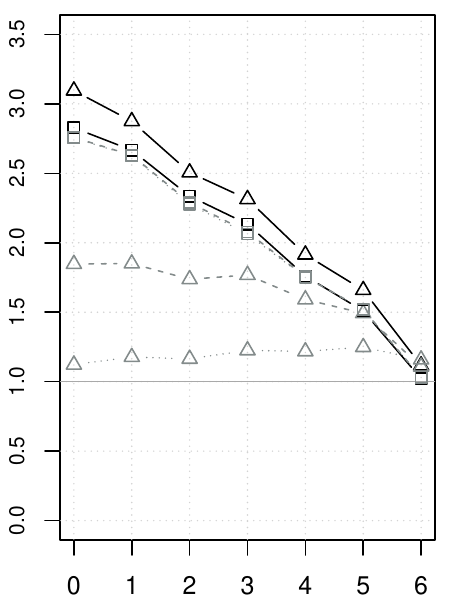}   & 
\includegraphics[scale=0.54]{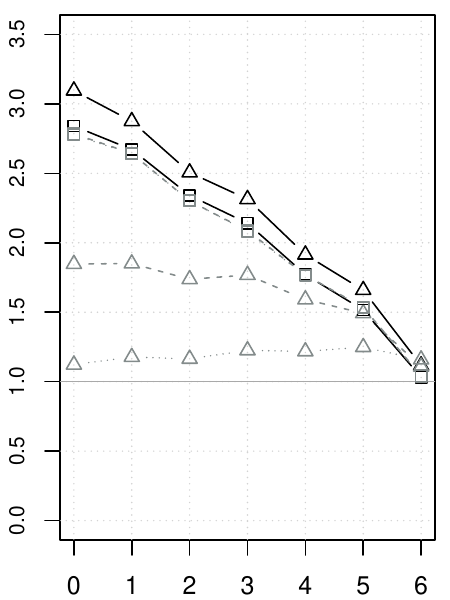} \\
\end{tabular}
\caption{ Sensitivity to tuning parameter in BAC and ACPME.
Parameter root MSE relative to an oracle OLS for the simulation scenarios described in Figure 1 
considering strong ($\alpha=1$), weak $(\alpha=1/3)$ and  no effect $(\alpha=0)$}
\label{fig:sensitivity_bac_acpme}
\end{figure}

\begin{figure}[htbp]
\centering
\begin{tabular}{ccc}
$\alpha = 1$ & $\alpha = 1/3$ & $\alpha = 0$ \\
\includegraphics[scale=0.54]{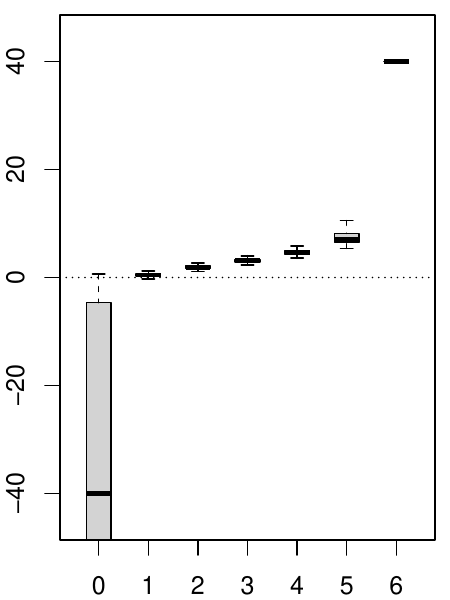} &
\includegraphics[scale=0.54]{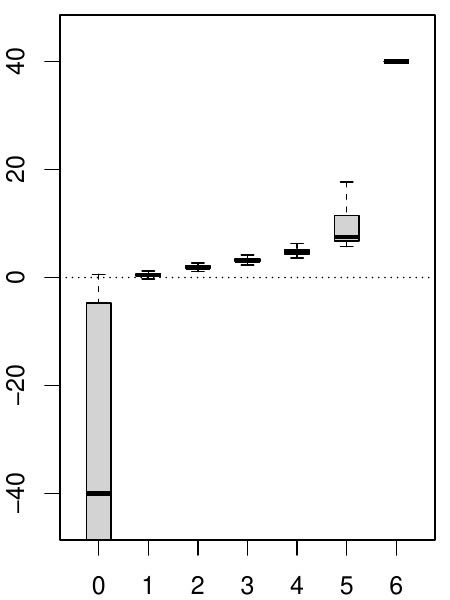} & 
\includegraphics[scale=0.54]{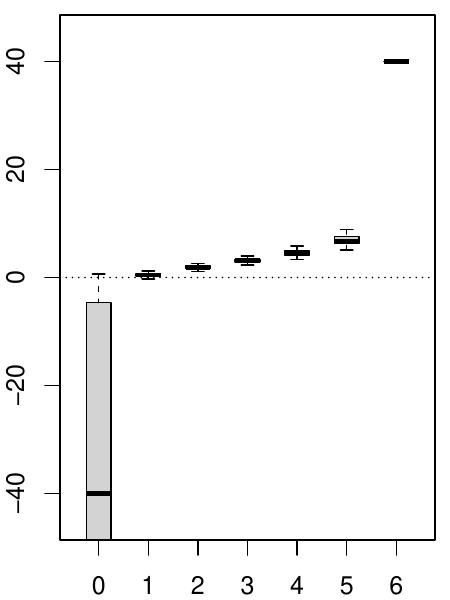}
\end{tabular}
\caption{ 
Distribution of the estimated CIL hyper-parameter $\hat{\theta}_1$ for the simulations described in Figure 1, 
considering strong ($\alpha=1$), weak $(\alpha=1/3)$ and  no effect $(\alpha=0)$}
\label{fig:cil_thetahat}
\end{figure}

\begin{figure}[htbp]
\centering
\begin{tabular}{ccc}
$\alpha = 1$ & $\alpha = 1/3$ & $\alpha = 0$ \\
\includegraphics[scale=0.54]{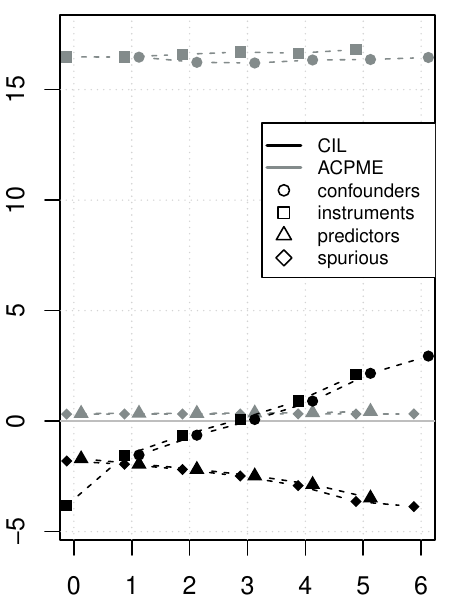} &
\includegraphics[scale=0.54]{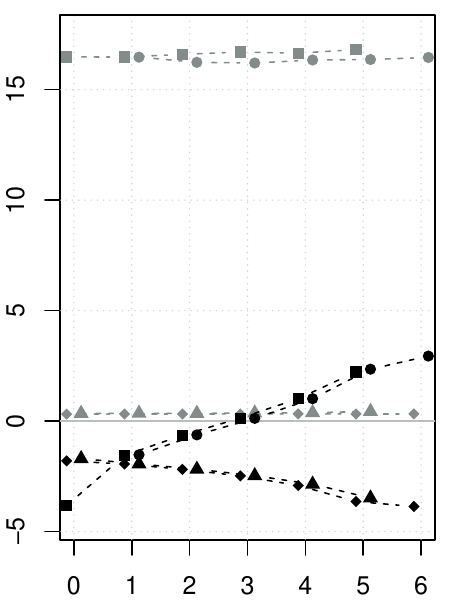} & 
\includegraphics[scale=0.54]{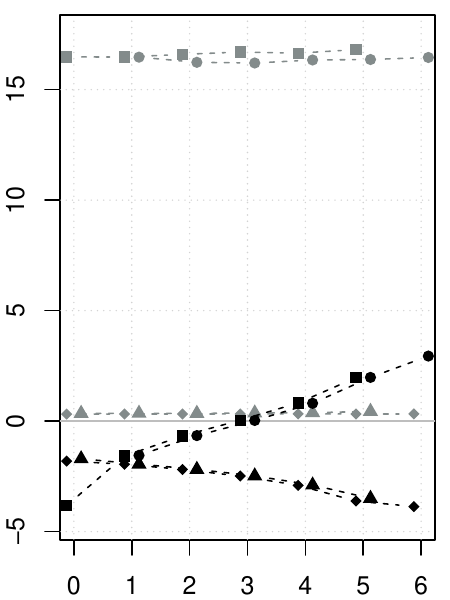}
\end{tabular}
\caption{ Comparison of the CIL and ACPME prior inclusion log-odds ($\textup{logit} \pi_j(\btheta)$) for the simulations described in Figure 1, 
considering strong ($\alpha=1$), weak $(\alpha=1/3)$ and no effect $(\alpha=0)$. The y-axis shows the mean $\textup{logit} \pi_j(\btheta)$ 
per covariate type (confounders, instruments, predictors ---active on the outcome but not on the treatment---, and spurious covariates), and the x-axis is the number of confounders (0 for no confounding, 6 for full confounding). 
Points in the figure are slightly offset on the $x$-axis when necessary to improve readability. }
\label{fig:compare_cil_acpme_logodds}
\end{figure}

\begin{figure}[htbp]
\centering
\begin{tabular}{ccc}
\multicolumn{3}{c}{Proportion of confounders} \\
$\alpha = 1$ & $\alpha = 1/3$ & $\alpha = 0$ \\
\includegraphics[scale=0.54]{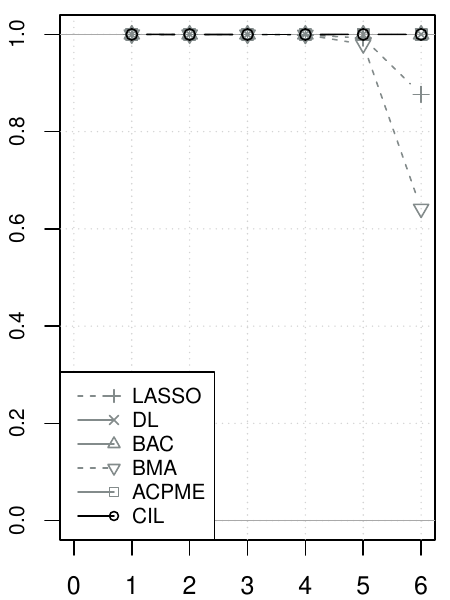}  & 
\includegraphics[scale=0.54]{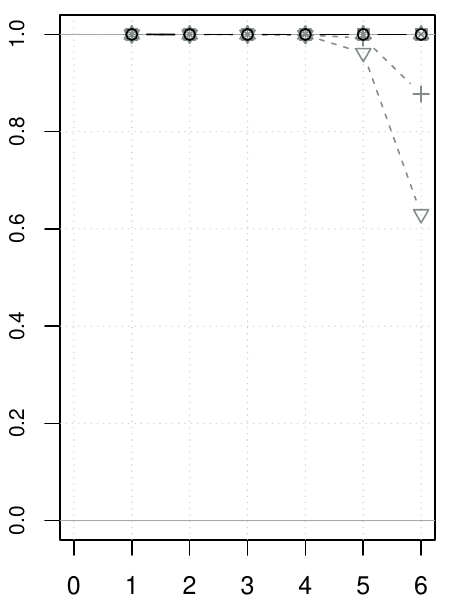}   & 
\includegraphics[scale=0.54]{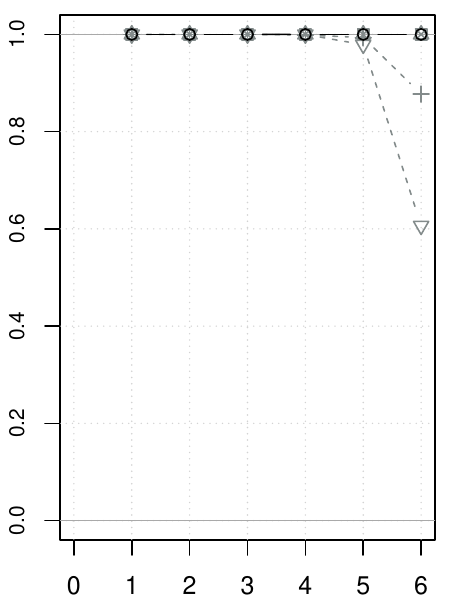} \\
\multicolumn{3}{c}{Proportion of instruments} \\
$\alpha = 1$ & $\alpha = 1/3$ & $\alpha = 0$ \\
\includegraphics[scale=0.54]{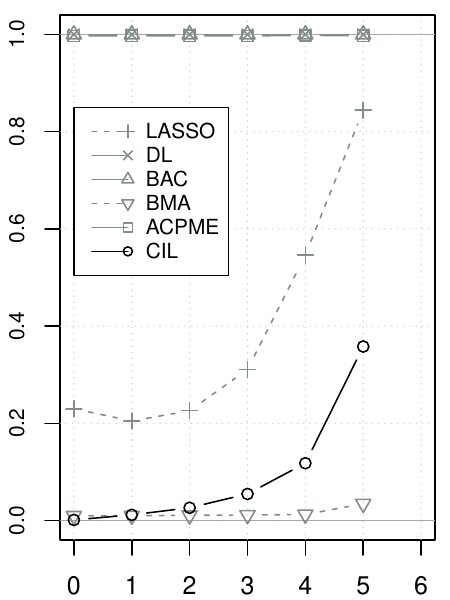}  & 
\includegraphics[scale=0.54]{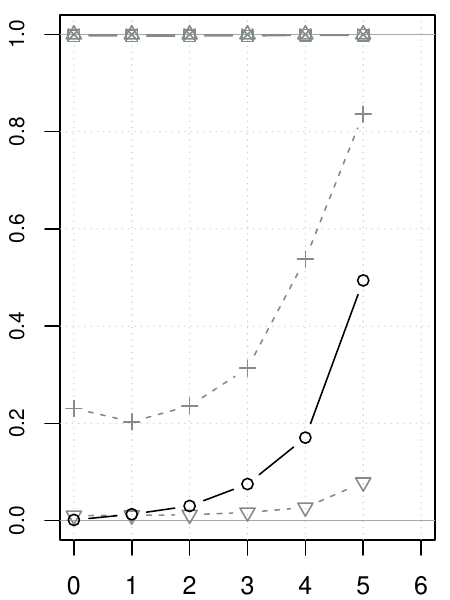}   & 
\includegraphics[scale=0.54]{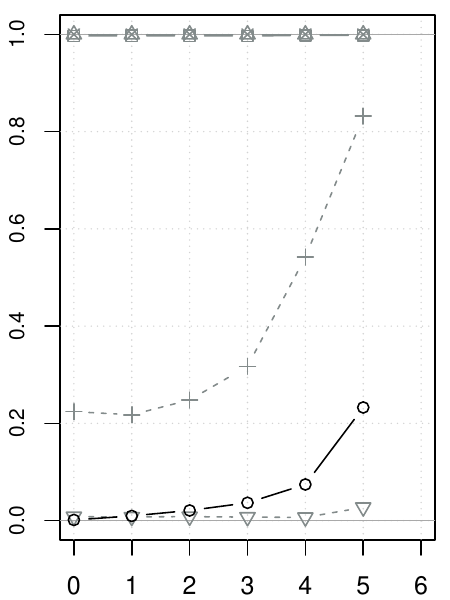} \\
\end{tabular}
\caption{Proportion of confounders (top panels) and instruments (bottom panels) selected by each method in the simulation scenarios described in Figure 1, 
considering strong ($\alpha=1$), weak $(\alpha=1/3)$ and  no effect $(\alpha=0)$.
For Bayesian methods, we report the average marginal posterior inclusion probability.
The x-axis quantifies the amount of confounding, measured by the number of covariates that are truly related to both the outcome and the treatment} 
\label{fig:prop_instruments_confounders}
\end{figure}

\begin{figure}[h]
\centering
\begin{tabular}{ccc}
$\alpha = 1$ & $\alpha = 1/3$ & $\alpha = 0$ \\
\includegraphics[width=0.31\textwidth]{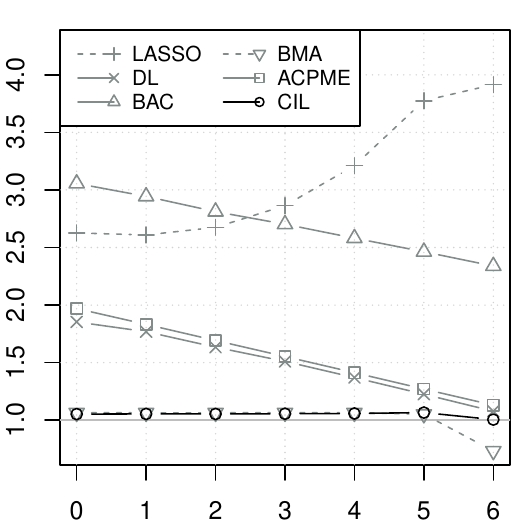} &
\includegraphics[width=0.31\textwidth]{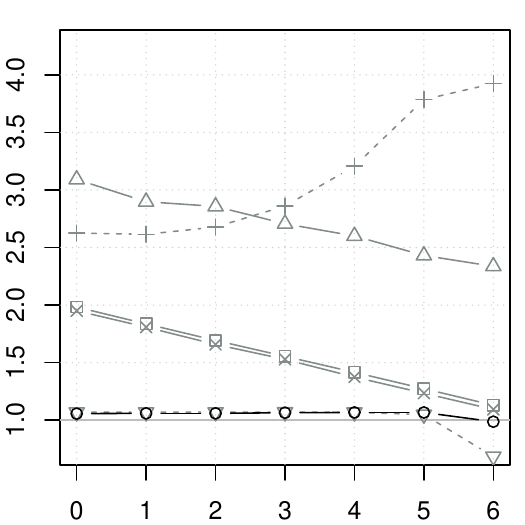} &
\includegraphics[width=0.31\textwidth]{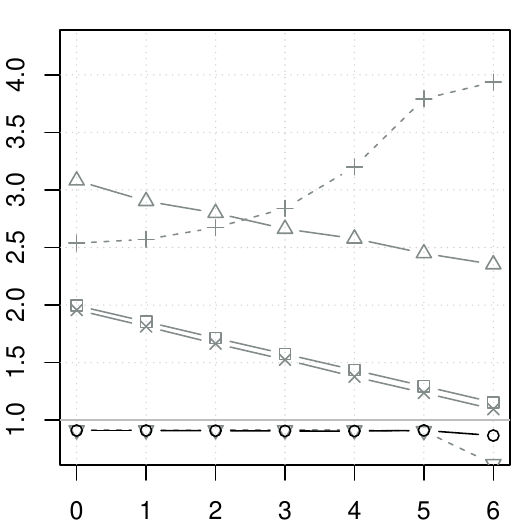} \\
\includegraphics[width=0.31\textwidth]{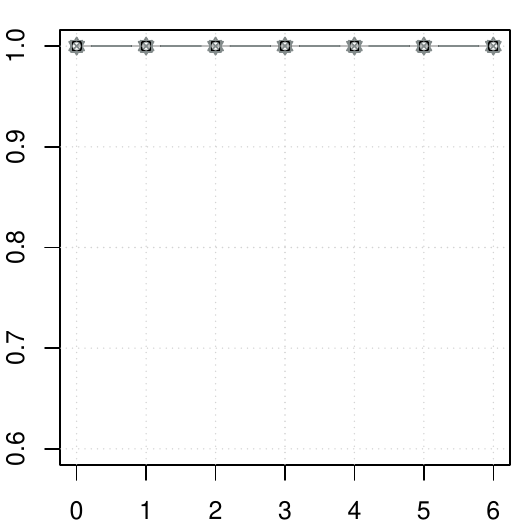} &
\includegraphics[width=0.31\textwidth]{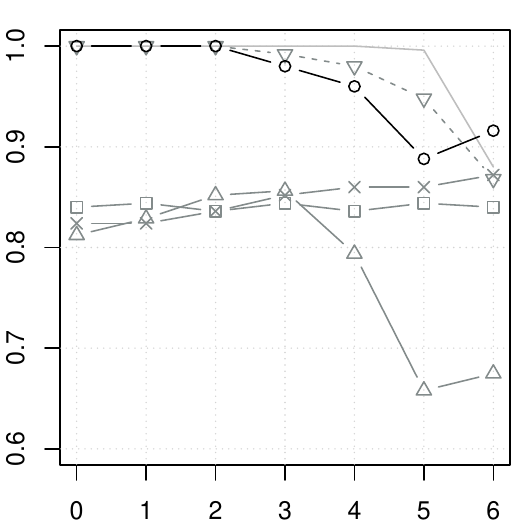} &
\includegraphics[width=0.31\textwidth]{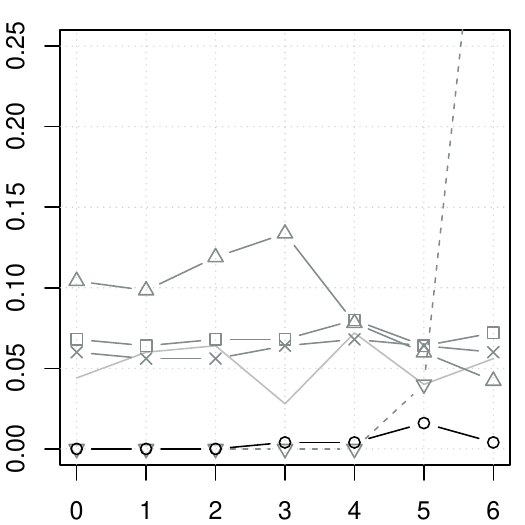} \\
\vspace{0.2cm}
\end{tabular}
\caption{To be read vertically in relation to Fig. 1. 
The top panels show the average outcome model size across levels of confounding, divided by the true model size (i.e. 1 indicates that it matches the true model size). The bottom panels show the probability of selecting the treatment using a 0.05 $p$-value cut-off for DL, and for Bayesian methods the treatment is included when marginal posterior inclusion probability is $>$1/2. The LASSO does not appear in these panels as its not designed for inference.}
\label{fig:fig1b}
\end{figure}

 In this section we expand upon the single treatment simulation results shown in Section 5.1.
Figure \ref{fig:sim_bias_variance} decomposes the mean squared errors of all methods shown in Figure 1 
into the corresponding squared bias and variance. Standard high-dimensional methods like LASSO and BMA suffer from high bias and variance. In contrast, specialized treatment effect methods like BAC, ACPME and double LASSO show little bias but suffer from higher variance, particularly in low confounding scenarios.

Figure \ref{fig:sensitivity_bac_acpme} assesses the sensitivity of BAC and ACPME to their respective hyperparameters.
The performance of BAC is very sensitive to its tuning parameter.
For BAC, setting the tuning parameter to $\omega=\infty$ means that, for any covariate found to be associated with the treatment, one forces its inclusion into the outcome model. $\omega=1$ means that inclusion in the treatment and outcome models is independent a priori, whereas $\omega=10$ represents a middle ground between the two other hyper-parameter choices.
Regarding ACPME, its performance was fairly robust to its tuning parameter choice (related to the use of eigenvalues, correlations or a projection to measure associations).

Figure \ref{fig:cil_thetahat} shows the distribution of the CIL hyper-parameter $\hat{\theta}_1$ for the same simulation scenarios. Recall that $\theta_1>0$ is interpreted as high confounding, $\theta=0$ as neutral confounding, and $\theta_1<0$ as no confounding. As expected, regardless of the treatment effect size $\alpha$, CIL estimates $\hat{\theta}<0$ when there is no overlap between covariates that truly affect the outcome and those that truly affect the treatment (no confounding), and it estimates larger $\hat{\theta}_1$ as said overlap increases (up to full confounding, when all 6 truly active covariates in both equations overlap).

Figure \ref{fig:compare_cil_acpme_logodds} compares prior inclusion probabilities between CIL and ACPME, to illustrate their key distinction: the former adapts to the true amount of confounding (by using the outcome data) and the latter does not. 
ACPME favors equally the inclusion of confounders and of instruments, relative to predictors (covariates only associated to the outcome) and spurious covariates. 
Critically, this occurs to the same extent regardless of whether there truly is no confounding or high confounding.
CIL, on the other hand, adapts to the true level of confounding. 
Under high confounding where there are more confounders than instruments (x-axis $\geq 5$ in Figure \ref{fig:compare_cil_acpme_logodds}), inclusion of these two covariate types is encouraged.
Under low confounding, where there are more instruments than confounders (x-axis $\leq 1$), their inclusion is disencouraged.
Note also that CIL discouraged the inclusion of covariates unrelated to the treatment (whether or not related to the outcome), particularly in high confounding, this is because CIL also learns the overall level of sparsity via the intercept $\theta_{0}$.  


Figure \ref{fig:prop_instruments_confounders} (top panels) shows the proportion of confounders selected by each method. In high-confounding scenarios, BMA and to a lesser extent LASSO fail to include a fraction of the confounders, explaining the higher bias and variance observed in Figure \ref{fig:sim_bias_variance}, whereas the remaining methods include all counfounders. In settings with less or no confounding, all methods successfully included all confounders.
The bottom panels show that double LASSO, BAC and ACPME included essentially all instruments, explaining their higher variance in Figure \ref{fig:sim_bias_variance}. LASSO also included a fair fraction of instruments, whereas CIL and BMA included less. BMA was particularly effective in this regard. Essentially, by inducing sparse solutions it includes less instruments, at the cost of missing some confounders.

Figure \ref{fig:fig1b} summarizes model selection results for the simulations in Figure 1. 

\subsection{Simulations under growing dimensionality ($T=1$)}

\begin{figure}[h]
\centering
\begin{tabular}{ccc}
$n = 50$ & $n = 100$ & $n = 100$ \\
$J+T = 25$ & $J+T = 100$ & $J+T = 200$ \\
\includegraphics[width=0.31\textwidth]{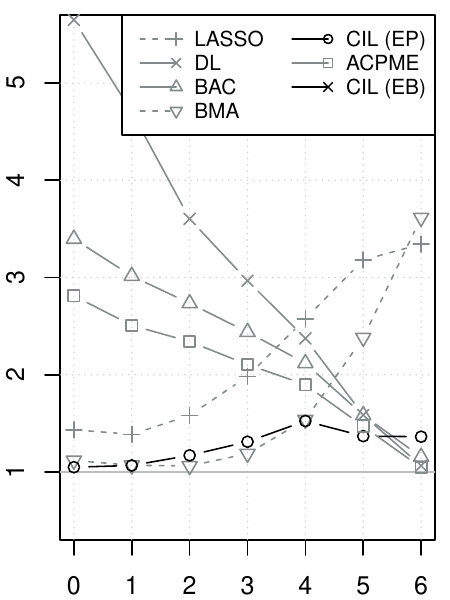} &
\includegraphics[width=0.31\textwidth]{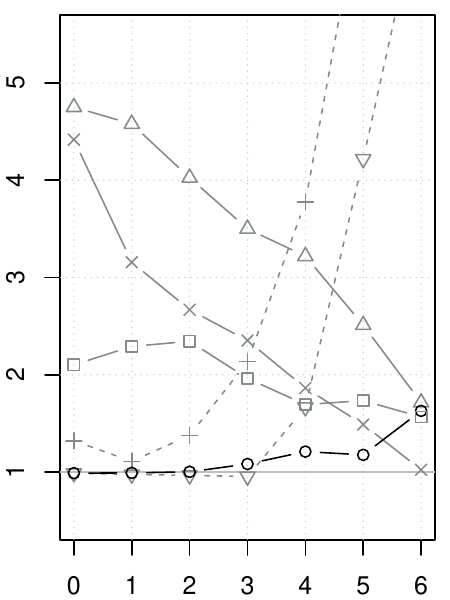} &
\includegraphics[width=0.31\textwidth]{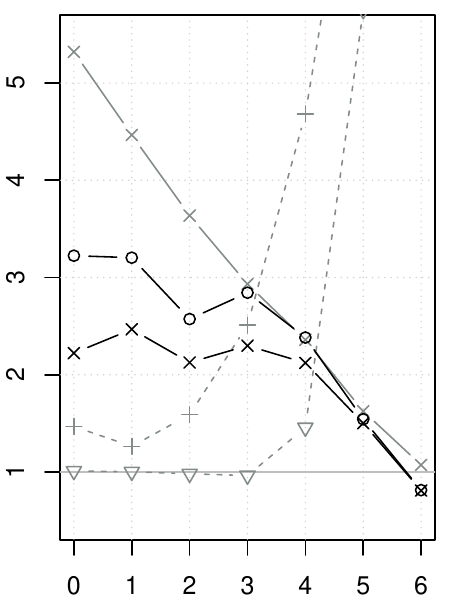} \\
\vspace{0.2cm}
\end{tabular}
\caption{Single treatment parameter RMSE (relative to oracle OLS) based on $R=250$ simulated datasets for each level of confounding.
In all panels, $\alpha = 1$ and $|\bgamma|_{0} = 6$.
We show the empirical Bayes version CIL only in the right panel, for the other panels results are undistinguishable relative to EP.}
\label{fig:singletreat_growingdim}
\end{figure}
Figure \ref{fig:singletreat_growingdim} studies the effect of growing number of covariates on inference, specifically for $J+T=25$, 100 and 200.

\subsection{Testing CIL to different amounts of confounders for $T=1$}

\begin{figure}[h]
\centering
\begin{tabular}{ccc}
$\norm{\bgamma}_{0} = 6$ & $\norm{\bgamma}_{0} = 12$ & $\norm{\bgamma}_{0} = 18$ \\
\includegraphics[width=0.31\textwidth]{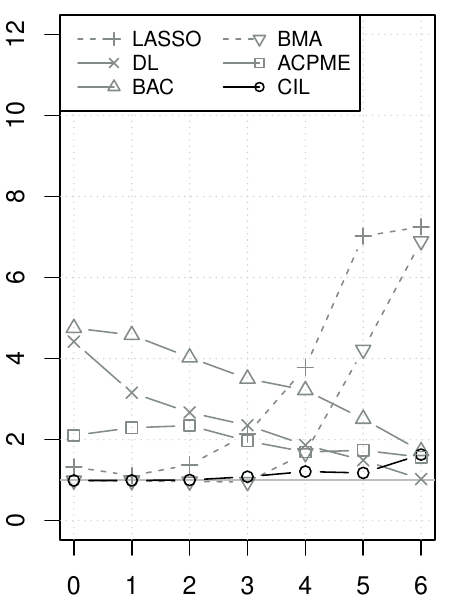} &
\includegraphics[width=0.31\textwidth]{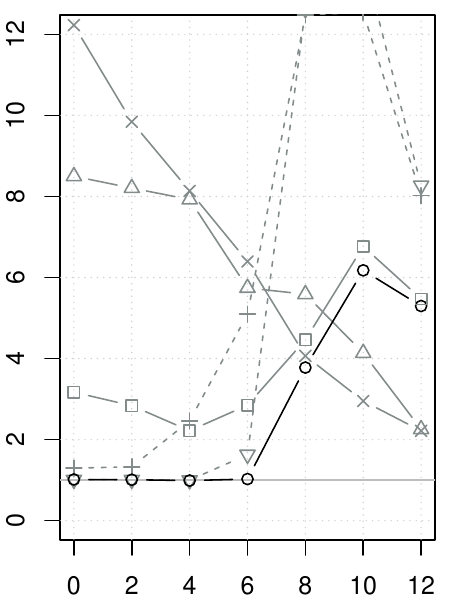} &
\includegraphics[width=0.31\textwidth]{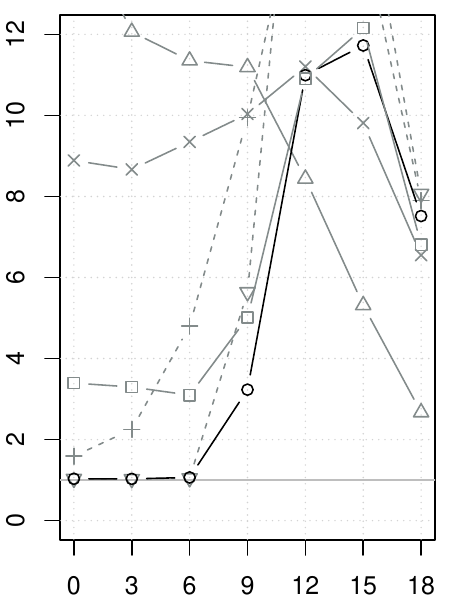} \\
\vspace{0.2cm}
\end{tabular}
\caption{Single treatment parameter RMSE (relative to oracle OLS) based on $R=250$ simulated datasets for each level of confounding reported, as described in Figure \ref{fig:intro2}. In all panels, $n=100$, $J+T=100$ and $\alpha = 1$. Sudden general improvement at the right end of center and right panels is due to a sharper deterioration of oracle OLS RMSE at complete confounding relative to other methods.}
\label{fig:fig3}
\end{figure}

Figure \ref{fig:fig3} shows the effect of having various amounts of active confounders. The results look consistent to the effects reported in Figures 1 
and \ref{fig:singletreat_growingdim}, which are magnified for large amounts of active confounders. These are really challenging situations to tackle since the tested methods aim at model sparsity, while the true model size is relatively large. Although our method still performed at oracle rates in low-confounding scenarios, its relative performance is compromised for the highest levels of confounding. This occurred in part because accurate point estimation in (2.4) 
became increasingly harder as the correlation between covariates strengthened, which in turn influenced the ability of the algorithm to calibrate $\btheta$ reliably. Even in these hard cases, however, its performance is not excessively far to the best competing method, while it clearly outperformed BMA on all of them.

\bibliographystyle{plainnat}
\bibliography{references}

\end{document}